\documentclass[twocolumn]{aastex63}

\usepackage{amssymb}
\usepackage{amsfonts}
\usepackage{amsmath}
\usepackage{multirow}
\usepackage[varg]{txfonts}
\usepackage{natbib}
\usepackage{color} 
\usepackage{graphicx}
\usepackage{hyperref}
\usepackage{multirow}
\usepackage{lineno}

\newcommand{\trm}[1]{\textrm{#1}}

\newcommand{\dg}[0]{$^\dagger$}
\newcommand{\mrow}[2]{\multirow{#1}{*}{#2}}
\newcommand{\kmps}{km\,s$^{-1}$}

\graphicspath{}

\received{?????, 2020}
\accepted{?????, 2020}

\begin{document}

\title{Supernova fallback as origin of neutron star spins and spin-kick alignment}
\shorttitle{Supernova fallback and spin-kick alignment}

\correspondingauthor{Hans-Thomas Janka}
\email{thj@mpa-garching.mpg.de}

\author[0000-0002-0831-3330]{Hans-Thomas~Janka}
\affiliation{Max Planck Institute for Astrophysics, Karl-Schwarzschild-Str.~1, 
85748 Garching, Germany}

\author{Annop Wongwathanarat}
\affiliation{Max Planck Institute for Astrophysics, Karl-Schwarzschild-Str.~1, 
85748 Garching, Germany}

\author{Michael Kramer}
\affiliation{Max-Planck-Institut f{\"u}r Radioastronomie, Auf dem H{\"u}gel~69,
D-53121 Bonn, Germany}
\affiliation{Jodrell Bank Centre for Astrophysics, University of Manchester, M13 9PL, UK}

\shortauthors{H.-Thomas~Janka, Annop Wongwathanarat, and Michael Kramer}

\begin{abstract}
Natal kicks and spins are characteristic properties of
neutron stars (NSs) and black holes (BHs). Both offer valuable clues to
dynamical processes during stellar core collapse and explosion.
Moreover, they influence the evolution of stellar multiple systems and the
gravitational-wave signals from their inspiral and merger.
Observational evidence of possibly generic spin-kick alignment
has been interpreted as indication that NS spins are either
induced with the NS kicks or inherited from progenitor rotation, which thus
might play a dynamically important role during stellar collapse. 
Current three-dimensional supernova simulations suggest that NS kicks are 
transferred in the first seconds of the explosion, mainly by anisotropic
mass ejection and, on a secondary level, anisotropic neutrino emission.
In contrast, the NS spins are only determined minutes to hours later by
angular momentum associated with fallback of matter that does not
become gravitationally unbound in the supernova.
Here, we propose a novel scenario to explain spin-kick alignment as a
consequence of tangential vortex flows in the fallback matter that is 
accreted mostly from the direction of the NS's motion. For this effect the
initial NS kick is crucial, because it produces a growing offset of the
NS away from the explosion center, thus promoting onesided accretion.
In this new scenario conclusions based on traditional concepts are
reversed. For example, pre-kick NS spins are not required, and
rapid progenitor-core rotation can hamper spin-kick alignment.
We also discuss implications for natal BH kicks and the possibility
of tossing the BH's spin axis during its formation.
\end{abstract}


\section{Introduction}\label{sec:intro}

Neutron stars are born with estimated typical rotation periods between around
10 milliseconds and hundreds of milliseconds 
\citep{Chevalier2005,Popov+2012,Igoshev+2013,Noutsos+2013}.
The origin of these spins is still unclear.
Various different explanations have been proposed: Inherited angular momentum
from the collapsing iron core of the progenitor star 
\citep[e.g.,][]{Ott+2006,Heger+2005}, possibly amplified by inward transport of angular
momentum and associated spin-up of slowly rotating stellar cores
through internal gravity waves during the progenitor evolution 
(\citealt{Fuller+2014}, \citealt{Fuller+2015}, \citealt{Ma+2019}, but see
\citealt{McNeill+2020} for counter-arguments);
spin-up of the newly formed proto-neutron star during
the pre-explosion phase by anisotropic impacts of accretion downflows
\citep{Wongwathanarat+2013}; spin-up or spin-down by triaxial instabilities 
such as spiral modes of the stalled supernova shock due to the standing accretion 
shock instability (spiral SASI; \citealt{Blondin+2007,Fernandez2010,Rantsiou+2011,Guilet+2014,Kazeroni+2016,Moreno+2016})
or low-$T/W$ spiral waves \citep{Kazeroni+2017};
or the accretion of angular momentum associated with turbulent mass motions in the 
infalling convective layers of the collapsing progenitor star
\citep{Gilkis+2014,Gilkis+2015,Gilkis+2016}.

It is extremely difficult to quantify any of the mentioned phenomena, 
individually or in combination, as origins of neutron star angular momentum or 
as contributing effects, because none of them is sufficiently well understood
in its exact consequences. In order to be quantitatively predictive,
the angular momentum evolution in aging stars including magnetic fields, for 
example, still requires a lot more theoretical work.
\citet{Heger+2005}, considering angular momentum transfer due to the Tayler-Spruit 
dynamo in single stars, came up with angular momentum estimates between 
$5\times 10^{47}$\,erg\,s and $4\times 10^{48}$\,erg\,s (corresponding to spin 
periods between 15\,ms and 3\,ms) for neutron stars in progenitors between 
12\,M$_\odot$ and 35\,M$_\odot$. Asteroseismic measurements, however, suggest 
that the Tayler-Spruit dynamo might considerably underestimate the angular 
momentum transport and the corresponding angular momentum loss in evolving 
stars. Accordingly, these and other observations reveal slower rotation
of stellar cores and white dwarfs than theoretically predicted, see, e.g.,
discussions by \citet{Fuller+2019}, \citet{Eggenberger+2019}, and 
\citet{TakahashiLanger2021} and references therein. On the other hand, 
counter-arguments against efficient angular momentum transport between the 
stellar cores and the radiative envelopes in massive stars have recently been
inferred from black hole spins deduced from gravitational-wave measurements of
binary black hole mergers \citep{Qin+2021}. These arguments, in turn, may 
be weakened by the suggestion that black holes might be spun up by the accretion
of random angular momentum from convective hydrogen envelopes~\citep{Antoni+2021}.

SASI spiral modes are another widely discussed source of neutron star angular
momentum. Early works concluded that neutron star angular momenta of up to a few
$10^{47}$\,erg\,s (corresponding to spin periods of 50--60\,ms) might be
reached for initially non-rotating or slowly rotating progenitor cores
\citep{Blondin+2007,Fernandez2010}. Later, \citet{Kazeroni+2017} found that
SASI spiral modes may spin neutron stars up to periods as low as 10\,ms 
(transferring angular momentum of up to $\sim$\,$10^{48}$\,erg\,s) if the rotation
inherited from the collapsing stellar core is slower than this value, and
to spin down neutron stars that receive more angular momentum than this limit
from the core rotation of their progenitor stars.
The presence of SASI spiral modes, however, is not an ubiquitous phenomenon in
collapsing stellar cores prior to the onset of an explosion. In particular
when the explosion sets in rather quickly (i.e., within only some 100\,ms after 
core bounce) and without a preceding phase of shock contraction, SASI
spiral modes are not common. SASI activity is weak and shock revival
occurs fairly rapidly in 3D supernova simulations 
that will be presented in our paper, compatible with simulation results reported 
in other publications \citep[e.g.,][]{Nordhaus+2010,Wongwathanarat+2010,
Wongwathanarat+2013,Burrows+2020,Powell+2020,Chan+2020}. Under such circumstances 
the angular momentum transferred to the neutron star before and during the 
onset of the explosion is relatively low, only a few times $10^{46}$\,erg\,s. In
contrast, post-explosion accretion and fallback accretion may deliver up to over
$10^{49}$\,erg\,s to the newly formed neutron star, as will be discussed in 
detail in our paper.

Here, we will therefore argue, based on recent results in the 
literature \citep{Powell+2020,Chan+2020,Stockinger+2020} and additional
ones presented in this work, that an important (possibly even the
dominant) process determining the birth spins of most neutron stars is likely
to be the anisotropic accretion of angular momentum associated with matter 
that is initially ejected during the early stages of the supernova explosion 
but does not get gravitationally unbound and thus falls back to the compact 
remnant. While all the phenomena
mentioned in the previous paragraphs may play a role under certain circumstances
and during some stages of the evolution of the collapsing stellar core and
new-formed proto-neutron star, the angular momentum of the fallback matter is 
large enough to overrule the early-time effects and to reset the neutron star spin 
to its final, potentially observable birth value. Although the amount of the
fallback matter is usually rather small (between some $\sim$\,$10^{-4}\,$M$_\odot$ 
and a few 0.1\,M$_\odot$) ---unless fallback leads to black hole formation
\citep[e.g.,][]{Ertl+2020,Woosley+2020,Chan+2018,Chan+2020}---
the angular momentum connected 
with the fallback material can be up to nearly $10^{50}$\,g\,cm$^2$\,s$^{-1}$, 
as we will show. It would thus be sufficient to lead to
millisecond and even sub-millisecond neutron star spin periods, if entirely 
accreted onto the neutron star. Such an outcome, however, would neither be 
compatible with the estimated rotation periods of new-born neutron stars
mentioned above, nor has the corresponding huge amount of spin-down energy
(up to several $10^{52}$\,erg) been observed in the majority of supernovae.
Therefore we will reason that the accretion is 
either incomplete or that the fallback is overestimated in the present spherically
symmetric (1D) or three-dimensional (3D)
simulations. We hypothesize that the cause for this overestimation is the 
fact that the neutron star in these simulations is fixed to the center of the
explosion, which coincides with the center of the computational grid, 
instead of being able to move out of the grid center because of its initial kick
velocity. 

This kick is imparted to the neutron star through
anisotropic mass ejection (hydrodynamic kick) and anisotropic 
neutrino emission (neutrino-induced kick). Consistent
with linear momentum conservation, the neutron star receives a momentum in
the opposite direction of the momentum carried away by the asymmetrically
ejected matter and anisotropically escaping neutrinos. The acceleration of
the neutron star takes place over several seconds at the beginning of the
supernova explosion, as we will also show for the 3D models discussed in this
paper. During roughly the first second, both hydrodynamic forces and 
gravitational forces accomplish the momentum exchange between ejecta and 
neutron star \citep{Scheck+2004,Scheck+2006,Nordhaus+2010,Nordhaus+2012,
Wongwathanarat+2010,Wongwathanarat+2013},
whereas on longer time scales the long-range and long-lasting gravitational
forces play the main role in transferring momentum from the ejecta 
to the neutron star. For this reason the acceleration that produces the hydrodynamic
neutron star kicks can also be considered as a ``gravitational tug-boat mechanism''
\citep{Wongwathanarat+2013}. In massive progenitors with high densities in their
collapsing cores and high mass accretion rates of the new-born neutron stars, the 
hydrodynamic kicks typically dominate the neutrino-induced kicks 
\citep[see][]{Bollig+2020}, whereas in low-mass progenitors with low mass 
accretion rates the neutrino-induced kicks can provide the leading contribution 
\citep[see][]{Stockinger+2020}. The kick velocities obtained in the simulations
are viable to explain the space velocities of young neutron stars and radio 
pulsars inferred from observations \citep[see, e.g.,][]{Lyne+1994,Lorimer+1997,
Cordes+1998,Arzoumanian+2002}

Fallback in supernovae has previously been considered as a source of 
post-explosion emission of electromagnetic radiation \citep{Chevalier1989},
neutrinos \citep{Houck+1991,Chevalier1995}, and gravitational waves 
\citep{Piro+2012,Sur+2021}; as an origin of disk
formation around neutron stars and black holes in successful and failed
supernovae and of associated light emission in the form of observable transients
\citep{Dexter+2013,Perna+2014,Quataert+2019,Moriya+2018,Moriya+2019};
as a power-source of supernovae in connection with accreting magnetars 
\citep{Piro+2011};
as a process to explain the trimorphism of neutron stars in rotation-powered 
pulsars, central compact objects, and magnetars \citep{Zhong+2021}; 
as kick mechanism of new-born black holes in fallback supernovae 
\citep{Janka2013,Chan+2020};
as a physical process leading to the spin-up of black holes formed
during the collapse of non-exploding supergiant stars \citep{Antoni+2021};
as a possible production site of r-process elements \citep{Fryer2006,Fryer+2006};
and as a mechanism of magnetic field amplification in the
surface layers of the new-born neutron star \citep{Soker2020}. However, to our 
best knowledge, the dynamics of anisotropic fallback and its dependence on the
neutron star kick in an asymmetric supernova explosion have not been discussed 
before. 

We emphasize that in the context of our paper, different from assumptions in 
some previous works \citep[e.g.,][]{Perna+2014}, the angular momentum associated 
with the anisotropic fallback is {\em not} mainly connected to the possible
rotation of the progenitor, but it is a consequence of the asymmetric mass ejection
in the supernova explosion, which again is a consequence of violent hydrodynamic 
instabilities that precede and accompany the onset of the supernova blast.
We also stress that the fallback phenomena in our focus here are distinctly different
from the disk formation, magnetic field amplification, and possible jet formation 
discussed by \citet{Gilkis+2014,Gilkis+2015,Gilkis+2016}, \citet{Quataert+2019},
\citet{Soker2020}, and \citet{Soker2021}, respectively.
Those authors considered the stochastic angular momentum accretion 
{\em prior to} (or in the absence of) the onset of a supernova explosion,
when the angular momentum is associated to fluctuating mass motions in the 
convective pre-collapse burning shells or in the envelope of slowly or nonrotating 
progenitors. In contrast, we discuss the accretion of some fraction of the
initial ejecta that remains gravitationally bound to the compact remnant 
{\em during} the developing supernova and therefore eventually turns around 
and falls back. Of course,
there may be dependencies between pre- and post-explosion accretion, because
the asymmetry of the mass ejection and the subsequent anisotropic fallback 
may be influenced by the pre-collapse asymmetries in the convective burning 
shells of the progenitor.

A second motivation of our work comes from the long-standing question of spin-kick 
alignment of young pulsars
\citep[e.g.,][]{Spruit+1998,Lai+2001,Johnston+2005,Romani+2003,Ng+2007}.
While in the cases of the Crab
and Vela pulsars the alignment of the spin axis and direction of proper motion 
is apparent and projection effects might play a role, further observational
evidence was presented by \citet{Johnston+2005,Wang+2006,Johnston+2007,Noutsos+2012}, 
and \citet{Noutsos+2013}. In the most recent analyses 
\citep{Noutsos+2012,Noutsos+2013}, considering large samples of young pulsars
with reliably determined kinematic ages, the alignment between spin and kick
directions was found to be typically within several 10$^\circ$, but the 
observed distribution of the spin-velocity offset angles is broad with a 
standard deviation around 30$^\circ$.

Particularly interesting is the association of the
pulsar PSR~J0538+2817 with the young supernova remnant S147 and an age 
of only 30$\pm$4\,kyr
\citep{Kramer+2003,Ng+2007a,Dincel+2015}, because the neutron star has a high
transverse velocity around 400\,km\,s$^{-1}$ \citep{Kramer+2003,Ng+2007a,
Chatterjee+2009} and also 
exhibits near proximity of its spin axis and velocity vector in two dimensions
\citep{Romani+2003,Ng+2007a,Johnston+2007}. 
Recent high-sensitivity interstellar scintillation and polarization 
observations of PSR~J0538+2817 using the Five-hundred-meter Aperture Spherical 
radio Telescope (FAST) confirm these previous results, reveal a total space
velocity of the pulsar of more than 400\,km\,s$^{-1}$, and demonstrate, 
for the first time, 3D alignment of the spin axis and space velocity of a pulsar
\citep{Yao+2021}. The 3D angle between spin axis and velocity vector
inferred from the FAST polarization results has a clear peak around 6$^\circ$
and a 68\% probability of being less than 28$^\circ$; and from X-ray torus
modeling a peak around 10$^\circ$ is deduced with a 68\% probability that the
angle is smaller than 23$^\circ$.

The young age of PSR~J0538+2817, its relatively long initial spin period of 
139\,ms \citep{Kramer+2003,Romani+2003}, its
relatively low surface magnetic field of $7.3\times 10^{11}$\,G \citep{Kramer+2003},
and high kick velocity discard a large variety of suggested scenarios that invoke 
extremely strong magnetic fields to produce large kicks
\citep[e.g.][]{Bisnovatyi-Kogan1993,Bisnovatyi-Kogan1996,Horowitz+1998,Arras+1999a,
Arras+1999b,Kusenko+1999,Fuller+2003,Farzan+2005,Socrates+2005,Murayama+2012,
Yamamoto+2021} and spin-kick alignment. 
The Harrison-Tademaru mechanism \citep{Harrison+1975,Tademaru+1975}, for example,
is based on asymmetric electromagnetic radiation emitted from an off-centered 
rotating magnetic dipole. However, this electromagnetic rocket effect requires 
the neutron star to have an initial rotation period of about a 
millisecond and an extremely strong magnetic dipole field (or order $10^{16}$\,G)
at the surface. Only then the kick can be as strong as measured for PSR~J0538+2817 
on a short time scale \citep{Xu+2021}. If these conditions are not fulfilled, the
acceleration is a weak and/or slow process that acts only over long periods of time
after the supernova explosion \citep[see][]{Lai+2001}, implying that the strong 
surface magnetic fields must survive for such long time spans. This, however, 
conflicts with the observed properties of PSR~J0538+2817.

Therefore, it remains a puzzle, where spin-kick alignment might come from.
With fallback being likely to be the dominant spin-up mechanism of most new-born
neutron stars, as we argued above, the question immediately pops up whether fallback
could also be responsible for alignment between neutron star kicks and spins. 
Previous hydrodynamical supernova simulations, partly including the fallback phases 
that extend over hundreds of seconds or longer after the onset of the explosion
\citep{Stockinger+2020,Chan+2020,Powell+2020,Mueller+2018,Mueller+2019,Wongwathanarat+2013},
showed no alignment of neutron star kicks and spins. In our paper we hypothesize that 
also for this result the reason is the neglect of the neutron star's motion in all
of the simulations. 
Rather than drifting out of the center of the explosion due to the kick obtained
in the first seconds of the developing supernova, the neutron star is fixed at the 
grid center for numerical reasons.

The neutron
star receives its natal kick by accelerating forces over several seconds, whereas
it gets spun up subsequently by accretion of matter with angular momentum from
fallback happening on time scales of tens of seconds to hours.
In the present paper we will discuss possible reasons how spin-velocity
alignment can be the outcome when the neutron star is displaced
away from the explosion center because of its birth kick. The initial kick
of the neutron star and its later spin-up will be crucial for the spin-kick
alignment. 
In the context of our proposed scenario it is therefore {\em not} the initial 
(pre-kick) rotation rate of the neutron star that determines the alignment
of kick velocity and spin axis, and it is also {\em not} a mechanism that imparts
spin and kick simultaneously to the neutron star as suggested by \citet{Spruit+1998}
in their multiple off-center-kick scenario. The hypothesis of a mechanism that
delivers kick and spin to the neutron star in combination, or the assumption
that a pre-kick rotation of the neutron star plays a crucial role,
was later adopted also by others, for example by \citet{Lai+2001}, 
in their discussions of theoretical neutron star kick and spin-up scenarios. 
Such a picture was further elaborated by \citet{Ng+2007} in their momentum-thrust 
scenario based on the neutrino-cooling emission of the proto-neutron star, and it was
also employed by \citet{Wang+2006} as a framework to interpret observational
spin-kick alignment of isolated pulsars and misalignment of neutron stars
in binaries, respectively. These authors drew conclusions on the duration of
the kick time scale or pre-kick spin period in order to obtain spin-kick alignment
in dependence on the initial rotation of neutron stars. In the statistical
argument by \citet{Spruit+1998},
analogously also applied by \citet{Ng+2007} for their
neutrino momentum-thrust scenario, it is an existing spin axis of the
neutron star that defines a preferred direction for a kick alignment
due to rotational averaging of the off-axis directions.

In contrast, our new scenario introduced in the present paper reverses this causal 
relation:
Here, it is the initial kick and associated displacement of the neutron star from the 
explosion center that defines the preferred direction and causes the
spin-velocity alignment. Spin-velocity correlation becomes likely for
neutron stars with high kick velocities, where the neutron star accretes fallback
matter mainly from the hemisphere facing the direction of neutron star motion.
But such an alignment is not an ubiquitous phenomenon. The neutron star rotation 
vector is expected to be randomly oriented relative to the kick direction for 
neutron stars with low kicks or for supernovae with no relevant accretion
of fallback matter onto the neutron star, or in cases where little angular momentum 
is associated with the fallback material accreted by the neutron star. 

Our paper is structured as follows. In Section~\ref{sec:observations} the 
current observational evidence of spin-kick alignment in young pulsars is summarized
based on existing publications. In Section~\ref{sec:simulations} we present
results from a large set of 3D long-time supernova simulations, all of which include 
fallback until a day or longer. The simulations serve as a motivation for introducing
a revision of the hydrodynamical scenario. We do this in Section~\ref{sec:scenario} 
by considering possible effects associated with a kick-induced displacement of the
neutron star from the center of the explosion, in particular the possibility
of fallback-triggered spin-kick alignment. In Section~\ref{sec:discussion} we discuss
implications and observational consequences of our revised scenario of spin-kick alignment,
and in Section~\ref{sec:conclusions} we conclude with a summary of our main results.

\section{Observational evidence for spin-kick alignment}
\label{sec:observations}

The realisation that the observed velocities of pulsars are large \citep{Gunn+1970,Lyne+1982}
raised the question about their origin early on (e.g.~\citealt{Tademaru+1975}). In order to
decide between different scenarios, especially between natal or post-natal origins, efforts 
were made to determine the direction of motion relative to the spin orientation 
\citep{Morris+76,Anderson+83}, but no correlation was found. However, a number of observational 
challenges may have led to misleading measurements or may have prevented a significant 
correlation to be found, even if it exists in principle. It is useful to recall these 
challenges.

Firstly, it is usually not possible to measure the radial velocity of pulsars (even though rare exceptions
exist, e.g., \citealt{Yao+2021}). Instead, what can be measured is the transversal motion on the plane
of the sky, i.e.~a 2D motion. Secondly, this motion is a {\em proper motion}, i.e.~an angular motion on the sky.
In order convert it into a velocity, one needs to use a distance measurement. Distances can be estimated
from the measurement of the so called Dispersion Measure, 
defined as the observed column density of free electrons along the line
of sight. Given a model for the Galactic free electron distribution (e.g.~\citealt{NE2001,YMW17}),
distances can be estimated with a typical uncertainty of 20\%, though deviations from the real
distance of individual cases can be significantly larger. 
Ideally, one can infer distances reliably by means of a parallax measurement,
either via pulsar timing or interferometric imaging -- which are also the two methods 
usually used to obtain proper motions.

The first way to determine proper motions is via 
pulsar timing measurements (e.g.~\citealt{handbook}). A proper motion manifests itself 
as a characteristic signature in the timing residuals, i.e.~in the comparison of the measured
pulse times-of-arrival (ToAs) with a so-called timing model that describes the rotation of the pulsar.
The signature comprises a sinusoid with an annual period and an amplitude that increases with time
(see e.g.~\citealt{Kramer+2003}). Unless the pulsar is located in the ecliptic, the two-dimensional proper motion 
vector, $\mu = (\mu_\alpha,\mu_\delta)$, relative to our local reference system can then be measured. 
We note that for a pulsar near the ecliptic, the variation of ToAs due to proper motion are
small in direction of ecliptic latitude, so that effectively only a 1D velocity (in direction
of ecliptic longitude) can be measured accurately (see e.g.~\citealt{handbook} for details).
A further complication arises for young pulsars, which often exhibit irregularities in their
spin-down, known as {\em timing noise}, which can be attributed to rotational instabilities caused by the
interior of the neutron star or magnetospheric re-configurations \citep{handbook,Lyne+2010}. Timing noise
may occur on similar timescales as the proper motion signature in the timing data, so it has been
traditionally difficult to separate both effects and to determine the velocity of young pulsars
reliably. A technique that makes use of the proper motion's annual periodicity in the timing residuals was
presented by \cite{Hobbs+2005}, which suddenly gave access to proper motion measurements
for a large sample of pulsars, especially young ones.

A second way to determine the astrometric properties of a pulsar is via interferometric imaging. High-spatial
resolution observations made at several epochs can reveal a motion of the pulsar on the sky caused by
proper motion and/or parallax (e.g.~\citealt{psrpi}). Interferometric measurements do not suffer
from the same difficulties as timing measurements, i.e.~they do not depend on the ecliptic position and they
are not affected by timing noise. In contrast, however, pulsars need to be bright enough to be detected in
imaging observations, while ionospheric effects and the availability of suitable telescope networks 
play an essential role in whether a proper motion (and parallax) measurement can be obtained. Recent
improvements in the applied techniques have nevertheless increased the accessible sample considerably
\citep{psrpi}.

Even though techniques exist now to measure the velocities of pulsars with a certain accuracy, there are also
observational challenges in determining the spin directions. As for the velocity, also the spin direction 
can usually only be inferred as a 2D projection on the plane of the sky. (Again, notably exceptions exist;
see e.g.~the recent examples by \citealt{Yao+2021} or \citealt{Guo+2021}). 
This is typically done via
polarisation measurements by inspecting the variation of the position angle  (PA) 
of the linearly polarised component of the pulsar radio emission as a function of pulse phase 
(also called the ``PA swing''). In textbook examples, like in the Vela pulsar, the swing 
exhibits an S-like shape, motivating the Rotating Vector Model \citep[RVM;][]{RVM} where this
shape is explained by purely geometrical effects. The centroid of the S-like swing defines
both a fiducial pulse phase and a fiducial PA value: 
the pulse phase marks the location of the ``fiducial plane'', where magnetic axis, spin axis and viewing vector to
the observer all fall into one plane. The fiducial value of the PA 
corresponds to the projection angle of the pulsar spin onto the plane of the sky \citep[measured
in a North-East coordinate system relative to North; see also, e.g.,][]{Gunn+1970}. 
However,
not all PA swings in pulsars can be readily described by a RVM (e.g. due to distortion in the PA swing by
plasma propagation effects, e.g.~\citealt{handbook}), but one may still be able to identify
the location of the magnetic spin axis relative to the observed pulse, so that the fiducial PA 
can be measured regardless \citep{Johnston+2005}. In any case, 
one needs to make sure that the derived fiducial
PA corresponds to an ``absolute'' PA value, namely one that would
be measured at infinite frequencies. PA values observed at a certain frequency are
affected by Faraday rotation in the interstellar medium or the Earth's ionosphere
(see \citealt{Johnston+2005,Johnston+2007} for a detailed discussion). Hence, it is crucial that any
measurement of an absolute position angle that is used to infer the (projected) spin direction is
accompanied by careful measurements of the ``Rotation Measure'' that describes the rotation of the
PA with frequency, so that Faraday rotation can be corrected for
\citep{handbook}.

From the above, it is clear that obtaining reliable information on the spin and velocity direction
is not trivial. Moreover, in order to judge the presence or lack of a correlation between the two
quantities, one needs to take also astrophysical considerations into account. A movement of the pulsar
in the Galactic gravitational potential will constantly alter the velocity direction with time, for which
reason the present-day orientation may have little resemblance with the direction possibly imprinted at birth.
Any existing correlation may therefore be washed out if pulsars are too old, or if the external velocity
component due to Galactic motion is significantly larger than any intrinsic velocity.

With these caveats in mind, \cite{Johnston+2005} made use of the \cite{Hobbs+2005} sample and revisited
the question of a possible correlation between pulsar spin and velocity directions. As described in
Section~\ref{sec:intro}, evidence for a correlation was found, spawning further and renewed
interest. A number of
further studies followed \citep{Wang+2006,Johnston+2007,Rankin2007,Noutsos+2012,Noutsos+2013,Rankin2015}.
Overall, all these studies concluded in favour of spin-velocity alignment,  which is also supported
by the special cases of certain pulsars associated with
supernova remnants,  i.e.\ Crab and Vela, by conclusions drawn from pulsar wind torus fitting \citep{Ng+2004,Ng+2008},
and especially by the mentioned recent evidence for even 3D alignment in
PSR J0538+2817 \citep{Yao+2021} and similarly by a recent study of PSR J0908$-$4913 by \cite{0908}.
Only the sample studied by 
\cite{Noutsos+2012,Noutsos+2013} was large enough to separate the pulsars by age, accounting
for possible motion in the Galactic potential. As mentioned,  the
observed distribution of the spin-velocity offset angles, $\Psi$, is broad with a 
standard deviation around $\Delta \Psi ~\sim 30^\circ$. For the youngest group, however, the offset angle 
appears to be somewhat smaller (see e.g.~Figure 1 in \citealt{Noutsos+2013}), but the number of pulsars
is also smaller.

From those studies, it seems established that for the sample of interest, i.e.~young (or relatively young) 
pulsars, the distribution of angular separations between spin and velocity, $\Psi$, is not random, but exhibits a clear
preference for alignment. However, establishing a more reliable estimate for the typical offset is desirable
and subject of ongoing work. Apart from the intention to improve on the solutions for overcoming
the explained observational challenges,
the motivation is twofold. Firstly, pulsar polarisation is often emitted in one of two orthogonal
modes \citep{handbook}, and it is not immediately clear which of the two modes should the velocity direction
be compared with \citep{Johnston+2005,Noutsos+2012,Rankin2015}. 
In other words, is the relevant offset angle $\Psi$ or
$(90^\circ-\Psi)$? Furthermore, in the previous studies, it was not attempted to separate the sample also in terms
of velocity magnitude, which is, as we discuss in the following, a relevant parameter in our theoretical scenario presented in Sections~\ref{sec:scenario} and \ref{sec:discussion}.
An improved detailed study of the observed distribution of angular separations, with samples optimised to look for
observational signatures that we derive in the following, including a significant amount of new data,
is indeed ongoing, and the results of this study will be presented elsewhere.

\begin{longrotatetable}
\label{tab:3dmodels}
\setlength{\tabcolsep}{2.2pt}
\begin{deluxetable*}{lcccccccccccccccc}
\tablehead{
\mrow{2}{Model} & $t_\trm{early}$ &
$t_\trm{end}$ & $v_\trm{NS}(t_\trm{early})$ & $v_\trm{NS}(t_\trm{end})$ & $M_\trm{NS}(t_\trm{end})$ &
$M_\trm{NS,g}(t_\trm{end})$ & $E_\trm{exp}(t_\trm{end})$ &
$\alpha_\trm{ej}(t_\trm{early})$ & $\alpha_\trm{ej}(t_\trm{end})$ & $M_\trm{ej}(t_\trm{end})$ & $J_\trm{NS,46}(t_\trm{early})$ &
$J_\trm{NS,46}(t_\trm{end})$ & $\theta_\trm{sk}(t_\trm{early})$ & $\theta_\trm{sk}(t_\trm{end})$
& $M_\trm{fb}(t_\trm{end})$ & $T_\trm{spin}(t_\trm{end})$ \\
 & [s] & [s] & [\kmps] & [\kmps] & [M$_\odot$] & [M$_\odot$] & [B] & [\%] &
     [\%] & [M$_\odot$] & [$\trm{g}\,\trm{cm}^2\,\trm{s}^{-1}$] &
     [$\trm{g}\,\trm{cm}^2\,\trm{s}^{-1}$] & [$^\circ$] & [$^\circ$] & [$10^{-2}$\,M$_\odot$] & [ms]
}
\startdata
W15-1-cw       & 1.3 & 432000 &  331 &  740 & 1.39 & 1.26 & 1.49 &  7.93 & 2.41 & 14.04 & 1.51 &  197.60 & 115 &  16 &  5.67 &   3.95\\
W15-2-cw       & 1.3 & 431995 &  405 &  793 & 1.40 & 1.27 & 1.50 &  9.59 & 2.59 & 14.02 & 1.55 &  181.10 &  60 & 108 &  7.05 &   4.37\\
W15-3-pw       & 1.3 & 431995 &  267 &  511 & 1.50 & 1.35 & 1.11 &  6.35 & 2.09 & 13.94 & 1.17 &  312.27 & 105 & 114 & 14.15 &   2.74\\
W15-6-cw       & 1.3 & 432003 &  437 & 1034 & 1.46 & 1.32 & 1.25 & 12.84 & 3.88 & 13.95 & 1.02 &  149.47 & 128 &  82 & 10.65 &   5.57\\
L15-1-cw       & 1.4 & 431993 &  161 &  338 & 1.53 & 1.38 & 1.78 &  5.01 & 1.13 & 13.49 & 1.89 &   94.64 & 150 & 110 &  1.41 &   9.32\\
L15-2-cw       & 1.4 & 420609 &   78 &  131 & 1.42 & 1.28 & 2.80 &  1.58 & 0.32 & 13.61 & 1.04 &    2.14 &  63 &  88 &  0.03 & 373.81\\
L15-3-pw       & 1.4 & 432007 &   31 &   80 & 1.78 & 1.58 & 0.85 &  1.33 & 0.46 & 13.23 & 1.55 &  483.67 & 125 &  74 & 16.12 &   2.19\\
L15-5-cw       & 1.4 & 431991 &  267 &  725 & 1.78 & 1.58 & 0.95 & 15.56 & 3.94 & 13.22 & 1.71 & 1137.48 &  66 &  93 & 16.41 &   0.93\\
L15-5-pw       & 1.4 & 432002 &  267 &  404 & 2.11 & 1.83 & 0.58 & 15.56 & 3.41 & 12.87 & 1.71 &  864.01 &  66 & 112 & 45.00 &   1.53\\
N20-4-cw       & 1.3 & 184056 &   98 &  275 & 1.41 & 1.28 & 1.67 &  2.54 & 0.81 & 14.69 & 2.09 &  252.60 &  46 & 121 &  0.50 &   3.15\\
B15-1-cw       & 1.1 &  48416 &   92 &  119 & 1.15 & 1.06 & 2.59 &  2.36 & 0.24 & 14.31 & 1.03 &   11.98 & 155 &  18 &  0.01 &  52.17\\
B15-1-pw       & 1.1 &  61744 &   92 &  115 & 1.24 & 1.14 & 1.40 &  2.36 & 0.34 & 14.21 & 1.03 &   47.95 & 155 &  61 &  1.02 &  14.20\\
B15-3-pw       & 1.1 &  61214 &   85 &  134 & 1.25 & 1.14 & 1.15 &  2.55 & 0.44 & 14.20 & 0.44 &    2.68 & 148 &  80 &  0.04 & 256.07\\
W18-pw         & 1.3 & 175849 &  104 &  258 & 1.55 & 1.39 & 1.36 &  2.25 & 0.90 & 15.40 & 0.71 &  291.41 &  78 & 117 & 15.25 &   3.06\\
W20-pw         & 1.3 &  61779 &   87 &  198 & 1.55 & 1.39 & 1.45 &  2.03 & 0.62 & 17.86 & 0.75 &  260.57 & 109 &  62 &  5.42 &   3.43\\
W16-1-pw       & 1.3 & 172213 &  232 &  353 & 2.27 & 1.95 & 0.88 &  7.70 & 2.33 & 13.06 & 0.24 & 2352.23 &  62 & 111 & 67.76 &   0.62\\
W16-2-pw       & 1.3 & 153565 &  348 &  562 & 2.01 & 1.76 & 1.16 &  9.55 & 2.88 & 13.33 & 1.01 & 2460.39 &  82 &  91 & 44.29 &   0.50\\
W16-3-pw       & 1.3 & 141634 &  151 &  249 & 1.72 & 1.53 & 1.48 &  3.45 & 0.97 & 13.63 & 1.56 & 2918.77 & 146 & 111 & 18.38 &   0.35\\
W16-4-pw       & 1.3 & 130866 &  244 &  409 & 1.62 & 1.45 & 1.82 &  4.79 & 1.35 & 13.74 & 2.28 & 1177.52 &  31 &  59 & 10.99 &   0.80\\
W18r-1-pw      & 1.3 & 180323 &   33 &   95 & 1.48 & 1.34 & 1.06 &  0.89 & 0.36 & 15.59 & 1.08 & 1962.32 & 116 &  83 & 13.97 &   0.43\\
W18r-2-pw      & 1.3 & 165178 &   24 &   54 & 1.37 & 1.25 & 1.31 &  0.54 & 0.17 & 15.71 & 0.24 & 1065.83 & 129 & 117 &  5.25 &   0.72\\
W18r-3-pw      & 1.3 & 156979 &   45 &  105 & 1.32 & 1.20 & 1.59 &  0.88 & 0.29 & 15.76 & 1.04 &  157.67 &  70 & 119 &  2.26 &   4.66\\
W18r-4-pw      & 1.3 & 139696 &   16 &   46 & 1.28 & 1.17 & 1.91 &  0.27 & 0.11 & 15.80 & 0.58 &   70.96 & 123 &  69 &  0.65 &   9.96\\
W18x-1-pw      & 1.3 & 154941 &  137 &  238 & 1.66 & 1.48 & 1.21 &  3.78 & 0.92 & 15.87 & 1.48 & 2594.84 & 125 &  98 & 15.81 &   0.38\\
W18x-2-pw      & 1.3 & 157572 &   46 &   91 & 1.56 & 1.40 & 1.45 &  1.10 & 0.30 & 15.98 & 0.79 &  754.83 &  56 &  59 &  7.42 &   1.19\\
M15-7b\dg-3-pw & 1.3 &  86403 &  727 & 1096 & 1.66 & 1.48 & 1.44 & 17.96 & 3.39 & 19.41 & 1.97 &  605.38 &  41 & 123 & 16.44 &   1.61\\
M15-7b\dg-4-pw & 1.3 &  86399 &  694 & 1102 & 1.54 & 1.39 & 1.79 & 13.87 & 2.84 & 19.53 & 1.61 &  417.70 & 150 &  91 &  9.62 &   2.12\\
M15-8b\dg-2-pw & 1.3 &  86410 &   13 &   40 & 1.34 & 1.22 & 1.32 &  0.32 & 0.10 & 20.72 & 1.11 &  101.33 & 146 &  26 &  1.61 &   7.36\\
M15-8b\dg-3-pw & 1.3 &  90753 &   14 &   28 & 1.30 & 1.19 & 1.62 &  0.29 & 0.06 & 20.76 & 0.45 &   28.82 &  99 & 149 &  0.42 &  25.04\\
M16-4a\dg-1-pw & 1.3 &  86413 &   32 &   92 & 1.67 & 1.49 & 1.34 &  0.76 & 0.32 & 17.35 & 2.04 & 2183.44 &  55 & 156 & 14.22 &   0.45\\
M16-4a\dg-2-pw & 1.3 &  90027 &   64 &  177 & 1.58 & 1.42 & 1.66 &  1.34 & 0.53 & 17.44 & 0.52 &  178.00 & 115 & 170 &  7.83 &   5.14\\
M16-4a\dg-3-pw & 1.3 &  89984 &   14 &   35 & 1.51 & 1.36 & 1.93 &  0.26 & 0.09 & 17.51 & 2.27 &   77.79 & 100 &  58 &  3.19 &  11.12\\
M17-7a\dg-2-pw & 1.3 & 193880 &   60 &  139 & 1.65 & 1.47 & 1.64 &  1.18 & 0.39 & 21.19 & 0.46 & 1104.85 &  27 & 109 & 15.12 &   0.87\\
M17-7a\dg-3-pw & 1.3 & 172744 &   61 &  140 & 1.56 & 1.40 & 1.95 &  1.07 & 0.34 & 21.28 & 1.46 &   80.73 &  21 &  67 &  8.78 &  11.16\\
M15-7b-1-pw    & 1.3 &  86390 &  469 &  767 & 1.58 & 1.41 & 1.41 & 11.92 & 2.28 & 19.48 & 0.81 &  214.58 & 122 &  41 & 10.34 &   4.25\\
M15-7b-2-pw    & 1.3 &  86402 &  310 &  483 & 1.56 & 1.40 & 1.43 &  7.09 & 1.40 & 19.49 & 2.24 &  122.29 & 114 & 124 &  9.47 &   7.34\\
M15-7b-3-pw    & 1.3 &  86395 &  478 &  751 & 1.59 & 1.42 & 1.43 & 11.01 & 2.22 & 19.46 & 3.47 &  585.55 & 117 &  77 & 12.46 &   1.57\\
M15-7b-4-pw    & 1.3 &  86397 &  601 &  901 & 1.49 & 1.35 & 1.78 & 11.48 & 2.26 & 19.55 & 0.90 &  314.30 &  63 &  85 &  6.66 &   2.71\\
M15-8b-1-pw    & 1.3 &  86410 &   40 &   71 & 1.32 & 1.20 & 1.57 &  0.84 & 0.16 & 20.72 & 0.48 &   96.02 &  92 & 167 &  0.81 &   7.65\\
M15-8b-2-pw    & 1.3 &  86396 &   16 &   35 & 1.38 & 1.26 & 1.12 &  0.44 & 0.10 & 20.66 & 0.48 &  191.17 &  23 & 127 &  2.62 &   4.07\\
M16-4a-1-pw    & 1.3 &  86412 &  171 &  352 & 1.67 & 1.49 & 1.56 &  3.65 & 1.13 & 17.34 & 2.41 &  256.64 &  10 & 116 & 13.77 &   3.81\\
M16-4a-2-pw    & 1.3 &  86394 &  332 &  592 & 1.90 & 1.67 & 1.07 &  8.82 & 2.61 & 17.10 & 2.73 & 2954.22 &  59 &  84 & 32.25 &   0.39\\
M16-7b-1-pw    & 1.3 &  86406 &  100 &  216 & 1.54 & 1.38 & 1.17 &  2.76 & 0.68 & 20.44 & 0.30 &  129.86 & 140 &  40 &  9.12 &   6.81\\
M16-7b-2-pw    & 1.3 &  86400 &   65 &  139 & 1.47 & 1.32 & 1.41 &  1.54 & 0.38 & 20.51 & 1.42 &  226.02 & 130 & 166 &  4.36 &   3.69\\
M17-7a-1-pw    & 1.3 &  86417 &   97 &  220 & 1.69 & 1.51 & 1.52 &  2.10 & 0.66 & 21.12 & 1.66 &  798.71 &  44 & 100 & 16.38 &   1.25\\
M17-7a-2-pw    & 1.3 &  86407 &   47 &   82 & 1.68 & 1.50 & 1.56 &  0.96 & 0.24 & 21.13 & 0.84 &  576.19 &  90 &  29 & 15.17 &   1.71\\
M17-8a-3-pw    & 1.3 &  86390 &  476 &  715 & 2.27 & 1.95 & 1.08 & 13.18 & 3.24 & 21.53 & 2.14 & 8536.20 & 122 & 117 & 52.60 &   0.17\\
M17-8a-4-pw    & 1.3 &  86412 &  445 &  732 & 2.09 & 1.81 & 1.22 & 11.34 & 2.88 & 21.72 & 2.11 & 5521.53 &  69 & 118 & 36.63 &   0.24\\
\enddata
\end{deluxetable*}
        \tablecomments{
	        Explosion and neutron star properties for all investigated 3D models. 
		Neutron stars are defined by the mass at densities $\rho\ge 10^{11}$\,g\,cm$^{-3}$.
		The model names consist of the progenitor's name (see Section~\ref{sec:progenitors}),
		a counter for the explosion run, and an extension ``cw'' for a constant neutrino-driven wind and ``pw'' for a power-law wind condition at the
		inner grid boundary;
        daggers indicate an older version of the corresponding binary BSG progenitor
		(see also Section~\ref{sec:SNmodeling}).
        The tabulated quantities are:
        $t_\mathrm{early}$ is the time when the central neutrino ``engine'' was 
        replaced by the spherical neutrino-wind boundary,   
		$t_\mathrm{end}$ is the termination time of the simulations, 
		$v_\mathrm{NS}(t_\mathrm{early})$ and $v_\mathrm{NS}(t_\mathrm{end})$ are the neutron star kick velocities at both times, 
		$M_\mathrm{NS}(t_\mathrm{end})$ and $M_\mathrm{NS,g}(t_\mathrm{end})$ the final
		baryonic and gravitational masses of the neutron star,
		$E_\mathrm{exp}(t_\mathrm{end})$ the final explosion energy,
		$\alpha_\mathrm{ej}(t_\mathrm{early})$ and $\alpha_\mathrm{ej}(t_\mathrm{end})$ are
		the momentum asymmetry parameters of the (postshock) ejecta at both times (defined in Equation~(\ref{eq:alpha}) and surrounding text),
		$M_\mathrm{ej}(t_\mathrm{end})$ is the final ejecta mass of the supernova,
		$J_{\mathrm{NS},46}(t_\mathrm{early})$ and $J_{\mathrm{NS},46}(t_\mathrm{end})$ are 
        the early and final values of the angular momentum of the neutron star, if the latter accretes the entire fallback mass,
		$\theta_\mathrm{sk}(t_\mathrm{early})$ and $\theta_\mathrm{sk}(t_\mathrm{end})$ are
		the relative angles between neutron star spin and kick at both times, 
		$M_\mathrm{fb}(t_\mathrm{end})$ is the final fallback mass, and
		$T_\mathrm{spin}(t_\mathrm{end})$ is the estimated spin period of the neutron star,
		assuming that the neutron star accretes the total fallback matter and has a final 
        radius of 12\,km and gravitational mass $M_\mathrm{NS,g}(t_\mathrm{end})$.
		}
\end{longrotatetable}

\begin{figure*}[tb]
\begin{center}
	\includegraphics[width=\textwidth]{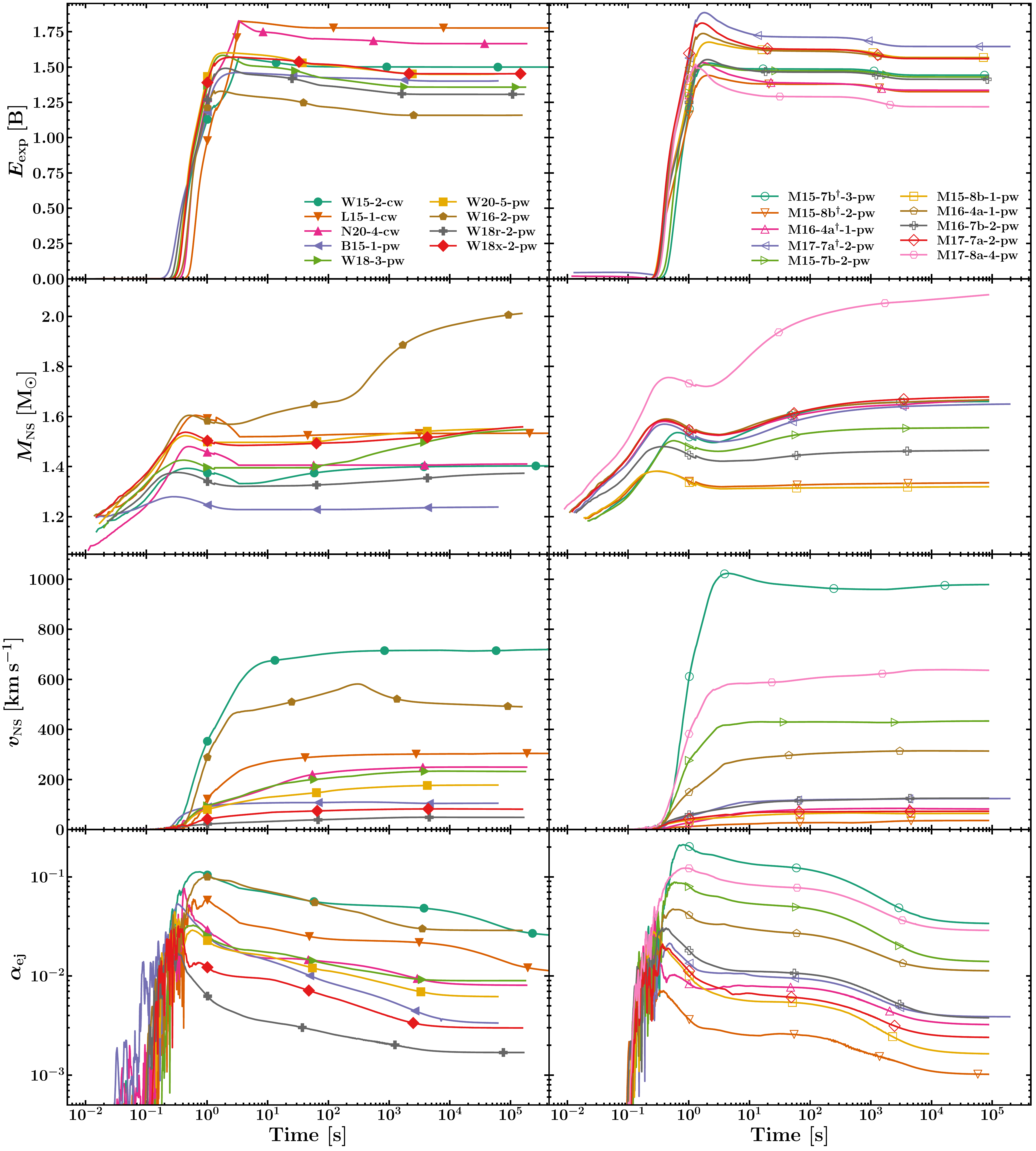}
	\caption{Time evolution of explosion and neutron star properties for a subset of
	the 3D supernova models of Table~\ref{tab:3dmodels}. Results for single-star 
	RSG and BSG progenitors are shown in the left panels, for binary-merger BSG
	progenitors in the right panels. $E_\mathrm{exp}$ is the
	(diagnostic) explosion energy, $M_\mathrm{NS}$ and $v_\mathrm{NS}$ the baryonic 
	mass and kick velocity of the neutron star, respectively (assuming all of the 
	fallback matter gets accreted onto the neutron star), and $\alpha_\mathrm{ej}$ 
	is the momentum asymmetry parameter of the ejected postshock matter (defined in
	Equation~(\ref{eq:alpha})).}
\label{fig:exp-vs-t}
\end{center}
\end{figure*}

\begin{figure*}[tb]
\begin{center}
        \includegraphics[width=\textwidth]{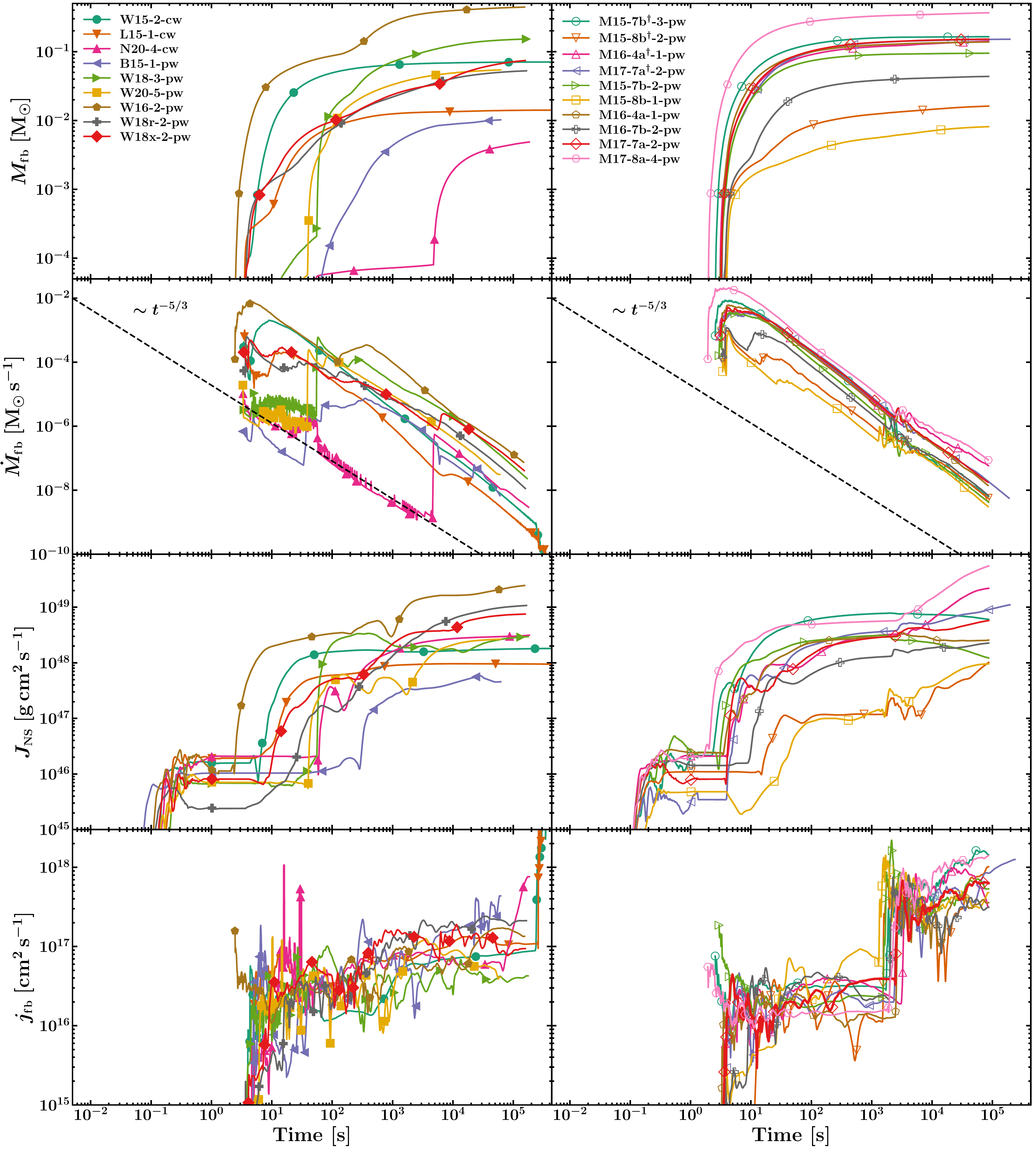}
        \caption{Fallback properties as functions of time for a subset of
	the 3D supernova models of Table~\ref{tab:3dmodels}. Results for single-star 
        RSG and BSG progenitors are shown in the left panels, for binary-merger BSG
        progenitors in the right panels. $M_\mathrm{fb}$ is the time-integrated 
	fallback mass, $\dot M_\mathrm{fb}$ the rate of mass-fallback
	\citep[the dashed line displays the asymptotic scaling $\propto t^{-5/3}$ 
	for the spherical case;][]{Chevalier1989}, $J_\mathrm{NS}$ is the
	angular momentum of the neutron star, assuming that all fallback matter
	with its angular momentum gets accreted
	by the compact object, and $j_\mathrm{fb}$ is the specific angular momentum
	associated with the mass-fallback rate $\dot M_\mathrm{fb}$.}
\label{fig:fallback-vs-t}
\end{center}
\end{figure*}

\section{Results of long-time 3D supernova simulations}
\label{sec:simulations}

Over the past couple of years one of the authors has built up a library of
nearly 50 3D simulations of neutrino-driven supernova explosions,
following the evolution from a few milliseconds after core bounce through shock
revival by neutrino heating until roughly one day or more later. 
The model set is mainly based on blue supergiant (BSG) single-star and 
binary-merger progenitors, which were investigated for phenomenological 
analysis in connection to Supernova~1987A (SN~1987A), but it includes
also two red supergiant (RSG) progenitors, all of them
with pre-supernova masses between $\sim$15\,M$_\odot$ and $\sim$24\,$M_\odot$.

\subsection{Progenitor stars}
\label{sec:progenitors}

A list of all 3D explosion models for these progenitors with the values
of the most relevant quantities characterizing the explosion and neutron
star properties is provided in Table~\ref{tab:3dmodels}.
The set of 3D supernova simulations includes single-star progenitors with
and without rotation, namely nonrotating RSGs from
\citet{Woosley+1995} (stellar model s15s7b2, explosion models W15-xx; the
naming convention of the supernova runs will be explained below) and  
\citet{Limongi+2000} (L15-xx), both with a zero-age-main-sequence (ZAMS) mass
of 15\,M$_\odot$; a nonrotating BSG progenitor with ZAMS mass of 20\,M$_\odot$ 
\citep[][N20-xx]{Shigeyama+1990}, nonrotating BSG progenitors of 
15\,M$_\odot$ \citep[][B15-xx]{Woosley+1988} and 20\,M$_\odot$ 
\citep[][W20-xx]{Woosley+1997}
ZAMS mass, as well as rotating BSG models with ZAMS masses of
16\,M$_\odot$ (W16-xx) and 18\,M$_\odot$ (W18-xx, W18r-xx, and W18x-xx)
(see \citealt{Sukhbold+2016} and \citealt{Utrobin+2019} for all four 
rotating progenitors).
Moreover, 3D explosion simulations of BSG progenitors of SN~1987A from 
binary-merger calculations by \citet{Menon+2017} are considered:
M15-7b-xx, M15-8b-xx, M16-4a-xx, M16-7b-xx, M17-7a-xx, and M17-8a-xx, where
the first and second numbers in the models names give the ZAMS masses 
(in M$_\odot$) of the 
two binary components. The name extensions xx denote different realizations
of the supernova calculations, either by choosing different
seed perturbations that were imposed to trigger the growth of nonradial
hydrodynamic instabilities at the onset of the explosion,
or by parametrically varying the blast-wave energy through different
choices of the time dependence of the explosion-driving neutrino engine.

\subsection{Supernova modeling}
\label{sec:SNmodeling}

Since the present paper is not focussed on the numerical modeling and 
simulation results, but these results are only taken as a motivation for
proposing the revised fallback scenario described in Section~\ref{sec:scenario},
we refrain from repeating a detailed description of the code and its inputs
here. All relevant information can be found in previous papers of some of the
authors \citep{Wongwathanarat+2013,Wongwathanarat+2015,Wongwathanarat+2017,
Utrobin+2015,Utrobin+2017,Utrobin+2019,Utrobin+2021} and in
Wongwathanarat \& Janka (in preparation).

The modeling strategy for the supernova explosions is based on the 
paradigm of the delayed neutrino-driven mechanism, whose viability
is now supported by self-consistent ``ab initio'' 3D simulations 
\citep[e.g.,][]{Takiwaki+2014,Melson+2015a,Melson+2015b,Lentz+2015,
Janka+2016,Mueller+2017,Mueller+2018,Burrows+2020,Bollig+2020}.
Since we are interested
in the long-time development of the supernova asymmetries but not in
the details of the physics that is relevant to obtain shock revival
and the onset of an explosion, the computationally
most expensive part of fully self-consistent simulations, namely the 
neutrino physics and transport, was approximated in our calculations
by assuming that the high-density core of the neutron star 
(the neutrino-opaque innermost 1.1\,M$_\odot$ with optical depth of more
than $\sim$10) is a black-body neutrino source, whose strength can be
chosen to serve our modeling needs. Exterior to this core we followed 
the neutrino transport and cooling around the neutrinosphere and the
neutrino heating behind the supernova shock by a simple gray scheme
\citep{Scheck+2006}. Parameters of the core model allowed us to regulate
the radiated luminosities such that the energy transfer by neutrinos 
produced explosions with desired energies. The neutrino energy deposition
in the gain layer triggers the growth of hydrodynamic instabilities, 
convective overturn \citep{Herant+1994,Burrows+1995,Janka+1996}
and the standing accretion shock instability \citep[SASI;][]{Blondin+2003,Blondin+2007}, 
that lead to shock deformation and an anisotropic onset of the explosion. 

Only the beginning of the explosion, in which the explosion asymmetries are
established, was simulated with the gray neutrino-transport approximation
combined with the core boundary condition for the time-dependent neutrino 
light-bulb. Again for reasons of computational efficiency, after the first 
1.1--1.4\,s ($t_\mathrm{early}$, which is model dependent and listed
in Table~\ref{tab:3dmodels}) 
the neutrino physics was switched off, the radius of the inner grid boundary
was moved farther out (typically to several 100\,km), and a spherical 
neutrino-driven wind was imposed there as an inflow boundary condition, by 
which we approximated the long-time energy input of the central engine to the
developing supernova blast.

The model names (Table~\ref{tab:3dmodels})
consist of the progenitor's name (with mass value), a number enumerating
the explosion run for a given progenitor, and the extension cw or pw.
This extension specifies whether a constant or a power-law prescription was
applied for the neutrino-wind boundary condition
\citep[for details, see][]{Wongwathanarat+2015}. Dagger symbols in the model
names mark some preliminary binary BSG progenitors that were later updated by 
slightly improved pre-collapse models.

\subsection{Explosion properties}

Large-scale and large-amplitude asymmetries are imprinted on the early 
ejecta due to hydrodynamic instabilities developing in the postshock 
layer before shock revival.
Since low-order spherical harmonics modes with significant contributions
by the dipole and quadrupole dominate the mass distribution, the neutron
star can receive a considerable recoil acceleration by the gravitational
tug-boat mechanism, producing natal kick velocities of several hundred up to more 
than 1000 kilometers per second, in agreement with the measured velocities of young
pulsars \citep{Scheck+2006,Wongwathanarat+2010,Wongwathanarat+2013,Nordhaus+2010,
Nordhaus+2012,Mueller+2017,Mueller+2018,Mueller+2019,Bollig+2020}.
When the supernova shock propagates outward through the star, the initial ejecta
asymmetries act as seed perturbations that instigate the efficient growth of 
secondary hydrodynamic instabilities in the form of Rayleigh-Taylor mixing and
associated Kelvin-Helmholtz shear instabilities. These instabilities grow in 
unstable shells where density and pressure gradients develop opposite signs after
the passage of the shock \citep[e.g.,][]{Chevalier+1978,Bandiera1984,Arnett+1989,
Benz+1990,Hachisu+1990,Fryxell+1991,Mueller+1991,Kifonidis+2003,Wongwathanarat+2015}.

The shock propagation is unsteady with alternating phases of acceleration 
and deceleration, depending on the steepness of the density profile
$\rho_\mathrm{star}(r)$ of the progenitor star.
The shock velocity at a radius $r = R_\mathrm{sh}$ roughly follows the relation
\begin{equation}
	v_\mathrm{sh}^2(R_\mathrm{sh})\approx
	  \frac{\gamma-1}{\frac{4\pi}{3}(1-\frac{1}{\beta})}\cdot
          \frac{E_\mathrm{th}}{\rho_\mathrm{star}(R_\mathrm{sh})\,R_\mathrm{sh}^3}\,,
\label{eq:vshock1}
\end{equation}
where $E_\mathrm{th}$ is the thermal energy in the postshock volume, 
$\gamma$ the adiabatic index of the postshock gas, and 
$\beta = \rho_\mathrm{post}/\rho_\mathrm{star}$ the compression
ratio of postshock to preshock density (for the derivation of 
Equation~(\ref{eq:vshock1}), see Section~\ref{sec:capture}).
Since $\gamma\sim \frac{4}{3}\,...\,\frac{5}{3}$ and the corresponding
$\beta\sim 7\,...\,4$, the coefficient in Equation~(\ref{eq:vshock1})
varies between 0.1 and 0.2, and one gets
\begin{equation}
	v_\mathrm{sh}(R_\mathrm{sh})\sim 3.5\times 10^8\,\mathrm{\frac{cm}{s}}
        \,\,\left(
	\frac{E_{\mathrm{th},50}}{\rho_{\mathrm{star},2}\,R_{\mathrm{sh},10}^3}
	\right)^{\!1/2} \,,
\label{eq:vshock2}
\end{equation}
when representative values of $E_{\mathrm{th},50} = E_\mathrm{th}/(10^{50}\,\mathrm{erg})$, 
$\rho_{\mathrm{star},2} = \rho_\mathrm{star}(R_\mathrm{sh})/(100\,\mathrm{g\,cm}^{-3})$,
and $R_{\mathrm{sh},10} = R_\mathrm{sh}/(10^{10}\,\mathrm{cm})$ are used for the
explosion and stellar parameters.
The unstable conditions form because the postshock matter gets compressed as
the shock passes the C+O/He and He/H composition interfaces and decelerates in
progenitor layers where the stellar density profile is flatter than $r^{-3}$,
i.e., $\rho_\mathrm{star}(R_\mathrm{sh})R_\mathrm{sh}^3$ increases with increasing
shock radius $R_\mathrm{sh}$. The secondary instabilities 
lead to partial fragmentation of the initially large ejecta asymmetries and
to efficient radial mixing of metal-core material (including radioactive 
species) into the He and H layers, as observed in SN~1987A and other supernovae
\citep[for a discussion of the cascade of hydrodynamic instabilities from core bounce 
to shock breakout, see][]{Kifonidis+2003,Kifonidis+2006,Wongwathanarat+2015}.

In our 3D simulations 
the initiation of the supernova blast by neutrino heating was not
modeled fully self-consistently, because the neutron star as a neutrino source
was described in an approximate manner. Therefore the detailed flow dynamics
around the compact remnant during the very first seconds can differ from ab initio 
models that follow the neutrino cooling of the neutron star in detail.
For example, instead of the longer-lasting period of accretion downflows and concomitant
buoyancy-driven outflows witnessed by \citet{Mueller+2017}, \citet{Stockinger+2020},
and \citet{Bollig+2020}, our models exhibit an early phase of
faster initial shock acceleration. A spherical neutrino-driven wind environment
develops around the neutron star within roughly one second, which pushes the 
shock and the asymmetric postshock ejecta. This points to an overestimation of
the strength of the neutrino energy deposition at the onset
of the explosion in our simplified engine model.\footnote{In simple terms,
the reason for this outcome is the use of exaggerated values for the neutrino
luminosities radiated by the high-density core of the forming neutron star. The neutrino
engine employed in our 3D simulations requires the prescription of (time-dependent)
neutrino luminosities at the inner grid boundary at an enclosed mass of 1.1\,M$_\odot$
as well as a choice of the time evolution of the corresponding radius. The radius
evolution of the boundary is supposed to mimic the contraction of the cooling
proto-neutron star, but was chosen to proceed more slowly and less strongly than
it happens for a realistic neutron star. This choice implied a less dramatic density increase
near the inner grid boundary and thus permitted larger numerical time steps because of a 
less constraining Courant-Friedrichs-Lewy condition. This, in turn, led to reduced demands of 
computational resources and allowed us to compute a large number of 3D models. However, it 
also implied an underestimation of the neutrino luminosity created in the accretion layer
of the proto-neutron star, which we had to compensate by imposing overestimated neutrino
luminosities at the inner grid boundary in order to obtain a desired final value
of the explosion energy. Enhanced boundary luminosities combined with reduced
cooling in the accretion layer led to an overactive neutrino-driven wind at the onset of
the supernova explosion and afterwards.}

The overestimation of the wind strength 
is mirrored by the rapid growth of the diagnostic explosion energy 
as a function of post-bounce time. The diagnostic explosion energy
is defined as the total energy (i.e.,
the internal plus kinetic plus gravitational energy) of all postshock matter
for which the sum of the three energy contributions is positive. It is displayed
for a selection of the single-star explosion models (left column) and
a subset of binary explosion models (right column) in the top panels of
Figure~\ref{fig:exp-vs-t}.
The initial rise to 1\,bethe ($=1\,\mathrm{B} = 10^{51}$\,erg) is steeper
than in current self-consistent models \citep[see, e.g.,][]{Bollig+2020}
and overshoots the final value. After
going through a local maximum, however, the explosion energy declines again,
because the outgoing shock sweeps up the gravitationally bound layers of
the oxygen and carbon shells, whose negative binding energy decreases
the diagnostic energy until it reaches the asymptotic value of the
explosion energy only after 
several seconds. At that time the energy input from the central power
source has ceased and the overburden of the remaining stellar shells ahead
of the shock has become negligibly small. In self-consistent supernova 
models that track the neutrino cooling of the neutron star in detail, the
explosion energy builds up to its terminal value more gradually
and more monotonically, but over a similar period of time 
\citep[e.g.,][]{Bollig+2020}.

Despite such differences in the neutrino energy deposition and initial 
blast-wave acceleration, the anisotropic mass ejection happens on the same
time scale and with similar morphological properties. Also the neutron star 
is kicked by the same physical mechanism. Of course, individual cases, 
comparing fully self-consistent 3D simulations with parametric explosion models,
for example on the basis of the same explosion energy for a given progenitor, must
be expected to differ in many properties (quantitatively and partly also
qualitatively), in particular of the innermost, neutrino-heated ejecta.
Nevertheless, the subsequent growth of secondary instabilities at the 
composition interfaces
and the long-time evolution of the supernova asymmetries do not depend
on fine details of the flow dynamics during the first few seconds of the
blast wave. We are therefore confident that our main conclusions based on the
3D explosion results are not jeopardized by
the modeling approximations employed in our set of supernova simulations. The
conclusions we can draw possess more general validity and apply also to fully 
self-consistent explosion calculations \citep[as, for example, presented 
in the long-time simulations of][]{Mueller+2018,Stockinger+2020}.

\subsection{Neutron star kicks and explosion asymmetry}

Figure~\ref{fig:exp-vs-t} also displays, as functions of time after core bounce,
the baryonic mass of the neutron star,\footnote{In post-processing the results
of our simulations, we defined the neutron star ---in line with previous publications 
of the Garching group--- by the mass that possesses a baryonic density of 
$\rho \ge 10^{11}$\,g\,cm$^{-3}$. Correspondingly, all neutron star quantities
listed in Table~\ref{tab:3dmodels} and displayed in Figures~\ref{fig:exp-vs-t} and 
\ref{fig:fallback-vs-t} (e.g., $J_\mathrm{NS}$, $T_\mathrm{spin}$, etc.) were evaluated 
for this mass. In the simulations discussed in the present paper, where the
high-density core of the neutron star with a baryonic mass of 1.1\,M$_\odot$
was replaced by an inner grid boundary, the neutron star consists of this core
mass plus all mass on the computational grid with a density higher than 
$10^{11}$\,g\,cm$^{-3}$.} its kick velocity, and the momentum
asymmetry parameter of the postshock ejecta. The neutron star's kick velocity is
computed by the requirement of linear momentum conservation from the momentum
of the ejected gas as 
\begin{equation}
	\pmb{v}_\mathrm{NS}(t) = -\,\frac{\pmb{P}_\mathrm{gas}(t)}{M_\mathrm{NS}(t)}\,,
	\label{eq:vns1}
\end{equation}
or
\begin{equation}
	v_\mathrm{NS}(t) = |\pmb{v}_\mathrm{NS}(t)| = \alpha_\mathrm{ej}(t)\,\,
	\frac{P_\mathrm{ej}(t)}{M_\mathrm{NS}(t)}\,,
        \label{eq:vns2}
\end{equation}
with $\alpha_\mathrm{ej}$ being the momentum asymmetry parameter defined by
\begin{equation}
	\alpha_\mathrm{ej}(t) = \frac{|\pmb{P}_{\mathrm{gas}}(t)|}{P_{\mathrm{ej}}(t)}\,,
	\label{eq:alpha}
\end{equation}
where
$\pmb{P}_\mathrm{gas}=\int_{R_\mathrm{gain}}^{R_\mathrm{sh}}\rho\,\pmb{v}\,\mathrm{d}V$ 
is the total linear momentum of the ejecta between the gain radius, $R_\mathrm{gain}$, 
and the supernova shock, $R_\mathrm{sh}$, and
$P_\mathrm{ej}=\int_{R_\mathrm{gain}}^{R_\mathrm{sh}}\rho\,|\pmb{v}|\,\mathrm{d}V$
is the total momentum stored in the ejecta, which becomes equal to the total radial 
momentum when the ejecta expand essentially radially.\footnote{Because in the discussed
3D simulations the neutrino transport is treated in a highly simplified way and the 
dense core of the neutron star is replaced by a 1D boundary condition, we do not consider
neutron star kicks associated with anisotropic neutrino emission. These kicks are typically
of the order of some 10\,km\,s$^{-1}$ up to about 100\,km\,s$^{-1}$ and therefore they are
usually subordinate contributions to the hydrodynamic kicks discussed here. We note that
a similarly small fraction of the ejecta momentum might be connected to anisotropic
neutrino absorption in the ejecta gas and should be compensated by a net momentum of the
escaping neutrinos in the opposite direction. For 3D simulations of supernova explosions, 
where these effects were carefully taken into account, see \citet{Stockinger+2020} and 
\citet{Bollig+2020}.} In Figure~\ref{fig:exp-vs-t} we
present the neutron star kick velocity computed with the baryonic mass of the 
neutron star, $M_\mathrm{NS}(t)$, used in Eq.~(\ref{eq:vns2}). Referring to the baryonic 
mass instead of the gravitational mass has two reasons. On the one hand, 
our simple neutrino emission model employed in
the 3D simulations does not permit a reliable, time dependent evaluation of the
gravitational mass of the neutron star, $M_\mathrm{NS,g}(t)=\,
M_\mathrm{NS}(t) - E_\nu(t)\,c^{-2}$, where $E_\nu(t)$ is the time-integrated neutrino energy 
loss of the compact remnant at time $t$. On the other hand, in the case that the ejecta 
momentum is constant, a neutron star that radiates neutrinos isotropically in its rest 
frame should not be further accelerated due
to the energy loss by the neutrino emission, despite the associated decrease of 
the neutron star's gravitational mass. Because the escaping neutrinos carry away
linear momentum in the observer frame, the linear gas momentum is shared between the
neutrinos and the neutron star, whose gravitational mass decreases but whose velocity
remains constant. Describing such
effects consistently in numerical simulations requires fully relativistic 
radiation hydrodynamics including the treatment of the neutron star motion on the 
computational grid instead of fixing the neutron star at the coordinate center. 

In order to also provide corresponding upper bounds of the neutron star kick 
velocities, we list in Table~\ref{tab:3dmodels} the final values of the kick 
velocities at the end of our simulations, $v_\mathrm{NS}(t_\mathrm{end})$, as computed
with the final gravitational masses of the cold neutron stars, which we obtained
from equation~(36) of \citet{Lattimer+2001} for the binding energy of a neutron
star with a baryonic mass of $M_\mathrm{NS}(t_\mathrm{end})$ and a radius of 12\,km.
It is evident that the results for $v_\mathrm{NS}(t_\mathrm{end})$ in Table~\ref{tab:3dmodels}
are at most about 10\% higher than the corresponding asymptotic values at the end of
the displayed evolution in Figure~\ref{fig:exp-vs-t}.

Figure~\ref{fig:exp-vs-t} shows that the mass of the neutron star reaches a local
maximum at several 100 milliseconds after bounce and then declines over some seconds by
a few hundredths of a solar mass ($\lesssim$\,0.08\,M$_\odot$) before it increases
again slightly or moderately by short-time (within several 100\,s) and long-time 
(over hours) fallback of initially ejected matter
that is unable to become gravitationally unbound. The early decrease of the
neutron star mass from the local maximum to a local minimum is a consequence
of the neutrino-driven wind that boosts the explosion energy with our neutrino-engine
model. This can be seen by the (approximate) correlation of the local mass minimum 
with the maximum of the explosion energy. Again, such a behavior is absent or less
pronounced in fully self-consistent supernova simulations with a detailed modeling of 
the neutrino emitting neutron star, where the neutron star mass and the explosion energy
approach their final values more monotonically (see \citealt{Bollig+2020}, 
\citealt{Mueller+2017}).

The kick velocities of the neutron stars (third row of Figure~\ref{fig:exp-vs-t})
asymptote to their final values at about 100\,s, i.e., after the first episode
of fallback (see Sect.~\ref{sec:fallback}), but near final values are reached
already after a few seconds post bounce. In our most extreme models $v_\mathrm{NS}$
reaches 1100\,km\,s$^{-1}$.
In the far majority of cases there is a continuous increase of the
kick velocity over time, because the fallback affects mainly
the slowest, innermost ejecta in the directions or in the hemisphere where
the blast wave is weaker. This implies that the faster part of the ejecta
escapes with an enhanced asymmetry and larger linear momentum
$|\pmb{P}_\mathrm{gas}|$ \citep{Janka2013}. However, in some of the plotted cases,
$v_\mathrm{NS}$ slightly declines after a peak value at several seconds.
This suggests some stochasticity in the fallback dynamics, and if the fallback
mass is relatively large ---all corresponding cases have $M_\mathrm{fb}\gtrsim
0.1$\,M$_\odot$--- also material in the bulk of the asymmetric ejecta can be 
affected by the fallback. With a part of this material returning back to the
neutron star, the remaining ejecta possess a smaller momentum asymmetry and 
the net kick of the neutron star, which is the opposite of the linear momentum
of the ejecta, is also reduced. This possibility was recently discussed in the
context of black-hole forming fallback supernovae by \citet{Chan+2020}. 

The momentum asymmetry parameter, $\alpha_\mathrm{ej}(t)$, evolves as expected
(bottom panels in Figure~\ref{fig:exp-vs-t}). It fluctuates stochastically before
the explosion sets in, because the violent hydrodynamic mass motions associated
with the hydrodynamic instabilities in the postshock region have no stable
direction. Only after shock revival
the asymmetry of the beginning explosion settles into its final shape and a well
defined momentum asymmetry develops. The $\alpha$-parameter reaches a maximum at
roughly 1\,s after bounce (with considerable case-to-case variation), which is close
to the time when the diagnostic explosion energy peaks in the considered models. 
This is also the time when the mass of the postshock ejecta that carries the imprints 
of the explosion asymmetry is minimal. After its peak value, $\alpha_\mathrm{ej}(t)$ 
declines basically 
monotonically as the outgoing shock sweeps up an increasing mass of the spherically 
stratified progenitor star. Since, apart from fallback, $|\pmb{P}_\mathrm{gas}|$
is conserved, Eq.~(\ref{eq:alpha}) implies that the value of $\alpha_\mathrm{ej}(t)$ 
must drop with growing postshock ejecta mass, because
$P_\mathrm{ej}=\int_{R_\mathrm{gain}}^{R_\mathrm{sh}}\rho\,|\pmb{v}|\,\mathrm{d}V
= \bar{v}_\mathrm{ej}M_\mathrm{ej} \sim \sqrt{2 E_\mathrm{kin}M_\mathrm{ej}}$
(where $\bar{v}_\mathrm{ej}$ is the average velocity of the ejecta). Once the 
explosion energy has saturated, the kinetic energy of the blast, $E_\mathrm{kin}$,
varies only within a factor of a few (being converted to internal energy
when the supernova shock slows down and rising again in phases of shock
acceleration), and the main time dependence of 
$\alpha_\mathrm{ej}(t)$ results from the growing mass $M_\mathrm{ej}$ of the 
postshock matter (and potentially from fallback effects).

\subsection{Neutron star spins from fallback}
\label{sec:fallback}

All supernova simulations in our model set were computed without taking
into account progenitor rotation, either because the stellar progenitors did 
not rotate or because their core and envelope rotation was so slow that it 
was dynamically irrelevant for the core collapse
and postbounce evolution. However, the hydrodynamic instabilities, convection
and SASI, which support the onset of the explosion and lead to neutron star kicks 
by asymmetric mass ejection, can also spin up the nascent neutron star
\citep[see, e.g.,][]{Blondin+2007,Fernandez2010,Wongwathanarat+2013,Guilet+2014,
Kazeroni+2016,Kazeroni+2017}.
The angular momentum is transferred when accretion flows hit the neutron star
off-center, and at the onset of the explosion the compact remnant has received
the negative angular momentum of the outward expanding ejecta. 

Because of weak SASI activity and a relatively quick onset of the explosion,
the net effect of this angular momentum separation between neutron star and 
ejecta is, however, relatively feeble in our models. This is in line with 
results of \citet{Rantsiou+2011}, and the neutron star angular
momentum $J_\mathrm{NS}(t_\mathrm{early})$ is at most a few times $10^{46}$\,erg\,s
at $t_\mathrm{early}= 1.1$--1.4\,s, which is the time when the ejecta have just 
separated from the neutron star \citep[Table~\ref{tab:3dmodels} and bottom panel of
Figure~\ref{fig:fallback-vs-t}; see also][]{Wongwathanarat+2013}. In the 
case of rigid rotation, angular momentum of this magnitude corresponds to spin 
periods of
\begin{align}
	T_\mathrm{spin} &= \frac{2\pi I_\mathrm{NS}}{J_\mathrm{NS}} \notag\\
	& \approx 1.09\,\,[\mathrm{s}]\,
        \left(\frac{M_\mathrm{NS}}{1.5\,\mathrm{M}_\odot}\right)
	\left(\frac{R_\mathrm{NS}}{12\,\mathrm{km}}\right)^{\! 2}
	\left(\frac{J_\mathrm{NS}}{10^{46}\,\mathrm{erg\,s}}\right)^{\!-1} \,,
	\label{eq:Tspin}
\end{align}
i.e., of typically hundreds of milliseconds to seconds.
Here, $I_\mathrm{NS}$ is the neutron star moment of inertia. 
In Equation~(\ref{eq:Tspin}) we use the idealization of a homogeneous 
density distribution ($\rho = \mathrm{const}$) in a spherical neutron star, 
which yields
$I_\mathrm{NS}\sim \frac{2}{5}M_\mathrm{NS}R_\mathrm{NS}^2 \approx
1.72\times 10^{45}\,\mathrm{g\,cm}^2\,(M_\mathrm{NS}/1.5\,\mathrm{M}_\odot)
(R_\mathrm{NS}/12\,\mathrm{km})^2$ and is a sufficiently good approximation
for rough estimates.\footnote{Interpreting $M_\mathrm{NS}$ in our 
expression for $I_\mathrm{NS}$ as the gravitating mass, the fit formula of
\citet{Lattimer+2005} yields a value for the neutron star moment of inertia 
that is 9.6\% lower than our approximate estimate for a neutron star with 
a mass of 1.5\,M$_\odot$ and an assumed radius of 12\,km.}
For the values of $T_\mathrm{spin}$ listed in
Table~\ref{tab:3dmodels} we employed the fit formula of 
\citet{Lattimer+2005} with the gravitational neutron star mass listed in
the table and 12\,km for the neutron star radius. 

In fully self-consistent explosion models of ultrastripped and low-mass
progenitors, respectively, \citet{Mueller+2018} and \citet{Stockinger+2020}
obtained $J_\mathrm{NS}(t_\mathrm{early})\sim 
10^{45}$--$10^{46}$\,erg\,s (spin periods of $\sim$1--10\,s), and 
\citet{Mueller+2017}, \citet{Mueller+2019}, \citet{Chan+2020}, and 
\citet{Bollig+2020} found
values of $J_\mathrm{NS}(t_\mathrm{early})$ up to several $10^{47}$\,erg\,s
for a wider range of progenitor masses. If conserved and redistributed, these
values would account for minimal early spin periods around 15--20\,ms for rigidly
rotating neutron stars of 12\,km radius. However, the initial angular momentum
is concentrated in a narrow accretion layer of relatively little mass near the
proto-neutron star surface, which spins with millisecond periods. Neutrino 
emission can very efficiently extract large amounts of angular momentum from 
such rapidly spinning near-surface layers of the proto-neutron star. The
neutrino-mediated loss of angular momentum, which can reach up to more
than $10^{48}$\,erg\,s \citep{Bollig+2020}, can thus drastically decelerate 
the initial neutron star rotation.
 
In all of these cases the initial rotation rates of the neutron stars just
after the onset of the explosion are therefore likely to be 
overridden by later fallback,\footnote{In the 3D simulations discussed
in this paper the fallback mass was determined by the matter that left the
inner grid boundary with negative velocities. The inner grid boundary during
the long-time simulations (i.e., after the calculations with neutrino physics
and after the neutrino-driven wind phase) was always
placed at a location where this outflow was supersonic, in order to avoid 
that perturbations and artifacts connected to the boundary condition could 
propagate back onto the computational grid. The robustness of the results for
the integrated mass of the outflow was tested by running simulations with the
inner grid boundary placed at different radii. The fallback rate of the matter,
$\dot{M}_\mathrm{fb}(t)$, and the specific angular momentum of this material,
$j_\mathrm{fb}(t)$, were determined for the mass flow leaving the grid at the
inner grid boundary as a function of time. As stated in the main text, for the
late-time neutron star quantities given in Table~\ref{tab:3dmodels} and in 
Figures~\ref{fig:exp-vs-t} and \ref{fig:fallback-vs-t},
we assumed that all of the thus determined fallback mass is
accreted by the neutron star. Of course, this is a working hypothesis and not
a result of detailed calculations.} which can carry up to more than
$10^{49}$\,erg\,s of angular momentum to the compact remnant. This is
typically 100--1000 more than is
transferred by the nonradial hydrodynamic flows in the early explosion phase. 
The effect is visible in Figure~\ref{fig:fallback-vs-t} (bottom panels), where the 
angular momentum of the neutron stars as a function of time rises steeply and 
grows by 2--3 orders of magnitude as soon as fallback sets in, {\em assuming} that
all of the fallback mass (top panels of Figure~\ref{fig:fallback-vs-t}) is accreted
by the neutron star. Similar results were obtained by \citet{Stockinger+2020} and
\citet{Chan+2020}. 

It should be noted that none of the existing 
3D simulations tracks the destiny of the fallback mass until it ends up
on the accretor. Instead, because of numerical time-stepping constraints,
inner grid boundaries are used where the inward directed mass flow leaves the 
computational grid and is counted as accreted fallback. Tests have verified 
that the results for the determined infalling mass are not sensitive to the 
exact positioning of the inner boundary radius
within numerically feasible settings \citep{Chan+2020,Ertl+2016}.
However, what exactly happens with this mass remains unclear and requires
a new generation of 3D models to follow the flow to the close proximity of
the gravitating compact object.

The fallback evolution differs strongly from case to case, see
Figure~\ref{fig:fallback-vs-t}, top and second rows, for the integrated fallback
mass, $M_\mathrm{fb}(t)$, and the rate of mass-fallback, $\dot M_\mathrm{fb}(t)$,
respectively. Higher fallback masses typically show a tendency to correlate with 
lower explosion energies (compare with the top panels of Figure~\ref{fig:exp-vs-t}
and see Table~\ref{tab:3dmodels}).
Despite the variations between the individual cases, three different fallback
episodes can basically be delineated. 
\begin{itemize}
\item[(1)]      The first fallback phase starts several seconds after core bounce
 		at the time when the central neutrino source abates and the 
		growth of the explosion energy slows down because the 
		neutrino-driven wind (or, more generally, the neutrino-driven
		mass outflow) becomes weaker. After the onset of the explosion,
		the longer-lasting supply of its energy
                is provided by a spherical neutrino-driven wind in the models
		discussed in this paper (with the wind imposed as an inner boundary 
                condition, see Section~\ref{sec:SNmodeling}). In the fully
		self-consistent explosion models of \citet{Mueller+2017}, 
		\citet{Stockinger+2020}, and \citet{Bollig+2020},
		it is accomplished by neutrino heating of persistent downflows,
		reversing them basically entirely to energy-loaded, buoyant outflows
		that lead to a continuous rise of the explosion energy over 
		several seconds. When the fueling of energy gets weaker, 
		a rarefaction wave sets in and pulls
		the innermost part of the ejecta back to the center \citep[discussed 
		for the first time by][]{Colgate1971}.
		This is seen as a steep increase of the mass-fallback rate
		in a first, early peak at $t\lesssim 10$\,s after bounce (see
		Figure~\ref{fig:fallback-vs-t} and Figure~25 in \citealt{Stockinger+2020}
		for low-mass explosion models). In self-consistent explosion models
		of massive progenitors, where a spherical neutrino-driven wind does
		not develop, the seconds-long period of simultaneous inflows and 
                outflows transitions gradually into this first episode of fallback.
\item[(2)]      The second fallback phase, which can connect or overlap with the 
                first phase, depending on the properties of the shell structure of the 
                progenitor star, is associated with the reverse shock that may form
		when the supernova shock is decelerated after passing the 
		CO/He-core interface \citep{Fryxell+1991}. 
                Moving back inward, the reverse shock amplifies
		a second wave of fallback, producing a local maximum in 
		$\dot M_\mathrm{fb}(t)$ between roughly 10\,s and several 100\,s 
                after bounce in some models, for extended RSG progenitors also
                later (see Figure \ref{fig:fallback-vs-t}).
		Some of the displayed cases exhibit only a flattening of the
		decline of the mass-fallback rate during this period (as visible
		also in the 9\,M$_\odot$ RSG case of \citealt{Stockinger+2020} and the 
		20.4\,M$_\odot$ RSG explosion model of \citealt{Ertl+2016}).
\item[(3)]      The third fallback phase is associated with the reverse shock from the
	        shock deceleration in the H-envelope after the shock has passed
		the He/H-interface \citep{Woosley1988,Shigeyama+1988,Chevalier1989}.
		This phase occurs typically at roughly $t \sim 10^3$--$10^4$\,s in the
		BSGs considered here and at $t \gtrsim 10^5$\,s in RSGs \citep[see 
		also][]{Ertl+2016}. In most cases it is barely visible in the integrated
                fallback mass, because the mass-fallback rate is much lower than in the
                other two phases (Figure \ref{fig:fallback-vs-t}).
\end{itemize}
In most cases shown in Figures~\ref{fig:exp-vs-t} and \ref{fig:fallback-vs-t}, the 
three phases connect continuously and monotonically with each other, whereas in a
smaller subset of cases the arrival of the reverse shock in the center coincides
with an abrupt increase of the fallback rate. 
The functional decline of the fallback rate exhibits great diversity, because it
depends on the details of the shell structure of the progenitor star, which 
differs between RSGs and BSGs as well as between single-star progenitors (left panels 
of Figures~\ref{fig:exp-vs-t} and \ref{fig:fallback-vs-t}) and binary-merger progenitors
(right panels of Figures~\ref{fig:exp-vs-t} and \ref{fig:fallback-vs-t}). Moreover,
in 3D the decrease of the fallback rate is more monotonic than in 1D, because the 
reverse shock and mass-infall are nonspherical and secondary shocks smooth the
flow.

During the first and second episodes of fallback, the angular momentum of the neutron star
increases from initially some $10^{46}$\,erg\,s to values between $\sim$$10^{47}$\,erg\,s
and several $10^{48}$\,erg\,s, provided all of the fallback mass gets
accreted onto the neutron star. This would correspond to a decrease of the neutron 
star rotation periods from an early value of some 100 milliseconds to seconds to 
later values of several milliseconds to tens of milliseconds (Equation~\ref{eq:Tspin}).

The third, late episode of fallback, although associated with very little mass,
can still carry high amounts of angular momentum. If the fallback matter were entirely
accreted by the neutron star, it would boost the remnant's angular momentum by 
another factor of 5--10 up to nearly $10^{50}$\,erg\,s in the most extreme cases
(Figure~\ref{fig:fallback-vs-t}, Table~\ref{tab:3dmodels}). Nominally,
this would speed up the neutron star rotation to millisecond or even to
sub-millisecond periods, i.e. close to or even beyond the Keplerian break-up 
frequency (Table~\ref{tab:3dmodels}).
These results point to interesting and relevant questions: What fraction of the 
fallback mass can make its way onto the neutron star and what happens with the 
rest?

\begin{figure}[tb]
\begin{center}
        \includegraphics[width=\columnwidth]{f3.pdf}
        \caption{Distribution of relative angles between spin and kick
	directions at early and late times for our set of 3D explosion
	models (Table~\ref{tab:3dmodels}). The binning is in intervals
	of 10 degrees. The 3D simulations discussed in this paper
        and in other recent publications do not exhibit any tendency of spin-kick
        alignment. Approximations in the current hydrodynamical models that could 
        be the reasons for this fact are mentioned in Section~\ref{sec:problems}.
	Physics effects that may lead to spin-kick alignment but are missed in 
	current 3D supernova simulations will be discussed in 
        Section~\ref{sec:scenario}, and the implications of these effects in 
        Section~\ref{sec:discussion}.}
\label{fig:histoskangle}
\end{center}
\end{figure}

Not unexpectedly, the spins and kicks of the neutron stars are basically 
randomly
oriented relative to each other at $t_\mathrm{early} = 1.1$--1.3\,s post
bounce, i.e., before the fallback, as well as after the fallback a day or 
more later, when our simulations were stopped (see Table~\ref{tab:3dmodels}).
Figure~\ref{fig:histoskangle} displays, for both instants, histograms of the
distributions of angles $\theta_\mathrm{sk}$ between spin axis and kick
direction for our entire set of models listed in Table~\ref{tab:3dmodels}.
These distributions reflect, roughly, the 
$\sin(\theta_\mathrm{sk})$-relation expected for isotropic 
conditions, possibly with a slight preference for spin-kick-angles
near 90 degrees.\footnote{The minimum around $\sim$105$^\circ$ and the
gap at $\sim$135$^\circ$ are probably fluctuations due to low-number 
statistics and are filled when plotting the distributions with coarser 
binning.} This is in line with findings reported by
\citet{Chan+2020}, \citet{Powell+2020}, and \citet{Stockinger+2020}. 

The near-isotropy of the
early distribution can be understood from the fact that the early
spin-up of the neutron star happens by randomly oriented off-center
impacts of asymmetric accretion downflows associated with the 
hydrodynamic instabilities in the postshock layer at around the 
onset of the explosion. In contrast, the 
neutron star kick is caused by the gravitational tug-boat mechanism
associated with the gravitational pull of the asymmetric ejecta,
acting over time scales of several seconds. Both effects are not correlated.
The near-isotropy of the distribution of the final neutron star spins relative
to the kick directions, on the other hand, is a consequence of the fact 
that the corresponding angular momentum is transferred in fallback that
happens only later. Again, kick and spin-up mechanisms have only a 
loose connection such that the fallback occurs preferentially from the
weakest side of the explosion, towards which the neutron star kick
is directed. But since the neutron star is artificially (for numerical
reasons) pinned to the center of the computational polar grid in our 
3D simulations and in all other works cited above in the context of 
neutron star kicks, fallback is swept towards the compact remnant from all
directions. Therefore, if this infalling matter gets accreted, there
is no reason to expect any preference of the neutron star's spin 
direction. In a nutshell, spin-kick alignment, as suggested by 
observations \citep[e.g.,][]{Johnston+2005,Ng+2007,Noutsos+2012,
Noutsos+2013,Yao+2021} remains unexplained.

\subsection{Implications and problems}
\label{sec:problems}

The results of our simulations, combined with those of other recent 3D long-time
simulations of supernove explosions \citep{Chan+2020,Powell+2020,Stockinger+2020}
demonstrate that fallback plays a dominant, so far underestimated role 
for the question how neutron stars obtain their birth spins. The 
angular momentum connected to the fallback matter is appreciable and the
formation of accretion disks around neutron stars, which requires a specific
angular momentum of 
\begin{equation}
j_\mathrm{Kepler} \gtrsim 1.55\times 10^{16}\,\mathrm{\frac{cm^2}{s}}\,\,
\left(\frac{M_\mathrm{NS}}{1.5\,M_\odot}\right)^{\! 1/2}
\left(\frac{R_\mathrm{NS}}{12\,\mathrm{km}}\right)^{\! 1/2} \,,
\label{eq:jkepler}
\end{equation}
may be quite a common phenomenon.

Our simulations indicate, however, that the accretion of the entire mass 
falling back after its initial expansion, or of a major part of it,
does not seem to be compatible with the long initial spin periods (tens
of milliseconds to hundreds of milliseconds) of neutron stars estimated
from observations \citep[e.g.,][]{Popov+2012,Igoshev+2013,Noutsos+2013}.
Most of the estimates of the final neutron star spin periods (after
the assumed fallback accretion) listed in the last column of 
Table~\ref{tab:3dmodels} are in the range of a few milliseconds only,
some even below one millisecond. In fact, if all of the fallback matter with 
its angular momentum ended up on the neutron stars, this would imply that
the neutron stars could be driven to their limit for tidal break-up.
This suggests that fallback accretion must be incomplete and only a smaller
fraction of the fallback matter (or of the angular momentum associated with
it) gets integrated into the compact remnants.

The underlying reason could be inefficient (disk) accretion by the 
neutron star, in which most of the inflowing mass is re-ejected 
instead of being added onto the neutron star. However, we hypothesize 
that another effect is likely to also play an important role,
namely the fact that the neutron star moves out of the explosion center
after having received its recoil momentum during the first seconds of the 
anisotropic supernova explosion. As we will argue below, this movement
of the neutron star could diminish its ability to gravitationally capture 
fallback matter, thus naturally reducing the amount of material that
reaches the immediate vicinity of the compact object to be potentially 
accreted.

Current supernova simulations neither follow the destiny of the 
fallback matter, nor do they include the gradual drift of the 
neutron star away from its birth location. Keeping the neutron star 
fixed at the center of the spherical polar grid and thus at the center 
of the explosion means that the neutron star
is treated as if it had an infinite inertial mass.
With this assumption its velocity remains zero although it adopts the  
inverse of the linear momentum that is carried away by anisotropic mass 
ejection and anisotropic neutrino emission.
The same numerical constraint and underlying assumption was applied in
all current supernova models that used spherical polar coordinates
\citep[e.g.,][]{Stockinger+2020,Chan+2020,Powell+2020}.
This will serve us as a motivation to assess the shortcomings connected
to such an approximation and to consider possible changes for
the fallback accretion by the neutron star when this approximation is
removed. We will argue that the movement of the kicked compact 
remnant may have important qualitative and quantitative consequences
and could, for example, lead to conditions that are more 
favorable for spin-kick alignment than found in existing 3D supernova 
models.

\begin{figure}[tb]
\begin{center}
        \includegraphics[width=\columnwidth]{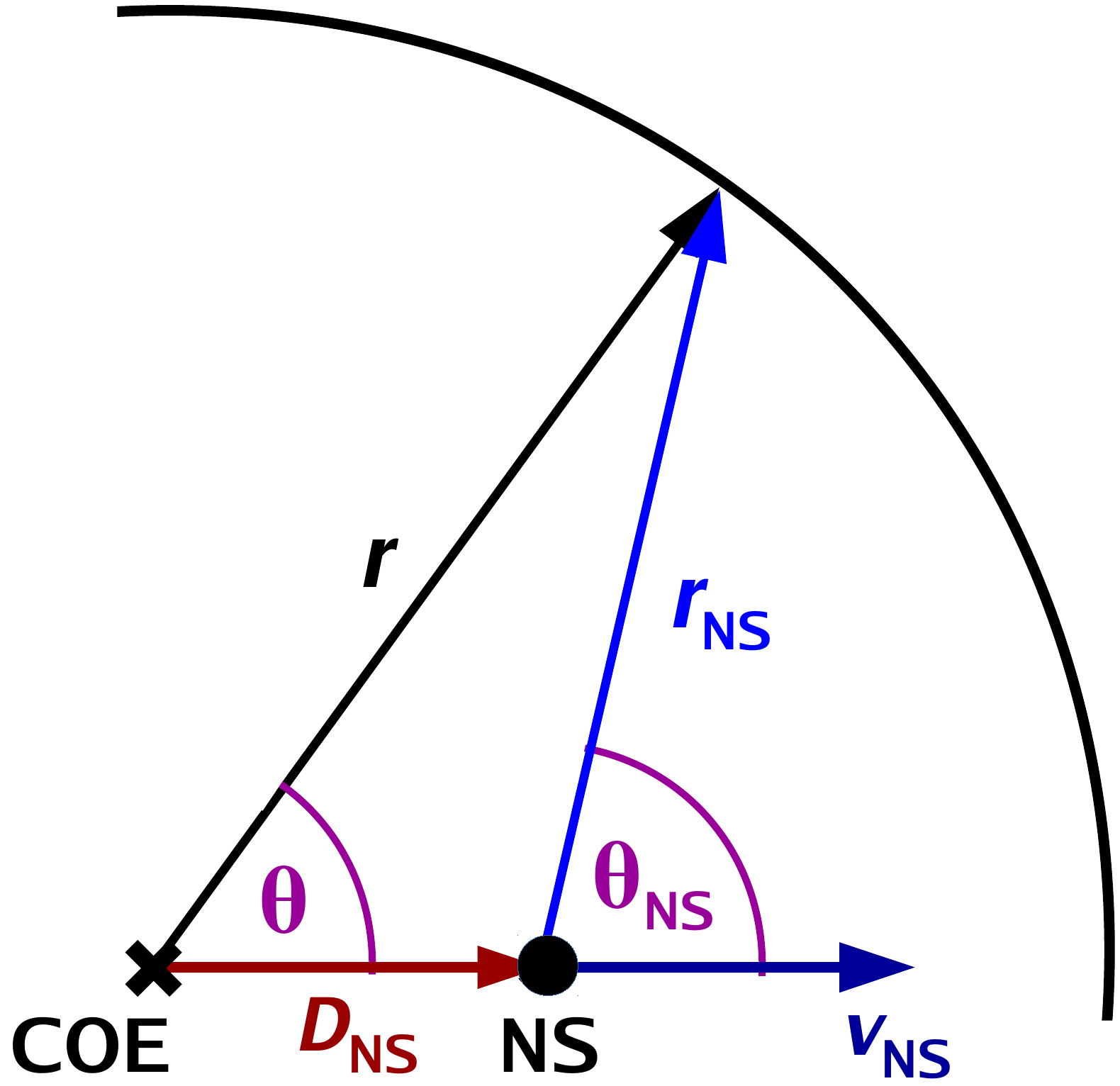}
        \caption{Basic geometry of the kicked neutron star (NS).
        Because of its natal kick velocity, $\pmb{v}_\mathrm{NS}$, the neutron
        star is displaced from the center of the explosion (COE) by a vector
        distance $\pmb{D}_\mathrm{NS}$. The vectors $\pmb{D}_\mathrm{NS}$ and
        $\pmb{v}_\mathrm{NS}$ are considered to have the same directions.}
\label{fig:Graphics1}
\end{center}
\end{figure}

\section{Fallback with neutron star migration and spin-kick alignment}
\label{sec:scenario}

The deficiencies of current 3D supernova simulations discussed in  
Section~\ref{sec:simulations}, in particular their inability to explain the
systematic trend of spin-kick alignment inferred for many young pulsars from
observations (see Section~\ref{sec:observations} for a summary), motivates us
to introduce a refined and revised picture of fallback accretion in supernovae 
in the following section. A crucial point in this revised scenario is the
inclusion of the neutron star's natal kick motion with velocity 
$v_\mathrm{NS}$, which is assumed to be determined quickly (i.e., within a few
seconds) after the onset of the explosion. The kick leads to a time-dependent 
displacement $D_\mathrm{NS}(t)$ of the neutron star from the center of the
explosion (COE). Figure~\ref{fig:Graphics1} displays the basic elements of the
emerging geometry.

\begin{figure*}[tb!]
\begin{center}
        \includegraphics[width=0.75\textwidth]{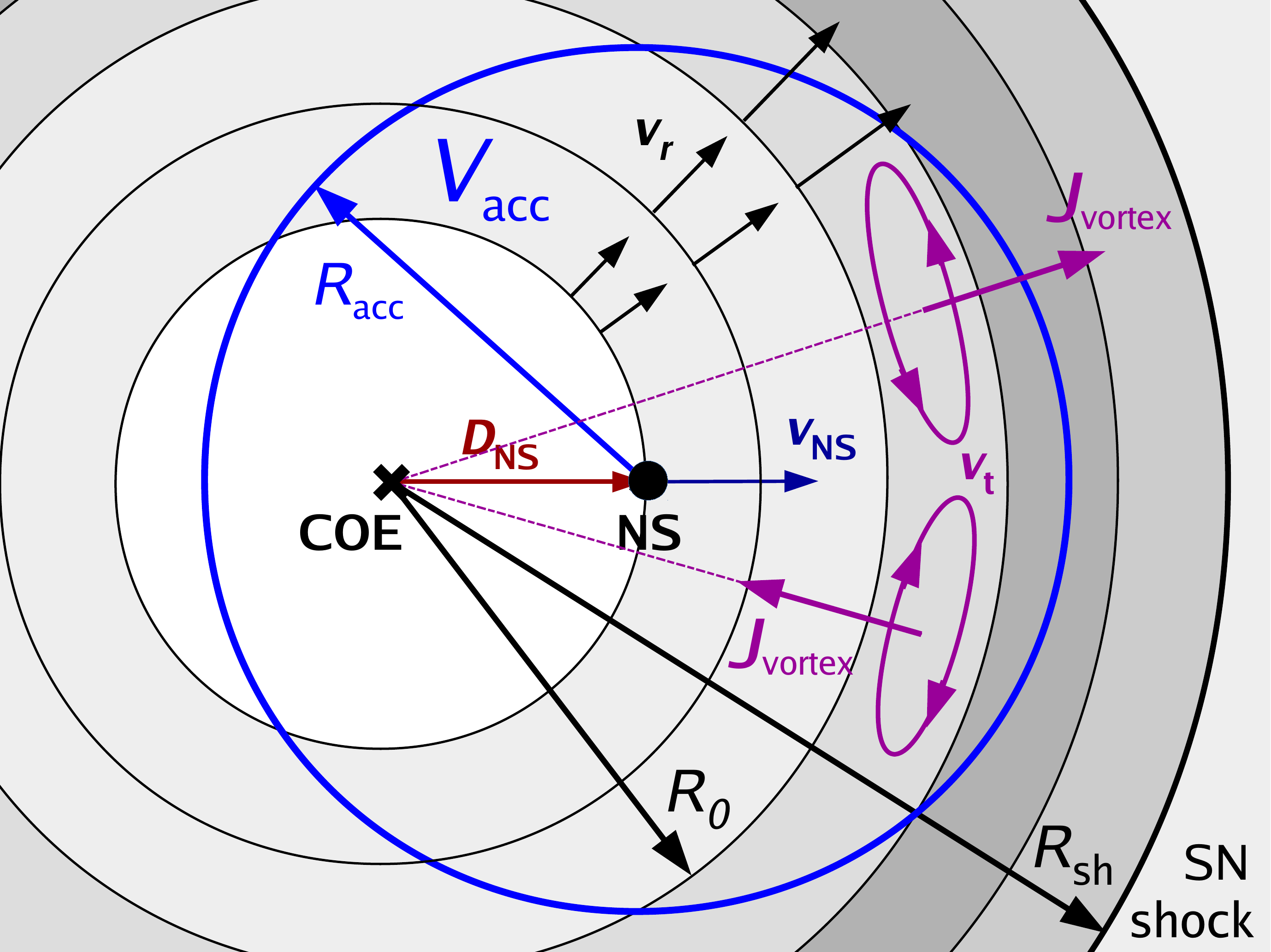}
        \caption{Expanding supernova ejecta with radial velocity field
        $\pmb{v}_r(\pmb{r})$, which is assumed to be spherically symmetric.
        For the purpose of illustration we display the non-extreme case
        where $D_\mathrm{NS}\sim R_\mathrm{acc}$, although the condition of
        Equation~(\ref{eq:ratioDR}) should hold frequently and is in our
        main focus of interest.
        The ejecta are considered to contain tangential flow vortices
        with velocities $\pmb{v}_\mathrm{t}(\pmb{r})$
        and corresponding angular momenta $\pmb{J}_\mathrm{vortex}$.
        Since these vortices lie in spherical shells around the COE
        and thus have velocities perpendicular to the radial directions,
        they should be imagined to be drawn in perspective, representing
        fluid motions in planes perpendicular to the vectors $\pmb{J}_\mathrm{vortex}$,
        which in turn are aligned (or anti-aligned) with the radius vectors from
        the COE to the vortex centers. The ejecta pile up in a dense shell
        (inner radius $R_0$) during deceleration phases of the supernova
        (SN) shock at radius $R_\mathrm{sh}$. The accretion volume
        $V_\mathrm{acc}$ of the displaced neutron star (NS) is assumed
	to be nearly spherical (radius $R_\mathrm{acc}$), which is 
        justified on grounds of the discussion in Section~\ref{sec:moredetails}.}
\label{fig:Graphics3}
\end{center}
\end{figure*}

Another ingredient of crucial importance is the accretion or capture
radius of the neutron star, $R_\mathrm{acc}$. This radius determines the neutron
star's accretion volume $V_\mathrm{acc}$, within which the gravitational potential
of the central mass influences the surrounding distribution of the supernova ejecta.
We shall introduce $R_\mathrm{acc}$ quantitatively in Section~\ref{sec:capture},
first by providing a rough estimate of the magnitude (Section~\ref{sec:roughviews}) 
and then by a more detailed consideration (Section~\ref{sec:moredetails}).
Accretion radius and corresponding accretion volume around the neutron star are
graphically visualized in Figure~\ref{fig:Graphics3}.

The radially expanding supernova ejecta are imagined to contain tangential vortex 
flows, which are imagined to be carried along with the radial flow that moves
with expansion velocity $\pmb{v}_r$; see Figure~\ref{fig:Graphics3} for a 
representation of the envisioned situation. These tangential vortices with 
velocities $\pmb{v}_\mathrm{t}$ can be relics of nonradial
(turbulent) gas motions in the initial explosion ejecta or of convective mass
motions in the inner burning shells of the progenitor. When the inner ejecta are
gravitationally captured by the neutron star, fall back, and are potentially
accreted by the compact remnant, their vorticity will add angular momentum
($\pmb{J}_\mathrm{vortex}$ for individual swirls) onto the neutron star.
We shall evaluate the angular momentum contained by the neutron star's
accretion volume (Figure~\ref{fig:Graphics3}) in Section~\ref{sec:alignment},
first for ejecta in spherical shells with constant density
(Section~\ref{sec:angmomaccvol}) and then for ejecta shells containing density 
inhomogeneities (Section~\ref{sec:densityvariations}). We shall see that this
angular momentum encompasses the contributions from all tangential vortex 
motions, and we shall argue that the asymmetric fallback accretion by a 
neutron star shifted out of the COE (measured by distance $D_\mathrm{NS}$) 
may yield a natural explanation for spin-kick alignment
(Section~\ref{sec:skalign}). We shall provide rough numerical estimates
of the corresponding angular momentum obtained by the neutron star in
Section~\ref{sec:estimates}.

Our discussion will thus be focussed on the question which consequences the 
neutron star's kick motion with velocity $v_\mathrm{NS}$ may have for the 
late-time fallback, i.e., for fallback at times $t_\mathrm{fb}$ when the
displacement of the moving neutron star from the COE,
\begin{equation}
        D_\mathrm{NS} \approx
        v_\mathrm{NS}\cdot t_\mathrm{fb} = 10^{4}\,\mathrm{km}\,\,
        \left(\frac{v_\mathrm{NS}}{500\,\mathrm{km\,s}^{-1}}\right)\cdot
        \left(\frac{t_\mathrm{fb}}{20\,\mathrm{s}}\right) \,,
\label{eq:NSshift}
\end{equation}
becomes significant, i.e., when the ratio $D_\mathrm{NS}/R_\mathrm{acc}$
is not much smaller than unity.

The considerations in the entire section will be based on a highly 
simplified and schematic picture of the conditions in exploding stars, as
sketched in Figure~\ref{fig:Graphics3}. Of course, this picture cannot capture
all of the complexity and enormous diversity of the real situation, which will
also vary considerably from case to case, depending, for example, on the 
progenitor properties. Our discussion is therefore only intended to illustrate
basic aspects and fundamental principles, but it is not suitable for an accurate
quantitative assessment. The latter will require future 3D simulations that
follow the time-dependent multi-dimensional dynamics by taking into account
all relevant physical effects.

\subsection{Fallback capture volume}
\label{sec:capture}

\subsubsection{Elementary considerations}
\label{sec:roughviews}

The accretion radius $R_\mathrm{acc}$ determines
the volume $V_\mathrm{acc}$ that is influenced by the neutron star's gravitational
attraction. It is therefore defined as the distance where the gravitating effects 
begin to dominate the kinetic motion of the gas, given by the condition:
\begin{equation}
	\xi\,\frac{G\,M_\mathrm{NS}}{R_\mathrm{acc}} = \frac{1}{2}\,v_\mathrm{kin}^2\,,
\label{eq:raccdef}
\end{equation}
where $\xi$ is a factor of order unity \citep{Frank+2002,Shapiro+1983}.
Here and in the following we assume that the fallback is governed by the gravitational
potential of the neutron star, i.e., the self-gravity of the fallback matter can be
ignored, which requires $M_\mathrm{fb}\ll M_\mathrm{NS}$.
Since we consider hydrodynamic accretion of pressure-supported gas that expands
relative to the neutron star moving with velocity $v_\mathrm{NS}$, the relevant
effective kinetic velocity $v_\mathrm{kin}$ is
\begin{equation}
	v_\mathrm{kin} = \left(c_\mathrm{s}^2 + v_\mathrm{rel}^2\right)^{1/2}
        \sim \left(c_\mathrm{s}^2 + v_\mathrm{exp}^2 + v_\mathrm{NS}^2\right)^{1/2}\,.
\label{eq:vkin}
\end{equation}
Here, $c_\mathrm{s}$ is the sound speed of the gas, $v_\mathrm{rel}$ the relative
velocity between neutron star and gas, which expands with velocity $v_\mathrm{exp}$,
and the second relation is meant to roughly represent the magnitude.\footnote{Here,
only the approximate scaling is relevant. In our more detailed discussion
in Section~\ref{sec:moredetails}, the accurate expression for $v_\mathrm{rel}$ of
Equation~(\ref{eq:vrel}) will be employed.}
Using typical values, $c_\mathrm{s} \approx 3000\,\mathrm{km\,s}^{-1}$, 
$v_\mathrm{exp}\approx 5000\,\mathrm{km\,s}^{-1}$, and 
$v_\mathrm{NS}\lesssim 1500\,\mathrm{km\,s}^{-1}$, one obtains
\begin{eqnarray}
	R_\mathrm{acc} &=& \xi\,\frac{2\,G\,M_\mathrm{NS}}{v_\mathrm{kin}^2} \nonumber\\
	&\sim& 
	1.1\times 10^4\,\mathrm{km}\,\,\xi\cdot 
	\left(\frac{M_\mathrm{NS}}{1.5\,\mathrm{M}_\odot}\right)
	\left(\frac{v_\mathrm{kin}}{6000\,\mathrm{km\,s}^{-1}}\right)^{-2}.
\label{eq:racc}
\end{eqnarray}

A crude estimate of the adiabatic sound speed $c_\mathrm{s}$ can be 
obtained by the consideration that the pressure in the postshock medium
exhibits only relatively little variation over a major fraction of the
postshock volume, apart from local, short-lived structures (such as, e.g.,
transiently existent reverse shocks). Therefore, with the postshock pressure,
$P_\mathrm{post}$, and the postshock density, $\rho_\mathrm{post}$, the 
sound speed follows from
\begin{equation}
	c_\mathrm{s}^2 = \gamma\,\frac{P_\mathrm{post}}{\rho_\mathrm{post}}
	= \gamma\,\frac{1}{\beta}\left(1-\frac{1}{\beta}\right)\,v_\mathrm{sh}^2 
	= \frac{\gamma}{\beta\,(1-\frac{1}{\beta})}\,\, v_\mathrm{post}^2 \,,
\label{eq:csound}
\end{equation}
where we used the postshock-to-preshock density ratio 
$\beta = \rho_\mathrm{post}/\rho_\mathrm{star}(R_\mathrm{sh})\gg 1$ 
and the shock-jump condition for the postshock pressure,
\begin{equation}
        P_\mathrm{post}(R_\mathrm{sh}) = \left(1 - \frac{1}{\beta}\right)\,
         	\rho_\mathrm{star}(R_\mathrm{sh})\,v_\mathrm{sh}^2(R_\mathrm{sh})\,,
\label{eq:ppost1}
\end{equation}
in the first transformation and 
\begin{equation}
v_\mathrm{post} = \left(1-\frac{1}{\beta}\right)\,v_\mathrm{sh}
\label{eq:vpost}
\end{equation}
for the relation between the velocities of the shock and of the postshock gas
in the second transformation. For the postshock pressure the approximate
relation 
\begin{equation}
	P_\mathrm{post}(R_\mathrm{sh}) \approx (\gamma-1)\,
                        \frac{E_\mathrm{th}}{\frac{4\pi}{3}R_\mathrm{sh}^3}
\label{eq:ppost2}
\end{equation}
also holds, which in combination with the shock-jump condition of 
Equation~(\ref{eq:ppost1}) yields Equation~(\ref{eq:vshock1}).
According to Equation~(\ref{eq:csound}), the sound speed in the postshock
medium ranges between $\sim$40\% and $\sim$70\% of the shock speed,
depending on the values of $\beta$ and $\gamma$.

Equations~(\ref{eq:NSshift}) and (\ref{eq:racc}) show that $D_\mathrm{NS}$
and $R_\mathrm{acc}$ are of similar size for average neutron star kick
velocities at times later than some 10\,s after the onset of the explosion.
This means that in the following we will be particularly interested in the
situation when 
\begin{equation}
    \frac{D_\mathrm{NS}}{R_\mathrm{acc}} \gtrsim 1\,.
\label{eq:ratioDR}
\end{equation}

\subsubsection{More detailed discussion}
\label{sec:moredetails}

In deriving the rough estimates in Section~\ref{sec:roughviews}, we 
ignored the velocity profile of the inner supernova ejecta. 
For a closer discussion, we consider the conditions in the central 
volume of the exploding star now in a more detailed, though still highly
idealized picture, referring to the schematic situation displayed in
Figure~\ref{fig:Graphics3}. We assume a homologous behavior of the 
expanding ejecta with a radial velocity proportional to radius
$r$ ($v_r \propto r$) up to a velocity $v_0$ at 
radius $R_0$, where the homology is broken by the collision of the
inner ejecta with the dense ejecta shell that accumulates behind the
supernova shock and follows the outward shock motion with postshock
velocity $v_\mathrm{post} \approx v_\mathrm{sh}$ (see 
Equation~(\ref{eq:vpost})).
This extremely crude and simplified picture is intended to capture the 
essential properties of the expanding ejecta that are most relevant for
the development of the fallback, although it ignores many details and 
complexities of the true ejecta structure. In reality, the exact ejecta 
profile and its time evolution differs from case to case, dependent on
the explosion energy and the progenitor structure.\footnote{From
a global perspective, the short dynamical time scale enables a 
quasi-homologous profile of the radial velocity to be quickly established 
or re-established, even though the explosion may start anisotropically,
inward travelling reverse shocks and their
outward reflections may lead to a transient destruction of the homology, and
multi-dimensional flows may create local perturbations and radial velocity
fluctuations. It is important to note that the typical time scale to erase
local radial velocity fluctuations on top of an underlying homologous velocity
profile in the absence of persistent perturbing effects is given by the 
inverse of the homology coefficient, i.e., by $R_0/v_0$ in our discussion.
It is thus of the same magnitude as the expansion time scale itself.}

The analysis in the present section has two goals: (1) With the mentioned
more detailed velocity profile (specified in Equations~(\ref{eq:vexp1})
and (\ref{eq:vexp2})) we aim at determining the direction 
dependence of the accretion radius. It will turn out that the 
accretion volume is nearly spherical around the location of the neutron
star. (2) We will show that Equation~(\ref{eq:ratioDR}) is well satisfied 
for typical neutron star kicks and common supernova conditions, because the 
radius $R_0$ easily fulfills the condition required for that. At the same
time, $R_0$ is sufficiently large that highly asymmetric fallback will 
affect slow matter in the central volume of the supernova, but 
the dense ejecta shell behind the supernova shock will not be 
globally affected by the gravitational attraction of the neutron star
(see the situation sketched in Figure~\ref{fig:Graphics3}).

For the reasons mentioned above, we assume in the following
that the expansion velocity of the gas in the central 
volume can roughly be represented by the following functional behavior:
\begin{eqnarray}
	v_\mathrm{exp}(r) &\approx& v_0\left(\frac{r}{R_0}\right) 
	\quad\mathrm{for}\quad r \le R_0\,, \label{eq:vexp1} \\
	v_\mathrm{exp}(r) &\approx& v_\mathrm{post}
        \quad\quad\ \ \mathrm{for}\quad R_0 < r \le R_\mathrm{sh}\,,
\label{eq:vexp2}
\end{eqnarray}
with $v_0 \lesssim v_\mathrm{post}$ and $R_0 \le R_\mathrm{sh}$.
$R_0$ is defined as the radius where deceleration
phases of the outgoing supernova shock lead to a slowdown and pileup of the
postshock ejecta in a dense shell (see Figure~\ref{fig:Graphics3}). 
Between $R_0$ and $R_\mathrm{sh}$ the expansion velocity
can be considered as being, very approximately, constant. In vector notation 
we thus have:
\begin{equation}
	\pmb{v}_\mathrm{exp} = \pmb{v}_r = \pmb{r}\cdot \frac{v_0}{R_0}
\label{eq:vr}
\end{equation}
and
\begin{equation}
	\pmb{v}_\mathrm{rel} = \pmb{v}_r - \pmb{v}_\mathrm{NS} \,.
\label{eq:vrel}
\end{equation}
Introducing the displacement vector of the neutron star from the COE,
$\pmb{D}_\mathrm{NS}$, and $\pmb{r}_\mathrm{NS}$ for the position vector 
centered at the neutron star (Figure~\ref{fig:Graphics1}), we obtain the 
relations
\begin{eqnarray}
	\pmb{r} &=& \pmb{D}_\mathrm{NS} + \pmb{r}_\mathrm{NS}\,,
                                              \label{eq:rrns1} \\
        r^2 &=& r_\mathrm{NS}^2 + D_\mathrm{NS}^2 + 
          	2r_\mathrm{NS}D_\mathrm{NS}\cos\theta_\mathrm{NS}\,,
	                                      \label{eq:rrns2} \\
        r\,\cos\theta &=& D_\mathrm{NS} + r_\mathrm{NS}\cos\theta_\mathrm{NS}\,,
				              \label{eq:rrns3}
\end{eqnarray}
where $\theta$ is the angle between $\pmb{r}$ and $\pmb{D}_\mathrm{NS}$,
and $\theta_\mathrm{NS}$ the angle between $\pmb{r}_\mathrm{NS}$ and 
$\pmb{D}_\mathrm{NS}$ (see Figure~\ref{fig:Graphics1}). 

\begin{figure}[tb!]
\begin{center}
        \includegraphics[width=\columnwidth]{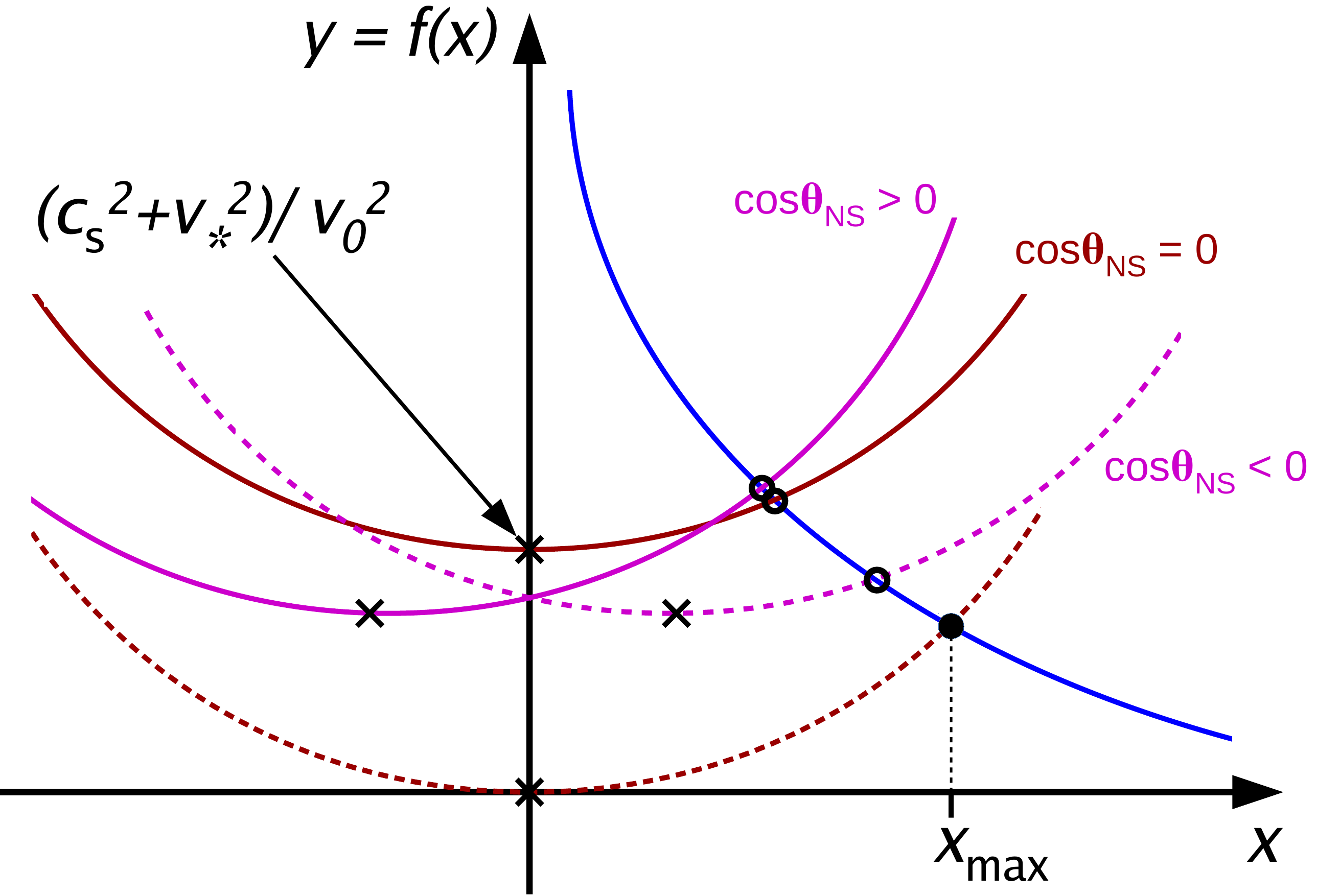}
        \caption{Solution of Equation~(\ref{eq:racc2}) as intersection
	    point of a hyperbola (blue) and a parabola (for the case of 
	    $v_\ast>0$). The solutions for different
        values of $\cos\theta_\mathrm{NS}$ are indicated by open circles,
        the upper bound (the solution of Equation~(\ref{eq:racc3}) with
        $c_\mathrm{s}^2/v_0^2$ ignored) is marked by a black bullet.}
\label{fig:Graphics2}
\end{center}
\end{figure}

In order to determine the direction dependent accretion radius 
$r_\mathrm{acc}(\theta_\mathrm{NS})$ of the neutron star, we set 
$r_\mathrm{NS} = r_\mathrm{acc}$ and employ Equation~(\ref{eq:raccdef})
with $r_\mathrm{acc}(\theta_\mathrm{NS})$ replacing the spherical
radius $R_\mathrm{acc}$. For the effective kinetic velocity on the
rhs of Equation~(\ref{eq:raccdef}) we 
write:\footnote{We remark here that we include in 
Equation~(\ref{eq:vkin2}) only the radial component of the fluid
velocity and ignore a tangential one. This nonradial component 
will become relevant in Section~\ref{sec:alignment}, but it is clearly
subsonic, for which reason we consider it as subdominant compared
to the other contributions to $\pmb{v}_\mathrm{kin}$.}
\begin{eqnarray}
	\pmb{v}_\mathrm{kin}^2 &=& c_\mathrm{s}^2 + \pmb{v}_\mathrm{rel}^2 =
			   c_\mathrm{s}^2 + (\pmb{v}_r - \pmb{v}_\mathrm{NS})^2
			   \nonumber\\
	&=& \left[c_\mathrm{s}^2 + v_\ast^2\left(1-\cos^2\theta_\mathrm{NS}\right)\right]
	    + \left(y + v_\ast\cos\theta_\mathrm{NS}\right)^2 .
\label{eq:vkin2}
\end{eqnarray}
Here, we have introduced 
\begin{eqnarray}
	y &=& \frac{v_0}{R_0}\,r_\mathrm{acc}  \label{eq:y}\\
	\mathrm{and}\quad \ 
	v_\ast &=& \frac{v_0}{R_0}\,D_\mathrm{NS} - v_\mathrm{NS}\,. \label{eq:vast}
\end{eqnarray}
We note that
$v_\mathrm{NS} < c_\mathrm{s}$ and $v_0 > c_\mathrm{s}$, and the
time the neutron star needs to cover a distance $D_\mathrm{NS}$ is
longer than the hydrodynamical time scale $R_0/v_0$, for which reason
usually $v_\ast \ge 0$. Now defining $x = y/v_0$, which implies
\begin{equation}
        r_\mathrm{acc}(\theta_\mathrm{NS}) = x\cdot R_0 \,,
\label{eq:x}
\end{equation}
we obtain from Equation~(\ref{eq:raccdef}), using Equation~(\ref{eq:vkin2}), 
the following relation to determine $x$ and thus
$r_\mathrm{acc}(\theta_\mathrm{NS})$:
\begin{eqnarray}
	&\xi&\,\frac{GM_\mathrm{NS}/R_0}{\frac{1}{2}\,v_0^2}\cdot\frac{1}{x} =
	\xi\,\frac{v_\mathrm{esc}^2(R_0)}{v_0^2}\cdot\frac{1}{x} \nonumber \\
	&=& \left[\frac{c_\mathrm{s}^2}{v_0^2} + \frac{v_\ast^2}{v_0^2}\,
	\left(1-\cos^2\theta_\mathrm{NS}\right)\right]
	    + \left(x + \frac{v_\ast}{v_0}\,\cos\theta_\mathrm{NS}\right)^2 ,
\label{eq:racc2}
\end{eqnarray}
where 
\begin{eqnarray}
	v_\mathrm{esc}(R_0) &=& \left(\frac{2\,GM_\mathrm{NS}}{R_0}\right)^{1/2}
	\nonumber\\
	&=& 2000\,\mathrm{\frac{km}{s}}\,
	\left(\frac{M_\mathrm{NS}}{1.5\,\mathrm{M}_\odot}\right)^{1/2}
	\left(\frac{R_0}{10^5\,\mathrm{km}}\right)^{-1/2}
\label{eq:vesc}
\end{eqnarray}
is the escape velocity at radius $R_0$. The accretion radius
$r_\mathrm{acc}(\theta_\mathrm{NS})$ is determined through Equation~(\ref{eq:racc2})
as the intercept point of a hyperbola and a parabola. If $v_\ast > 0$, 
the resulting $r_\mathrm{acc}$ is smaller in outward directions, i.e.,
for $\cos\theta_\mathrm{NS}>0$, than the value perpendicular to the radial direction
($\cos\theta_\mathrm{NS}=0$), because the expansion velocity increases according
to $v_r \propto r$. For the same reason, $r_\mathrm{acc}$ increases when
$\cos\theta_\mathrm{NS}<0$, i.e., on the side where the COE is located
(Figure~\ref{fig:Graphics2}). If $v_\ast < 0$ the situation is reversed.
Since typically $|v_\ast| \ll v_0$, Equation~(\ref{eq:racc2}) can be replaced by
\begin{equation}
	\xi\,\frac{v_\mathrm{esc}^2(R_0)}{v_0^2}\cdot\frac{1}{x} \approx
	\frac{c_\mathrm{s}^2}{v_0^2} + x^2
\label{eq:racc3}
\end{equation}
with good approximation, for which the solution of $r_\mathrm{acc}$ becomes
independent of $\theta_\mathrm{NS}$. An upper bound is obtained when we ignore
the term $c_\mathrm{s}^2/v_0^2$ on the rhs:
\begin{equation}
	r_\mathrm{acc} \approx 
        \xi^{1/3}\,R_0\,\left(\frac{v_\mathrm{esc}(R_0)}{v_0}\right)^{2/3}\,.
\label{eq:racc4}
\end{equation}
Since $c_\mathrm{s}$ is usually lower than $v_0$, but still cannot be neglected,
the accretion radius is smaller than the value of Equation~(\ref{eq:racc4}): 
\begin{equation}
R_\mathrm{acc} < r_\mathrm{acc} \approx  
        \xi^{1/3}\,R_0\,\left(\frac{v_\mathrm{esc}(R_0)}{v_0}\right)^{2/3}\,.
\label{eq:racc5}
\end{equation}

This discussion shows that the accretion volume around the moving neutron
star, which is displaced from the COE by a distance $D_\mathrm{NS}$, is 
nearly spherical (i.e., independent of $\theta_\mathrm{NS}$) 
with a radius $R_\mathrm{acc}$ that is constrained by
Equation~(\ref{eq:racc5}). If the fallback concerns only a 
small fraction of the ejecta\footnote{This excludes fallback supernovae from
our discussion, because they are defined as cases where the fallback mass 
amounts to a sizable fraction of the initial ejecta, in which case
---in contrast to our assumptions--- $M_\mathrm{fb}$ can be comparable
to the initial mass of the compact remnant or even exceed it.}
the condition $v_0 > v_\mathrm{esc}(r\gtrsim R_0)$ should hold, because
the mass exterior to $R_0$ (having $v_\mathrm{post}\gtrsim v_0$; see 
above) should be on escape trajectories.

Using Equation~(\ref{eq:vast}) and assuming, again, that $v_\ast$ is small
($v_\ast \sim 0$), we get
\begin{equation}
	D_\mathrm{NS} \cong R_0\,\frac{v_\mathrm{NS}}{v_0} = 
	R_0\, \frac{v_\mathrm{NS}}{v_\mathrm{esc}(R_0)}\,
	\left(\frac{v_\mathrm{esc}(R_0)}{v_0}\right) \,.
\label{eq:dns}
\end{equation}
Comparing this relation with Equation~(\ref{eq:racc5}) we see that
$D_\mathrm{NS}$ and $R_\mathrm{acc}$ fulfill the relation of 
Equation~(\ref{eq:ratioDR}) roughly when
$v_\mathrm{NS}\gtrsim v_\mathrm{esc}(R_0)$, i.e., for
\begin{eqnarray}
	R_0 &\gtrsim& \frac{2\,GM_\mathrm{NS}}{v_\mathrm{NS}^2} \nonumber\\
	&=& 4\times 10^5\,\mathrm{km}\,\,
        \left(\frac{M_\mathrm{NS}}{1.5\,\mathrm{M}_\odot}\right)
        \left(\frac{v_\mathrm{NS}}{1000\,\mathrm{km\,s}^{-1}}\right)^{-2}.
\label{eq:r0relation}
\end{eqnarray}
Such lower limits of $R_0$ are far below the typical pre-collapse radii 
of BSG and RSG stars (which range from $\sim$\,$10^7$\,km to over $10^9$\,km),
and therefore the relation of Equation~(\ref{eq:ratioDR}) should be
easily fulfilled for sufficiently large neutron star kicks and common
supernova conditions.
Also the velocity relation of $v_0 = v_\mathrm{exp}(R_0) > v_\mathrm{NS} 
> v_\mathrm{esc}(r\gtrsim R_0)$ represents conditions that can be commonly 
expected during the developing supernova explosions of such progenitors.
The geometrical situation that crystalizes out of the discussion of this 
section is visualized in Figure~\ref{fig:Graphics3} (where for the purpose
of illustration we have chosen to display the case of $D_\mathrm{NS}\sim 
R_\mathrm{acc}$, which is less extreme than the condition of 
Equation~(\ref{eq:ratioDR})). Since $R_0$ is easily larger than
both $D_\mathrm{NS}$ and $R_\mathrm{acc}$ (estimated in 
Equations~(\ref{eq:NSshift}) and (\ref{eq:racc}), respectively),
we expect fallback accretion to be highly asymmetric for 
typical neutron star kicks of several 100\,km\,s$^{-1}$, and to 
affect the matter in a volume in the deep interior of the supernova,
but not in the main ejecta shell.

\begin{figure*}[!]
\begin{center}
        \includegraphics[width=0.75\textwidth]{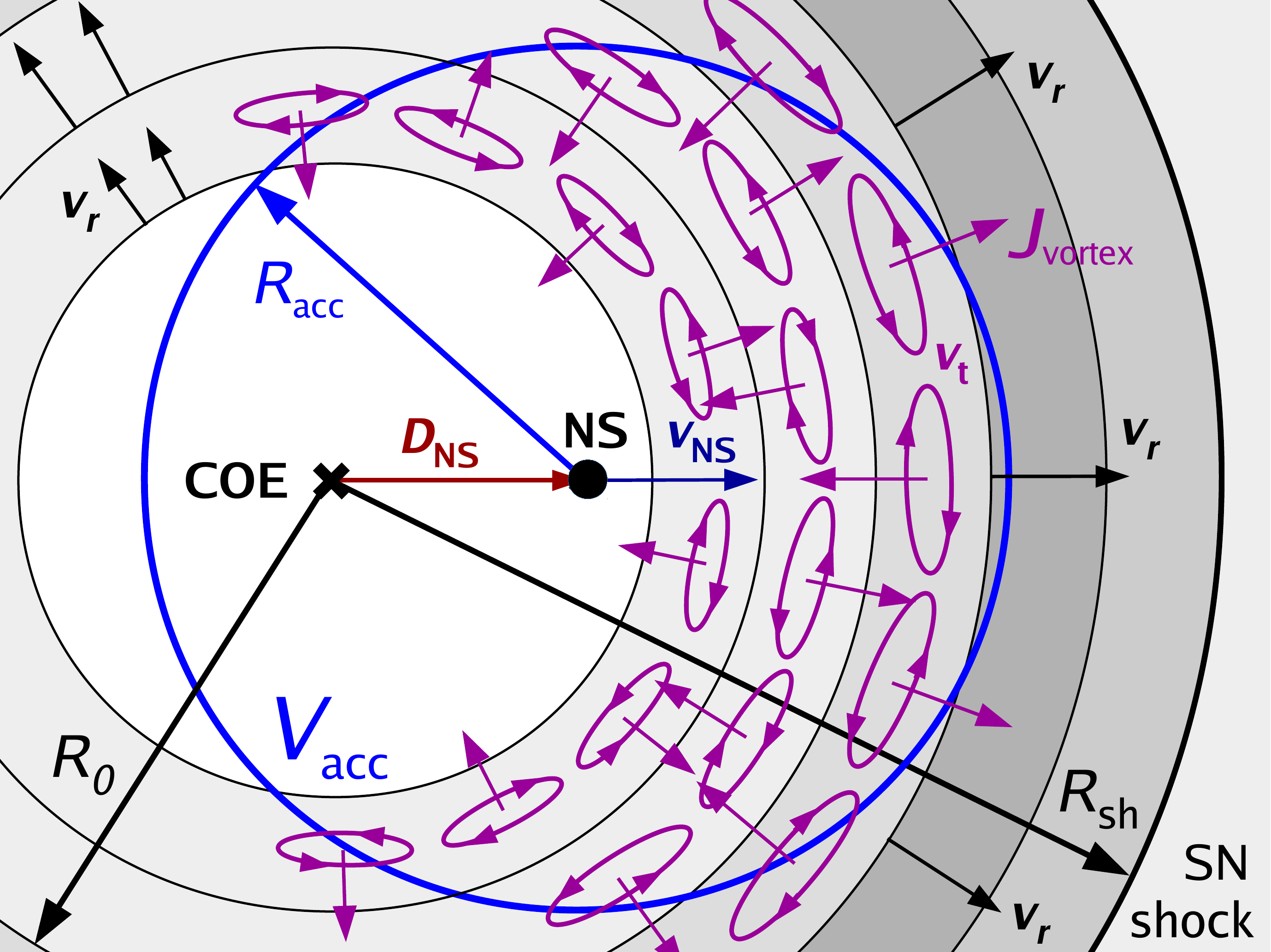}
        \caption{Analogue to Figure~\ref{fig:Graphics3}, but visualizing
	the ejecta kinematics underlying the discussion in 
	Section~\ref{sec:angmomaccvol}.}
\label{fig:Graphics4}
\end{center}
\end{figure*}

\subsection{Fallback accretion of angular momentum\\ and spin-kick alignment}
\label{sec:alignment}

Spin-kick alignment can occur only if some physical effect defines a 
special or preferred direction of the system. In previously suggested
explanations for spin-kick alignment this effect was the rotation of the 
progenitor star and of its degenerate core, which was assumed to play the
crucial role in setting the spin period and determining the spin direction 
of the relic neutron star. 

In our scenario the direction-defining effect is the kick motion of the 
new-born neutron star. Our revised picture is motivated by the results of
3D simulations presented in Section~\ref{sec:simulations} and in previous 
recent papers \citep{Mueller+2019,Stockinger+2020,Chan+2020}, which show 
that the spin of neutron stars born in the collapse of slowly rotating 
stellar cores is likely to originate mainly from the angular momentum 
accreted through fallback matter.

Provided this is the main cause of neutron star spins, how could such
a situation lead to spin-kick alignment?
In the following we will argue that asymmetric accretion, connected to
the displacement of the neutron star from the COE as discussed in 
Section~\ref{sec:capture}, offers possibilities to explain spin-kick
alignment quite naturally.

\subsubsection{Angular momentum in the accretion volume}
\label{sec:angmomaccvol}

Let us consider the situation visualized in Figure~\ref{fig:Graphics4}.
The (nearly) spherical accretion volume $V_\mathrm{acc}$ with radius 
$R_\mathrm{acc}$ around the neutron star, which is shifted by a vector distance
$\pmb{D}_\mathrm{NS}$ from the COE, is imagined to contain a large number of
tangential vortex flows with individual angular momenta $\pmb{J}_\mathrm{vortex}$. 
While the total angular momentum of all vortices in the entire volume of
the star will add up to zero (unless the star possesses global rotation), we 
are interested here in estimating the angular momentum falling back to the 
neutron star with the matter in the accretion volume. We will see that this
angular momentum is given by the sum of the angular momenta 
$\pmb{J}_\mathrm{vortex}$ of all swirls in the accretion volume.

In order to compute the angular momentum associated with the 
matter in volume $V_\mathrm{acc}$, we
split the gas velocity $\pmb{v}$ into a radial component $\pmb{v}_r$ parallel 
to the radius vector $\pmb{r}$ and a tangential component $\pmb{v}_\mathrm{t}$
perpendicular to $\pmb{r}$:
\begin{equation}
	\pmb{v}(\pmb{r}) = \pmb{v}_r(\pmb{r}) + \pmb{v}_\mathrm{t}(\pmb{r}) \,.
\label{eq:vsplit}
\end{equation}
The nonradial flow component is subsonic, $|\pmb{v}_\mathrm{t}| \ll c_\mathrm{s}$,
and typically (but not strictly) one has $|\pmb{v}_\mathrm{t}| < |\pmb{v}_r|$
in the radially moving flow of supernova explosions (unless reverse shocks
stall the flow).
The angular momentum received by the neutron star from the gas of density
$\rho(\pmb{r})$ in the accretion volume, assuming that all matter ends up on 
the neutron star, is given by
\begin{eqnarray}
	\pmb{J}_\mathrm{NS,acc} &=& \int_{V_\mathrm{acc}} \mathrm{d}V\,
	\rho(\pmb{r})\cdot\left[\pmb{r}_\mathrm{NS}\times\left(\pmb{v}_r + 
	\pmb{v}_\mathrm{t} - \pmb{v}_\mathrm{NS}\right)\right] \nonumber \\
	&=& \int_{V_\mathrm{acc}} \mathrm{d}V\,\rho\left[\pmb{r}_\mathrm{NS}
	\times \pmb{v}_r\right] -
	\int_{V_\mathrm{acc}} \mathrm{d}V\,\rho\left[\pmb{r}_\mathrm{NS}
	\times \pmb{v}_\mathrm{NS}\right]\nonumber \\ 
	&+& \int_{V_\mathrm{acc}} \mathrm{d}V\,\rho\left[\pmb{r}_\mathrm{NS}
        \times \pmb{v}_\mathrm{t}\right] \nonumber \\
	&=& \pmb{J}_r + \pmb{J}_v + \pmb{J}_\mathrm{t} \,\,,
	\label{eq:angmomacc}
\end{eqnarray}
where the second to fourth lines represent the definition of the components
$\pmb{J}_r$, $\pmb{J}_v$, and $\pmb{J}_\mathrm{t}$ associated with the radial
gas motion, neutron star motion, and tangential gas motion, respectively.

For simplicity, we first consider the situation that the density distribution
and radial velocity of the ejecta are spherically symmetric, i.e.,
$\rho(\pmb{r})= \rho(r)$ and $\pmb{v}_r(\pmb{r})=\pmb{v}_r(r)=v_r(r)\cdot\pmb{r}/r$
(a homologous radial velocity profile is a special case). With this assumption
$\pmb{J}_r$ can be evaluated, using Equation~(\ref{eq:rrns1}), as follows:
\begin{eqnarray}
	\pmb{J}_r &=& \int_{V_\mathrm{acc}} \mathrm{d}V\,\rho(r)\left[\pmb{r}_\mathrm{NS}
	\times \pmb{v}_r(r)\right] \nonumber\\
	&=& \int_{V_\mathrm{acc}} \mathrm{d}V\,\rho(r)
	\left[\pmb{r} \times \pmb{v}_r(r)\right] - 
	\int_{V_\mathrm{acc}} \mathrm{d}V\,\rho(r)
	\left[\pmb{D}_\mathrm{NS}\times\pmb{v}_r(r)\right]\nonumber \\
	&=& 0\,\, .
\label{eq:angmomr}
\end{eqnarray}
The first term vanishes because $\pmb{r}\parallel\pmb{v}_r$, and the second
term disappears for symmetry reasons because the volume $V_\mathrm{acc}$
is rotationally symmetric around the vector of $\pmb{D}_\mathrm{NS}$ and the 
density and radial velocity are assumed to be spherically symmetric around the COE.
Similarly, for $\pmb{J}_v$ one gets
\begin{equation}
	\pmb{J}_v = -\int_{V_\mathrm{acc}} \mathrm{d}V\,\rho(r)
	\left[\pmb{r}_\mathrm{NS} \times \pmb{v}_\mathrm{NS}\right] = 0\,\,.
	\label{eq:angmomv}
\end{equation}
It is zero, too, again because of rotational symmetry around 
$\pmb{D}_\mathrm{NS}$, which we assume to be parallel to $\pmb{v}_\mathrm{NS}$.
For $\pmb{J}_\mathrm{t}$ we also apply Equation~(\ref{eq:rrns1}) for a splitting
into two contributions:
\begin{eqnarray}
	\pmb{J}_\mathrm{t} &=& \int_{V_\mathrm{acc}} \mathrm{d}V\,\rho(r)
	\left[\pmb{r}_\mathrm{NS} \times \pmb{v}_\mathrm{t}\right] \nonumber\\
        &=& \int_{V_\mathrm{acc}} \mathrm{d}V\,\rho(r)
	\left[\pmb{r} \times \pmb{v}_\mathrm{t}\right] - 
        \int_{V_\mathrm{acc}} \mathrm{d}V\,\rho(r)
	\left[\pmb{D}_\mathrm{NS}\times\pmb{v}_\mathrm{t}\right] .
\label{eq:angmomt}
\end{eqnarray}
Let us imagine the nonradial component of the flow to consist of an 
arrangement of individual vortex structures revolving around the radial
directions with tangential velocities $\pmb{v}_\mathrm{t}(\pmb{r})$
(Figure~\ref{fig:Graphics4}). We remind the reader of our underlying 
assumption that the ejecta have reached a state of homologous expansion
($v_r(r)\propto r$; Equation~(\ref{eq:vr})),
in which case variations of the radial velocity
are ``filtered out'' and only tangential vortex motions remain 
in the expanding medium. These vortices may either be relics of 
convection cells in the inner burning layers of the progenitor star, or 
they may be the results of turbulent flows connected to 3D hydrodynamic
instabilities at the onset of the supernova explosion. We can now
rewrite the volume integrals as sums over all vortex loops in volume 
$V_\mathrm{acc}$:
\begin{eqnarray}
	\pmb{J}_\mathrm{t} &=& \sum_\mathrm{vortices}
	\int_{V_\mathrm{vortex}} \mathrm{d}V\,\rho(r)
        \left[\pmb{r} \times \pmb{v}_\mathrm{t}\right] \nonumber\\
	&-& \sum_\mathrm{vortices}
	\int_{V_\mathrm{vortex}} \mathrm{d}V\,\rho(r)
        \left[\pmb{D}_\mathrm{NS}\times\pmb{v}_\mathrm{t}\right] \,\,.
\label{eq:angmomt2}
\end{eqnarray}
Integrals over closed vortices in the second sum vanish,
\begin{equation}
	\int_{V_\mathrm{closed vortex}} \mathrm{d}V\,\rho(r)
        \left[\pmb{D}_\mathrm{NS}\times\pmb{v}_\mathrm{t}\right] = 0\,\,,
\label{eq:vortex}
\end{equation}
because for the fixed radius vector $\pmb{D}_\mathrm{NS}$ the angular momenta 
of mass elements in the vortex moving on opposite sides of the vortex axis
cancel each other.
Therefore, if the flow in volume $V_\mathrm{acc}$ consists of a 
system of mostly closed vortex structures, one derives:
\begin{eqnarray}
	\pmb{J}_\mathrm{t} &\approx& \int_{V_\mathrm{acc}} \mathrm{d}V\,\rho(r)
        \left[\pmb{r} \times \pmb{v}_\mathrm{t}\right] \nonumber\\
        &=& \sum_\mathrm{vortices}
        \int_{V_\mathrm{vortex}} \mathrm{d}V\,\rho(r)
        \left[\pmb{r} \times \pmb{v}_\mathrm{t}\right] = 
	\sum_\mathrm{vortices} \pmb{J}_\mathrm{vortex} \,\,.
\label{eq:angmomt3}
\end{eqnarray}
Combining Equations~(\ref{eq:angmomr}), (\ref{eq:angmomv}), and 
(\ref{eq:angmomt3}) thus yields:
\begin{equation}
	\pmb{J}_\mathrm{NS,acc} \approx \int_{V_\mathrm{acc}} \mathrm{d}V\,\rho(r)
        \left[\pmb{r} \times \pmb{v}_\mathrm{t}\right] 
	= \sum_\mathrm{vortices} \pmb{J}_\mathrm{vortex} \,\,.
\label{eq:angmomacc2}
\end{equation}

\begin{figure}[tb]
\begin{center}
        \includegraphics[width=\columnwidth]{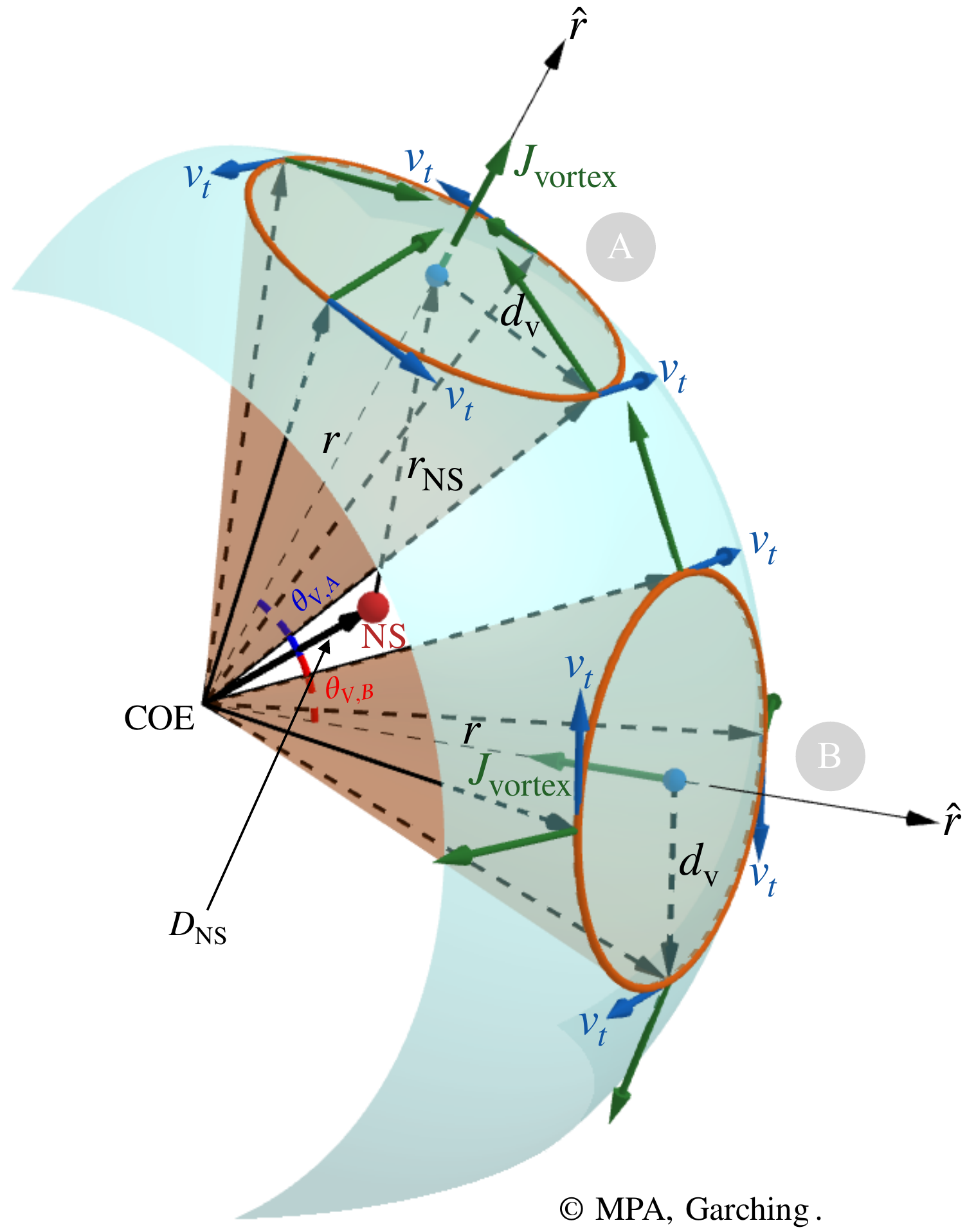}
	\caption{Basic geometry of tangential vortices and their angular
	momentum vectors $\pmb{J}_\mathrm{vortex}$. Note that in the integrand of
	Equation~(\ref{eq:angmomacc2}) each fluid element with tangential velocity
	$\pmb{v}_\mathrm{t}$ has an angular momentum perpendicular to the local 
	radius vector $\pmb{r}$ and with a (small) component parallel to the vortex
	axis. For symmetry reasons the integration over the
	volume of a vortex lying in a spherical shell of radius $r$ with constant 
	density $\rho(r)$ leads to a net angular momentum of the vortex pointing 
	inward or outward along the vortex axis, as shown in the figure.
	}
\label{fig:Graphics-vortices}
\end{center}
\end{figure}

The angular momentum acquired by the neutron star through accretion of
matter from the accretion volume can thus be viewed as the summed angular
momenta associated with tangential motions in the fluid vortices, 
with $\pmb{J}_\mathrm{vortex}$ being the angular momenta of the individual
vortices, which (for symmetry reasons) are parallel to the vortex axis
(see Figure~\ref{fig:Graphics-vortices}).
If accretion dominates the final rotation of the neutron star and
overrules the initial proto-neutron star spin right after its formation,
as suggested by the results of 3D simulations presented in 
Section~\ref{sec:fallback} and papers referenced there,
then the neutron star's angular momentum is only determined after fallback
accretion. This implies:
\begin{equation}
	\pmb{J}_\mathrm{NS} \approx \pmb{J}_\mathrm{NS,acc}
	\approx \sum_\mathrm{vortices} \pmb{J}_\mathrm{vortex} \,\,.
	\label{eq:angmomacc3}
\end{equation}
This relation has important implications for the discussion of possible
spin-kick alignment. In previously suggested scenarios for spin-kick alignment
it was the spin of the progenitor's degenerate core, which was inherited by 
the neutron star and decided its kick direction by rotational averaging 
(of recoil effects associated either with anisotropic neutrino emission in
super-strong magnetic fields or stochastic momentum thrusts or with 
hydrodynamical forces caused by impacts of anisotropic accretion downflows).
In contrast to those ideas, in the scenario discussed
here, the neutron star's kick is randomly directed and is set by processes in 
the supernova core over a short time scale of just a few seconds. The neutron 
star obtains its final spin through accretion only on much longer time scales 
of tens to thousands of seconds (or possibly even longer), and therefore 
spin-kick alignment would have to be a consequence of the properties of this
accretion process and the angular momentum obtained with the accreted mass.

We repeat that our underlying assumption in this context is that 
the expanding ejecta have reached a (quasi-)homologous state, i.e.,
$\pmb{v}_r(\pmb{r}) \propto \pmb{r}$ for the radial flow,\footnote{A 
dependence of the radial velocity on the distance $r$ from the COE,
i.e., $\pmb{v}_r(\pmb{r}) = \pmb{v}_r(r)$, is sufficient for our arguments.}
and therefore tangential mass motions (in lieu of radial ones) 
dominate the flow vorticity in the initial ejecta. We apply
this assumption also to the material that is later on affected by the fallback,
because fallback is likely to involve the slowest parts of the ejecta, which
lag behind the faster ejecta that have arranged themselves to locations at 
larger radii. This means that
we assume that vortex motions perpendicular to the radial direction (and thus 
within spherical shells around the COE) are much larger than local variations
in the radial velocity field, i.e., 
$v_\mathrm{t} \gg \delta v_r$, when $\delta v_r$ are radial velocity 
fluctuations associated with vortical motions in planes containing the
radius vectors. For this reason each of the vortices we imagine in the 
fallback material possesses mainly a radial component of the angular momentum
if the density is constant on concentric spheres ($\rho(\pmb{r}) =
\rho(r)$), and the nonradial components are far subdominant and can be 
neglected in our discussion. (When considering density inhomogeneities in 
Section~\ref{sec:densityvariations}, nonradial components of 
$\pmb{J}_\mathrm{vortex}$ will be taken into account, too.)

It is clear that this schematic picture 
is based on radically simplifying and idealizing assumptions. 
Validation will have to come from 3D simulations, which will have to 
reveal whether such conditions develop, at least approximately, in
the innermost supernova ejecta, which are later affected by fallback.
This concerns the question when the expelled material evolves towards 
a (nearly) homologously expanding outflow, and also the question how
reverse shocks affect the velocity field in the fallback material.
At the time being, since currently existing 3D supernova simulations
cannot provide conclusive evidence,\footnote{Besides not 
taking the neutron star motion into account, long-time explosion 
simulations use an inner grid boundary that is placed at a stepwise
increased radius in order to ease numerical time-stepping constraints.
This fact does not permit the models to track the destiny of the 
fallback matter in the central volume. Moreover, vorticity associated 
with convective mass motions prior to core collapse is so far included 
only in very few supernova simulations. And although these most
advanced supernova simulations have been started
from 3D initial models, the initial conditions themselves are still 
subject to severe approximations.} we have to refer to the considered
situation as a mere working hypothesis.

\begin{figure*}[!]
\begin{center}
        \includegraphics[width=0.75\textwidth]{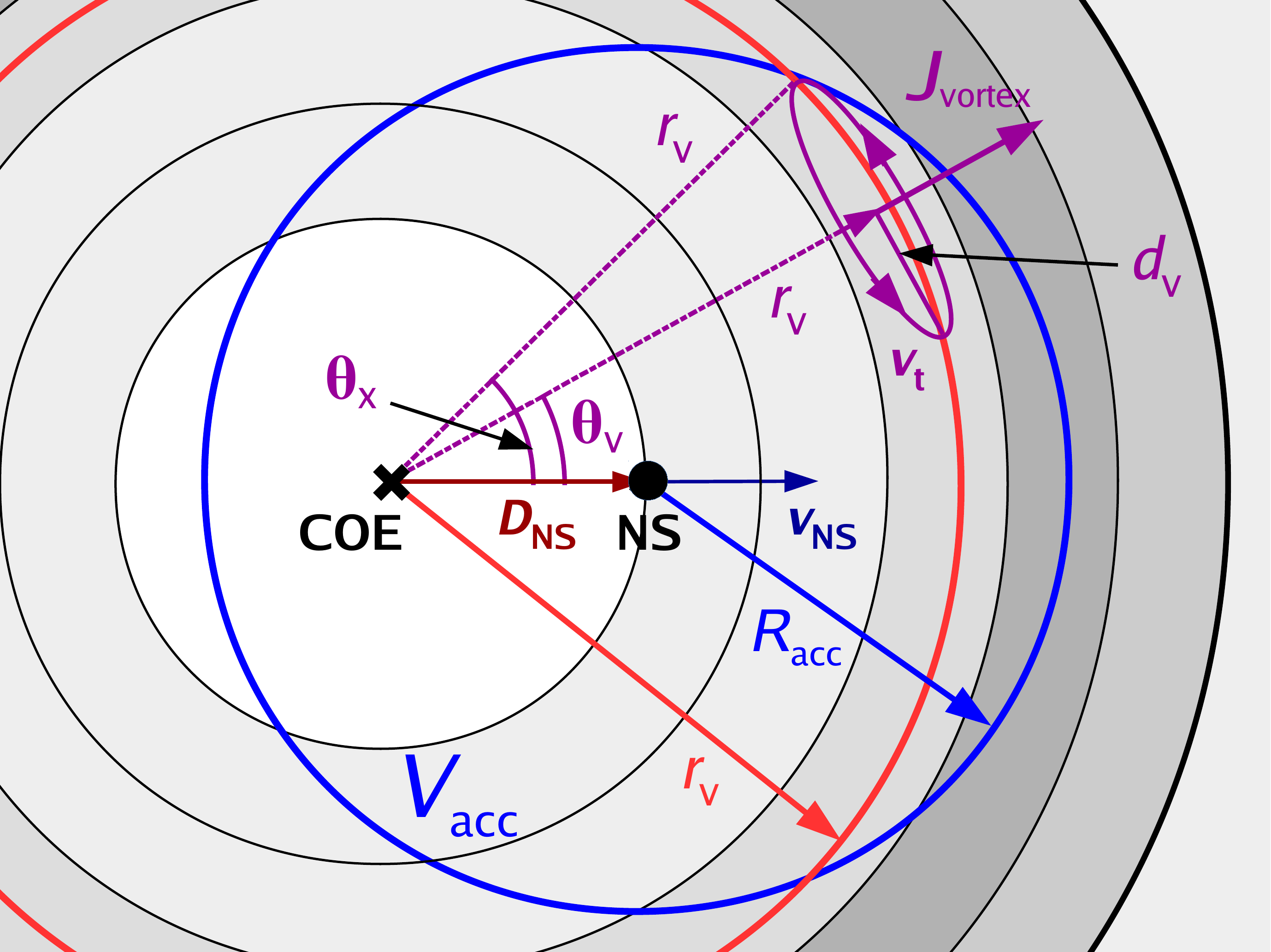}
        \caption{Geometry and structure elements of relevance for the 
        discussion in Section~\ref{sec:skalign}
        (Equations~(\ref{eq:costhetax})--(\ref{eq:costhetavmax})).}
\label{fig:Graphics5}
\end{center}
\end{figure*}

\subsubsection{Spin-kick alignment due to accretion of vortex flows}
\label{sec:skalign}

Spin-kick alignment requires that the neutron star's angular momentum component 
parallel to $\pmb{v}_\mathrm{NS}$ is larger than the angular momentum component
perpendicular to the velocity vector of the neutron star. Anti-alignment, obeying
an analogue criterion, shall explicitly be included here, because both cannot be
discriminated observationally \citep{Johnston+2005}.
In the context of our discussion we will thus talk only about alignment of
spin and kick axes, but not necessarily of the vector directions.
Assuming that the accreted angular momentum dominates the pre-accretion 
angular momentum of the neutron star by far, i.e., accretion determines the 
final neutron star spin (as suggested by the 3D simulations discussed in
Section~\ref{sec:fallback}), we can convince ourselves that the condition 
$|\pmb{J}_{\mathrm{acc},\parallel\,\pmb{v}_\mathrm{NS}}| >
|\pmb{J}_{\mathrm{acc},\perp\,\pmb{v}_\mathrm{NS}}|$ is quite
naturally fulfilled and therefore
\begin{equation}
	\pmb{J}_\mathrm{NS} \approx \pmb{J}_\mathrm{NS,acc} = 
	\pmb{J}_{\mathrm{acc},\parallel\,\pmb{v}_\mathrm{NS}} +
        \pmb{J}_{\mathrm{acc},\perp\,\pmb{v}_\mathrm{NS}} 
        \sim \pmb{J}_{\mathrm{acc},\parallel\,\pmb{v}_\mathrm{NS}}\,\,.
\label{eq:angmom}
\end{equation}
Let us consider the discussed arrangement of vortex flows visualized in
Figure~\ref{fig:Graphics4}.
Vortices with different helicities (which are both present because of angular 
momentum conservation in a nonrotating progenitor) contribute to the sum of 
Equation~(\ref{eq:angmomacc3}) with vectors in opposing directions. 
Adding up the angular momentum vectors of a large number of vortices 
will lead to cancellations of the components perpendicular to 
$\pmb{v}_\mathrm{NS}$, and effectively a residual vector 
$\pmb{J}_{\mathrm{acc},\parallel\,\pmb{v}_\mathrm{NS}}$ parallel or 
anti-parallel to $\pmb{v}_\mathrm{NS}$ will result.
In the case of one or a few big vortex structures, the sum vector 
$\pmb{J}_\mathrm{NS,acc}$ will also possess a major component along
the line defined by the vector of $\pmb{v}_\mathrm{NS}$. 

One can estimate the maximum angle between spin and kick in the case
of a single, dominant vortex by using Equations~(\ref{eq:rrns2}) and 
(\ref{eq:rrns3}) to compute the angle $\theta_\mathrm{x}$ (measured at
the COE) for the intercept circle of a sphere of radius $r_\mathrm{v}$,
which is imagined to contain the vortex center (Figure~\ref{fig:Graphics5}),
and the neutron star's accretion sphere with radius $R_\mathrm{acc}$:
\begin{equation}
	\cos\theta_\mathrm{x} = 
	\frac{r_\mathrm{v}^2+D_\mathrm{NS}^2-R_\mathrm{acc}^2}
					{2D_\mathrm{NS}r_\mathrm{v}} 
\label{eq:costhetax}
\end{equation}
(for $D_\mathrm{NS}\neq 0$).
Since the neutron star moves out of a bubble region left around the COE
and will therefore mainly accrete from the forward direction, where dense
shells build up during the deceleration phases of the supernova shock, 
it is reasonable to choose, for the most extreme case, the equality 
$r_\mathrm{v} \sim R_\mathrm{acc}$, which yields
\begin{equation}
	\cos\theta_\mathrm{x,max} \sim 
        \frac{D_\mathrm{NS}}{2R_\mathrm{acc}} \,.
\label{eq:costhetaxmax}
\end{equation}
Taking $R_\mathrm{acc}$ and $D_\mathrm{NS}$ to be of similar size, too, 
which is based on our arguments discussed in Section~\ref{sec:capture},
we get $\cos\theta_\mathrm{x,max} \sim \frac{1}{2}$ and thus
$\theta_\mathrm{x,max}\sim \frac{\pi}{3}$. Therefore, very approximately,
we can estimate for the maximum angle, $\theta_\mathrm{v,max}$,
between $\pmb{v}_\mathrm{NS}$ and the angular momentum axis of a vortex
with radius $d_\mathrm{v}$ located on a sphere of radius
$r_\mathrm{v} \sim R_\mathrm{acc} \sim D_\mathrm{NS}$:
\begin{equation}
	\theta_\mathrm{v,max} \sim \theta_\mathrm{x,max} - 
	\arcsin\left(\frac{d_\mathrm{v}}{R_\mathrm{acc}}\right)
	\sim \frac{\pi}{3} - \frac{d_\mathrm{v}}{D_\mathrm{NS}} \,.
\label{eq:costhetavmax}
\end{equation}
This result is a rough proxy of the maximum angle $\theta_\mathrm{sk,max}$
between neutron star spin axis and kick axis, provided the spin is connected
to the discussed accretion scenario. If a few big vortices dominate the
tangential flow, one expects $d_\mathrm{v}\sim \frac{1}{2}D_\mathrm{NS}$,
in which case $\theta_\mathrm{sk,max}$ should be less than about 30$^\circ$.
For sufficiently large natal neutron star kicks and correspondingly 
anisotropic fallback because of the neutron star's displacement from the 
COE, we therefore expect the neutron star spins to be distributed 
predominantly within roughly 30$^\circ$ around the kick direction.

One major vortex flow is likely to be predominant in the accretion volume,
for example, in the case of a rapidly rotating progenitor, if the neutron star
receives its natal kick along the progenitor's spin axis or nearly aligned with 
it. Also in this special case spin-kick alignment or anti-alignment after 
fallback accretion is expected. However, rapid progenitor rotation will
{\em inhibit} systematic spin-kick alignment in the discussed scenario, if the
neutron star kick direction is random instead of being tightly correlated with
the star's spin axis. If, for example, the kick is perpendicular to the 
rotation axis shared by the stellar core and the newly formed neutron star, 
the compact
remnant, moving away from the COE in the equatorial plane of the rotation, will 
mainly accrete fallback material with rotational angular momentum 
{\em perpendicular} to the kick-velocity vector. A statistically isotropic 
distribution of the kick directions will thus lead to a large majority 
of cases with neutron star spins and kicks being oriented perpendicular to
each other. In the context of the proposed scenario, the observational tendency
of spin-kick alignment \citep{Johnston+2005,Noutsos+2012,Noutsos+2013} therefore
suggests the interpretation that the progenitor stars {\em did not rotate
rapidly at the time of their collapse.} 

Addressing these aspects from an observational perspective, one needs to be 
aware of the caveats and challenges connected to the information provided by 
measurements, as described in Section~\ref{sec:observations}. 
In particular, because of the existence of ``orthogonal polarisation modes'' 
(OPMs) the linearly polarised pulsar radiation may be emitted in two different modes, 
separated by 90$^\circ$ (e.g.~\citealt{handbook}). In a pulsar where these OPMs are observed, 
it may hence not be possible to decide between an aligned or orthogonal situation 
for this individual case. Nevertheless, the interpretation of non-rapidly rotating 
progenitor stars in the framework of our spin-kick scenario is strong motivation
for further investigation. Once more, we encounter an interesting antipode to 
previously suggested explanations of spin-kick alignment for
neutron stars born in the collapse of rapidly spinning stellar cores.

\subsubsection{Influence of density inhomogeneities}
\label{sec:densityvariations}

Finally, we release our assumption of a spherically symmetric density
distribution used in Sections~\ref{sec:angmomaccvol} and
\ref{sec:skalign} and address now the consequences of density 
inhomogeneities in the fallback matter, i.e., $\rho(\pmb{r}) \neq \rho(r)$.
In this case the angular momenta $\pmb{J}_\mathrm{vortex}$
of vortices are not parallel to the radius vectors
any more, even if we assume the vortex flows to lie in spherical shells
around the COE and thus to have velocities $\pmb{v}_\mathrm{t}$ 
perpendicular to the radial directions. 
A detailed and quantitative assessment is difficult,
since it has to refer to a model of the structure of the density
variations. For the purpose of a matter-of-principle discussion we
consider a highly simplified, extreme situation: the accretion volume 
of the neutron star (imagined to be basically spherical but shifted 
by a distance $D_\mathrm{NS}$ relative to the COE, as argued before)
shall contain gas with masses $m_2 \ge m_1$ in the two hemispheres
left and right of the vector $\pmb{D}_\mathrm{NS}$, a total mass
$m = m_1+m_2$, and a hemispheric density contrast that leads to a
mass difference $\Delta m = m_2 - m_1$ between the two hemispheres.

Evaluating Equation~(\ref{eq:angmomacc}) for this special situation,
considering the two masses to be represented by their mass-centers,
we get:
\begin{eqnarray}
   \pmb{J}_r + \pmb{J}_v &\approx& m_1\,\pmb{r}_{\mathrm{NS},1}\times\pmb{v}_{r,1}
				+ m_2\,\pmb{r}_{\mathrm{NS},2}\times\pmb{v}_{r,2}
				                           \nonumber\\
				&-& \left(m_1\,\pmb{r}_{\mathrm{NS},1} +
				        m_2\,\pmb{r}_{\mathrm{NS},2}\right)\times
					\pmb{v}_\mathrm{NS} \nonumber\\
	      &=& \,-\,\Delta m \,\left( \pmb{D}_\mathrm{NS}\times \pmb{v}_{r,2}
	            + \pmb{r}_{\mathrm{NS},2}\times\pmb{v}_\mathrm{NS} \right)\,,
\label{eq:jrv}
\end{eqnarray}
where we assumed the centers of the two masses to be located at the same 
average distance from the COE and symmetrically to the vector 
$\pmb{D}_\mathrm{NS}$. This allowed us to apply the symmetry conditions
already exploited in Section~\ref{sec:angmomaccvol}. Moreover, we
used $m_2 = m_1 + \Delta m$ and again $\pmb{v}_r(\pmb{r}) = \pmb{v}_r(r)$
and Equation~(\ref{eq:rrns1}).
The vector $\pmb{J}_r + \pmb{J}_v$ is perpendicular to $\pmb{v}_\mathrm{NS}$
(which we again assume to be parallel to $\pmb{D}_\mathrm{NS}$).
For an order-of-magnitute estimate we take, as an extreme case that
maximizes the angular momentum, $r_\mathrm{NS} = R_\mathrm{acc}$
and for $v_r$ the escape velocity at this radius, i.e., 
$v_r = \sqrt{2\,GM_\mathrm{NS}/R_\mathrm{acc}}$, to 
write\footnote{It is important to consider the velocity directions 
{\em before} the deceleration by the reverse shock, because the reverse
shock moving across an inhomogeneous medium will destroy the spherically
symmetric profile $\pmb{v}_r(\pmb{r}) = \pmb{v}_r(r)$, which can be
assumed for the initially expanding, nearly homologous flow.}
\begin{eqnarray}
	J_r+J_v &\sim& \Delta m\,\left|\,-\,D_\mathrm{NS}\,
	\sqrt{\frac{2\,GM_\mathrm{NS}}{R_\mathrm{acc}}}\,
	\langle\sin\theta_m\rangle\right.  \nonumber\\
	&\phantom{+}& \left.\phantom{\sqrt{\frac{2\,GM_\mathrm{NS}}{R_\mathrm{acc}}}}
	+ R_\mathrm{acc}\,v_\mathrm{NS}\,
	\langle\sin\theta_{\mathrm{NS},m}\rangle\, \right| \,.
	\label{eq:jrv2}
\end{eqnarray}
Here, $\theta_m$ is the angle between $\pmb{D}_\mathrm{NS}$ and 
$\pmb{v}_{r,2}$, and $\theta_{\mathrm{NS},m}$ the angle between
$\pmb{r}_{\mathrm{NS},2}$ and $\pmb{v}_\mathrm{NS}$. Both are related
via Equation~(\ref{eq:rrns3}). The angle brackets indicate suitable
hemispheric averages corresponding to the mass-center of $m_2$. 
The minus sign in the first term on the right-hand-side of
Equation~(\ref{eq:jrv2}) results from
the fact that the two vector contributions in Equation~(\ref{eq:jrv})
point in opposite directions.

Applying the same basic approximations, we obtain for $\pmb{J}_\mathrm{t}$:
\begin{equation}
	\pmb{J}_\mathrm{t} \approx 
                           m_1\,\pmb{r}_{\mathrm{NS},1}\times\pmb{v}_{\mathrm{t},1}
                         + m_2\,\pmb{r}_{\mathrm{NS},2}\times\pmb{v}_{\mathrm{t},2}\,,
\label{eq:jt}
\end{equation}
which is a vector with components parallel and perpendicular to 
$\pmb{v}_\mathrm{NS}$. Referring to our symmetry assumptions, i.e.,
$r_{\mathrm{NS},1} = r_{\mathrm{NS},2} = R_\mathrm{acc}$ and 
$\theta_{\mathrm{NS},1} = \theta_{\mathrm{NS},2} = \theta_{\mathrm{NS},m}$, 
and using $v_{\mathrm{t},1} = v_{\mathrm{t},2} = v_\mathrm{t}$
for a closed, steady vortex flow, these two components become:
\begin{eqnarray}
	J_{\mathrm{t},\parallel\,\pmb{v}_\mathrm{NS}} &=&
	(m_1 + m_2)\,R_\mathrm{acc}\,v_\mathrm{t}\,\sin\theta_{\mathrm{NS},m}\,,
\label{eq:eq:jtparallel} \\
	J_{\mathrm{t},\perp\,\pmb{v}_\mathrm{NS}} &=&
        (m_2 - m_1)\,R_\mathrm{acc}\,v_\mathrm{t}\,\cos\theta_{\mathrm{NS},m}\,.
\label{eq:eq:jtperp}
\end{eqnarray}
Again, the minus sign in Equation~(\ref{eq:eq:jtperp}) is a consequence
of the opposing directions of the two vector components contributing to
$J_{\mathrm{t},\perp\,\pmb{v}_\mathrm{NS}}$.
It depends on the details of the mass distribution and the vortex helicity
whether the two angular momentum components perpendicular to $\pmb{v}_\mathrm{NS}$,
namely $J_r+J_v$ and $J_{\mathrm{t},\perp\,\pmb{v}_\mathrm{NS}}$, are rectified
or partially compensate each other. However, in any case it is clear that both
of these terms depend on the mass difference $\Delta m$, whereas the angular
momentum of the accreted fallback matter parallel to $\pmb{v}_\mathrm{NS}$, 
i.e., $J_{\mathrm{t},\parallel\,\pmb{v}_\mathrm{NS}}$, depends on the total
mass $m = m_1 + m_2$. It should thus dominate in size, and even in our quite 
extreme situation the main component of the neutron star spin obtained from
fallback accretion will lie along the axis defined by the vector of the 
neutron star kick velocity.

We conclude that also in the presence of density inhomogeneities, one can 
expect that fallback and mass accretion from a volume $V_\mathrm{acc}$ shifted
to an asymmetric location relative to the COE by a distance $D_\mathrm{NS}$, 
should favor spin-kick alignment (or anti-alignment) of the neutron star.

\subsubsection{Some simple estimates}
\label{sec:estimates}

Using the previous estimates and scales, we now derive an
order-of-magnitude estimate for the neutron star angular momentum,
assuming that the neutron star is able to accrete most or all of the mass in 
the fallback volume. We find this angular momentum to be compatible with the
angular momentum of new-born neutron stars deduced from observations, and 
also to be of similar magnitude as found in recent 3D simulations 
(see Section~\ref{sec:simulations}). According to our discussion in 
Section~\ref{sec:skalign}, we expect that the angular momentum vectors will 
be preferentially distributed around the kick direction within a cone of 
half-opening angle of $\sim$30$^\circ$ for neutron stars with sizable 
natal kick velocities.

For our following estimates we consider the neutron star to capture the 
fallback mass mostly from a spherical, reverse-shock decelerated
shell with inner radius $R_\mathrm{shell}$ within a wedge of semi-opening 
angle $\theta_\mathrm{w}$ (measured at the COE) around the neutron star's
direction of motion. For $\cos\theta_\mathrm{w}$ 
Equation~(\ref{eq:costhetax}) holds ($\theta_\mathrm{x}$ there is 
$\theta_\mathrm{w}$ here) with $r_\mathrm{v}$ being replaced by 
$R_\mathrm{shell}$ (Figure~\ref{fig:Graphics5}). 
As before, we set $R_\mathrm{shell}\sim R_\mathrm{acc}$
for an extreme case to get
\begin{equation}
	\cos\theta_\mathrm{w}^\mathrm{max} \sim \frac{D_\mathrm{NS}}{2R_\mathrm{acc}}\,.
\label{eq:costhetaw}
\end{equation}
The solid angle subtended by the wedge is $\Delta\Omega_\mathrm{w} = 
2\pi(1-\cos\theta_\mathrm{w}^\mathrm{max})$, and the fraction of the shell 
encompassed by the wedge is
\begin{equation}
	f_\mathrm{w} = \frac{\Delta\Omega_\mathrm{w}}{4\pi} 
	\sim \frac{1}{2}\left(1-\frac{D_\mathrm{NS}}{2R_\mathrm{acc}}\right)\,.
\label{eq:fw}
\end{equation}
For $D_\mathrm{NS}\sim R_\mathrm{acc}$ (referring to arguments in 
Section~\ref{sec:capture}) we obtain $\theta_\mathrm{w}^\mathrm{max} \sim \frac{\pi}{3}$,
$\Delta\Omega_\mathrm{w}\sim \pi$, 
and $f_\mathrm{w}\sim\frac{1}{4}$. At most a quarter of the spherical shell is
thus encompassed by the neutron star's accretion volume $V_\mathrm{acc}$.

An upper bound of the corresponding accretion of angular momentum can be estimated
by Equation~(\ref{eq:eq:jtparallel}) for the component parallel to the neutron star's 
line of motion. Employing the relation
\begin{eqnarray}
	R_\mathrm{acc}\sin\theta_\mathrm{NS,w}^\mathrm{max} &=& 
        R_\mathrm{shell}\sin\theta_\mathrm{w}^\mathrm{max} \nonumber\\
	&\sim& R_\mathrm{shell} \left(1-
        \frac{D_\mathrm{NS}^2}{4R_\mathrm{acc}^2}\right)^{1/2} ,
\label{eq:rsin}
\end{eqnarray}
where we made use of $\cos\theta_\mathrm{w}^\mathrm{max}$ from 
Equation~(\ref{eq:costhetaw}), we obtain
\begin{eqnarray}
	J_\mathrm{t,\parallel\,\pmb{v}_\mathrm{NS}} &<&
	M_\mathrm{acc}\,v_\mathrm{t}\,R_\mathrm{acc}\,\sin\theta_\mathrm{NS,w}^\mathrm{max}
	\nonumber\\
	&\sim& M_\mathrm{acc}\,v_\mathrm{t}\,R_\mathrm{acc}\left(1-
	\frac{D_\mathrm{NS}^2}{4R_\mathrm{acc}^2}\right)^{1/2} .
	\label{eq:jmax}
\end{eqnarray}
Here, we denoted the total accreted mass in the wedge by $M_\mathrm{acc}$
and made again the approximation of $R_\mathrm{shell}\sim R_\mathrm{acc}$.
This yields an upper limit of the accreted specific angular momentum of
\begin{eqnarray}
	j_\mathrm{t,\parallel\,\pmb{v}_\mathrm{NS}} &=& 
	J_\mathrm{t,\parallel\,\pmb{v}_\mathrm{NS}}\,M_\mathrm{acc}^{-1}
	\nonumber\\
	&<& 8\times 10^{16}\,\mathrm{\frac{cm^2}{s}}\,
	\left(\frac{R_\mathrm{acc}}{3\times 10^4\,\mathrm{km}}\right)
	\left(\frac{v_\mathrm{t}}{300\,\mathrm{km\,s}^{-1}}\right) ,
	\label{eq:jspecmax}
\end{eqnarray}
when $D_\mathrm{NS} \sim R_\mathrm{acc}$ is applied. The total angular momentum
for an accreted mass of $M_\mathrm{acc}\lesssim \frac{1}{4}\,M_\mathrm{shell}$,
with $M_\mathrm{shell}$ being the shell mass, is therefore limited by
\begin{eqnarray}
	J_\mathrm{t,\parallel\,\pmb{v}_\mathrm{NS}} &<&
	4\times 10^{47}\,\mathrm{erg\,s}\, \nonumber\\
	&\times& \left(\frac{M_\mathrm{shell}}{10^{-2}\,\mathrm{M}_\odot}\right)
	\left(\frac{R_\mathrm{acc}}{3\times 10^4\,\mathrm{km}}\right)
        \left(\frac{v_\mathrm{t}}{300\,\mathrm{km\,s}^{-1}}\right)\,,
	\label{eq:jmaxval}
\end{eqnarray}
corresponding to natal spin periods of the neutron stars between some ten
milliseconds to hundreds of milliseconds according to Equation~(\ref{eq:Tspin}), 
with the reference values used for the quantities in Equation~(\ref{eq:jmaxval}).
This is compatible with typical birth periods of neutron stars inferred from
observations \citep{Popov+2012,Igoshev+2013,Noutsos+2013}.%
\footnote{Considering a sample of 30 rotation-powered neutron stars proposed
to be associated with supernova remnants, 
\citet{Popov+2012} deduced a Gaussian distribution of the initial spin period 
---truncated at zero--- with a mean and standard deviation of 100\,ms. 
\citet{Noutsos+2013} constructed a somewhat larger sample of young radio pulsars, 
including moderately older pulsars, and used kinematic ages to derive a much 
broader range of initial spin 
periods, $T_\mathrm{spin}^0 = 63_{-35}^{+728}$\,ms (68\% confidence limits around 
the most probable value). The larger width of their distribution may be related
to the somewhat older pulsars, which currently have periods of several 100\,ms, or to the
different age estimator used. Overall, values of typically tens of
ms, but ranging up to (at least) 100--200\,ms, are a likely range of birth periods.}

The numerical value inserted for $R_\mathrm{acc}$ in 
Equations~(\ref{eq:jspecmax}) and (\ref{eq:jmaxval}) is guided by
Equation~(\ref{eq:racc}), $M_\mathrm{shell}$ is normalized by a value comparable 
to the fallback mass of an average case of
the 3D explosion models discussed in Section~\ref{sec:simulations}, and the 
value inserted for $v_\mathrm{t}$ is plausible because it is clearly subsonic
and of order 10\% of the typical shock velocity in most of the exploding star.
The numbers thus obtained for $j_\mathrm{t,\parallel\,\pmb{v}_\mathrm{NS}}$ and
$J_\mathrm{t,\parallel\,\pmb{v}_\mathrm{NS}}$ are in the ballpark of the 3D 
simulation results displayed in Figure~\ref{fig:fallback-vs-t}, although they 
need to be viewed as upper limits for the conditions of neutron stars
displaced from the COE by a distance $D_\mathrm{NS}\sim R_\mathrm{acc}$, whereas
in Figure~\ref{fig:fallback-vs-t} the fallback mass in the entire $4\pi$-domain
around the COE (which is appropriate for $D_\mathrm{NS} = 0$) 
was considered. But the possible range of the numerical estimates is
large, depending on the exact values inserted for the factors. In the hydrodynamical
supernova simulations the fallback rates of mass and angular momentum are highly 
time and case dependent, which is evident from Figure~\ref{fig:fallback-vs-t},
and the fallback mass (being a proxy of $M_\mathrm{shell}$ occurring in 
Equation~(\ref{eq:jmaxval})) can be several 0.1\,M$_\odot$.

\section{Implications of the fallback spin-kick scenario}
\label{sec:discussion}

A variety of implications and predictions can be concluded from the 
proposed new fallback scenario for spin-kick alignment or anti-alignment. 
These conclusions partly reverse those drawn from previously suggested
mechanisms.

\subsection{Dependence on neutron star kicks}

As discussed above, the possibility of spin-kick alignment hinges on the
displacement of the neutron star from the COE and should thus become more
likely for neutron stars with large kicks and for fallback accretion 
happening at late times.
Late fallback is associated with the highest angular momentum
(see Figure~\ref{fig:fallback-vs-t}) and the distance of the kicked neutron 
star from the COE is growing with time. Therefore fallback matter from the 
forward direction of the neutron star's motion will dominate the accretion
at late times when the neutron star has moved far away from the COE, and 
consequently late fallback is most conducive to spin-kick 
alignment.\footnote{We remark that usually fallback accretion, 
multi-directional as well as onesided, has a minor influence
on the neutron star kick because of the small amount of mass involved 
\citep[this, however, is different in the case of fallback supernovae;][]{Janka2013}.
Typically, it tends to {\em increase} the neutron star kick, because the 
slowest fraction of the supernova ejecta is most inclined to fall back,
thus leaving the remaining ejecta expanding with a higher linear momentum
in the opposite direction.}
Since angular momentum accreted by the neutron star from fallback is 
connected to tangential vortex flows in the accreted matter, the net vector
direction can point radially inward or outward. Spin-kick alignment or
anti-alignment are therefore expected to be equally probable possibilities.

Inversely, neutron stars with low kick velocities
are less likely to exhibit aligned or anti-aligned spins and kicks, because
they stay close to the COE and can accrete fallback from arbitrary directions. 
Because of this dependence on the neutron star kick, small relative angles
between the spin and kick axes are not expected to be ubiquitous and, in
the case of low-velocity neutron stars, should happen only accidentally
within the discussed fallback scenario.

Randomly scattered orientations of spin and kick for low neutron star
velocities should thus transition to more correlated orientations if the
neutron star kick is large. The velocity where this transition occurs 
is determined by the condition 
$R_\mathrm{acc}\sim D_\mathrm{NS}\sim v_\mathrm{NS}t_\mathrm{fb}$.
Here, $t_\mathrm{fb}$ is the time of the most relevant episode of fallback 
accretion with angular momentum transfer to the neutron star. It
depends on the fallback history, which varies from progenitor to progenitor 
and is also influenced by the explosion energy. Equation~(\ref{eq:racc}) 
shows that $R_\mathrm{acc}$ is of order $10^4$\,km or a few times this
value. According to Figure~\ref{fig:fallback-vs-t}, $t_\mathrm{fb}$ can 
vary over a wide
range from some 10\,s to many 1000\,s, depending on the progenitor. If
$t_\mathrm{fb}$ is 10--100\,s, spin-kick alignment can be expected only
for neutron star kicks larger than some 100\,km\,s$^{-1}$, whereas for
$t_\mathrm{fb}> 1000$\,s correlated spin and kick directions become
likely already when $v_\mathrm{NS}$ exceeds some 10\,km\,s$^{-1}$. 
Consequently, neutron stars with observed spin-kick alignment and 
measured kick velocity carry information about the time scale of the
spin-setting phase of fallback accretion.

Another aspect of variability is the effect of the initial angular momentum
of the neutron star at the time of its formation shortly after the launch of 
the supernova shock. Although fallback accretion should be the
leading effect that sets the final neutron star spin, there can be exceptions
when the accreted matter does not carry significant amounts of angular
momentum. In such cases, which are presumably statistically rare, the 
spin of the neutron star will either reflect the angular momentum inherited 
from the rotating progenitor core, or it will be determined by asymmetric 
flows associated with hydrodynamic instabilities during and shortly after 
shock revival \citep[see, e.g., numerical simulations by][]{Blondin+2007,
Fernandez2010,Kazeroni+2016,Kazeroni+2017,Bollig+2020}.
Also in such cases spin-kick alignment would be an accidental 
outcome, unless very rapid progenitor rotation imposes a preferred kick 
direction along the rotation axis of the stellar core and new-born neutron 
star.

\subsection{Influence of stellar rotation}

In the discussed fallback scenario for spin-kick alignment, the rotation of
the progenitor star and of the neutron star are no necessary
prerequisite that determines the kick direction. This is in stark contrast
to previously suggested mechanisms that could lead to spin-kick alignment.
All of the previous theories assume that the neutron star rotation determines
the kick direction that develops later on. In contrast, in our revised
picture the neutron star receives its kick first and acquires its
angular momentum subsequently by off-center accretion of vorticity in
fallback matter. Therefore, observed spin-kick
alignment sets no constraints on the rotation of the progenitor core or
of the neutron star prior to the time when the compact remnant received
its kick.

This also implies that spin-kick alignment of observed neutron stars cannot
be interpreted as indicative of any pre-collapse stellar spin, and the 
kick and spin directions are not systematically correlated with the 
rotation axis of the progenitor. As we argued in Section~\ref{sec:skalign}, 
in the fallback scenario of spin-kick alignment progenitor rotation will 
destroy the possibility of spin-kick alignment, if it is fast enough to
exceed the specific angular momentum connected to vortex flows.
This requirement implies that the stellar layers outside of the 
progenitor's iron core should possess a rotational angular momentum of
$j_\mathrm{rot} > j_\mathrm{vortex}$, where $j_\mathrm{vortex}$ for the
vortex flows in the fallback matter is of the order of several 
$10^{16}$\,cm$^2$\,s$^{-1}$ (Equation~(\ref{eq:jspecmax})). Observationally
established spin-kick alignment as a common property of young pulsars
would therefore suggest that the silicon and oxygen shells of pre-collapse
stars rotate more slowly than this limit. In our spin-kick scenario the 
actual situation is therefore opposite to what has been suggested in 
previous literature.

If the angular 
momentum of the fallback matter is dominated by the angular
momentum associated with stellar rotation instead of local tangential 
vortex flows, the accreting neutron star will exhibit spin-kick alignment 
only if its kick is along the progenitor's rotation axis, or
if the kick is small and the neutron star stays near the COE.
If, however, the neutron star kick has an arbitrary direction relative
to the progenitor's rotation axis, spin-kick alignment will get lost 
by the fallback accretion in rapidly rotating stars. In the case of kicks 
perpendicular to the progenitor's rotation axis, for example, fallback 
will lead to nearly orthogonal directions of spin and kick axes. 
In the hydrodynamic and neutrino kick scenarios discussed in our paper 
on grounds of our hydrodynamic explosion models and previous simulations, 
the neutron star kicks are found to
be randomly oriented unless rotation in the stellar core is extremely
rapid and thus determines the explosion dynamics and ejecta asymmetry.

The explosion dynamics and ejecta asymmetry begin to depend on the spin 
of the progenitor core if the neutron star (with a final radius of 
$\sim$12\,km) could attain a spin period of a millisecond or less (prior 
to fallback accretion) under the assumption of angular momentum conservation
in the collapsing stellar core. Such conditions require a pre-collapse
rotation rate of the iron core with an average angular frequency of about 
0.4\,rad\,s$^{-1}$ or more, corresponding to a mass-averaged specific 
angular momentum of at least $\sim$$4\times 10^{15}$\,cm$^2$\,s$^{-1}$.

\subsection{Binary neutron stars}

The physical mechanisms that are at work in the explosions of single stars 
also determine the formation properties of neutron stars in binary systems.
Binary neutron stars, in particular the second-born neutron star, offer
particularly interesting test cases of the proposed new scenario for
spin-kick alignment. 

If the second-born neutron star received a small
kick, which can be inferred from a small orbital eccentricity, the 
neutron star spin should have a random orientation. This is expected if
the spin is connected to fallback accretion and not to angular momentum
inherited from rapid rotation of the core of the progenitor star.
In contrast, if the binary-system parameters ---in particular a considerable 
orbital eccentricity--- suggest that the second-born neutron star received
a large natal kick, the fallback scenario favors spin-kick alignment.
In such cases the spin direction of the neutron star would
provide rough information of the kick direction. This kick direction
should be compatible with the orbital parameters of the binary system.
Known binary neutron star systems may be checked for this requirement.

\subsection{Partial accretion, disks, jets} 

In the context of fallback one needs to carefully discriminate between
fallback mass and the mass that is captured by the accretion volume of the
neutron star (with mean radius $R_\mathrm{acc}$), which is only a fraction
of the fallback mass. In turn, only a fraction of this captured mass might
finally end up on the neutron star in an accretion process that might not
be perfectly efficient. In the following we will briefly mention a few 
consequences of these facts and possibilities.

Because of the huge amounts of angular momentum associated even with
little fallback mass (see Table~\ref{tab:3dmodels}, 
Figure~\ref{fig:fallback-vs-t}), the disk criterion of
Equation~(\ref{eq:jkepler}) is easily fulfilled. This applies also to
the discussed case that the neutron star captures only a minor
fraction of the fallback material with the specific angular momentum
being constrained by Equation~(\ref{eq:jspecmax}). Therefore fallback
disks should be a widespread phenomenon, also in our revised scenario
where the neutron star moves away from the COE due to its natal kick.

When high-angular-momentum matter assembles into an accretion disk, 
magnetic fields may be rotationally amplified, which in turn could drive
collimated polar outflows, fed by a fraction of the disk material. Since
the escape velocity from radii very close to the accreting neutron star
is several ten percent of the speed of light, even
$10^{-3}$\,M$_\odot$ of these polar ejecta could carry an energy of up
to $\sim$\,$10^{50}$\,erg, scaling linearly with the ejected mass.
This value is far sufficient to explain the energy that has been
estimated for jet-like structures seen in nearby supernova remnants,
for example the high-velocity, wide-angle NE and SW features seen in 
Cassiopeia~A \citep[e.g.,][]{Fesen+2016}, the faint
protrusion or ``chimney'' that extends out from the northern rim
of the visible Crab Nebula
\citep[which is often called northern ejecta ``jet''; e.g.][]{Gull+1982,
Blandford+1983,Davidson+1985,Fesen+1993,Black+2015}, and, if these
structures are indeed connected to flows originating from the explosion
center, also some ear-like extensions that have been interpreted into
images of various supernova remnants (\citealt{Bear+2017, Bear+2018a}; perhaps
even in SN~1987A, \citealt{Soker2021}). All of these features might possibly 
be relics of collimated post-explosion outflows linked to the formation of 
fallback disks around the new-born neutron stars.

It is interesting to note that the northern funnel of Crab as well as the
Cas~A NE and SW jets are seen nearly perpendicular to the observationally
inferred direction of the neutron star's kick motion. Moreover, spin and
kick of the Crab pulsar are visibly aligned and thus both have an 
orientation that is nearly orthogonal to the northern funnel. (Nothing is 
known about the rotation of the compact object in Cas~A). These geometries
therefore seem to be in conflict with our proposed fallback scenario for
spin-kick alignment, which is a fact that deserves a comment.
The conflict, however, is only apparent. 

In our scenario the spin-kick
alignment is likely to be a consequence of the latest fallback, which
is associated with the highest specific angular momentum of the fallback
material (see Figure~\ref{fig:fallback-vs-t}) and the largest displacement
of the kicked neutron star from the COE. Both determine the vorticity
of the matter that is selectively accreted from the forward direction of 
the neutron star's motion. Earlier fallback, in particular at times when
the neutron star has not yet moved away from the COE by any relevant 
distance, is instead expected to produce spins that are randomly
oriented and thus preferentially perpendicular to the kick direction
(see Figure~\ref{fig:histoskangle}).
Therefore, in our proposed scenario, a near-orthogonality of jet axis
and neutron star kick vector points to a possible origin of the jet 
from an early fallback episode, which formed a transient neutron star disk
whose rotation axis had a large angle relative to the neutron star's
kick vector. The effect of this accretion phase on the neutron star's
spin was then overruled by later fallback, which enforced spin-kick
alignment because of its high angular momentum.\footnote{The small mass
of the late fallback and of a possibly re-ejected fraction of it would, 
however, not be energetic enough to produce strong post-explosion jets.
This would then be compatible with the lack of jet-kick alignment
inferred from the interpretation of morphological features in Cas~A 
and 11 other core-collapse supernova remnants with known directions 
of the neutron star kicks \citep{Bear+2018b}.}

Such an interpretation is supported by the relatively large estimates
of mass ($\sim$0.1\,M$_\odot$) and kinetic energy ($\sim$\,$10^{50}$\,erg)
in the NE jet and SW counterjet \citep{Fesen+2016}, 
which show strong emission from Si, S, Ar, 
and Ca, i.e., from chemical elements that stem from the inner layers of the
progenitor's metal core but not from the innermost supernova ejecta, which 
would be rich of iron-group elements formed in the first second(s) of the 
explosion. The fact that intermediate-mass elements are found in the
jets suggests that this material has not originated from the immediate
vicinity of the neutron star, where it would have been heated to nuclear
statistical equilibrium (NSE) and dissolved into free nucleons and alpha 
particles.
During the re-ejection and expansion cooling these nucleons and alpha
particles would have reassembled into iron-group material associated with
some remaining helium. Instead, fallback of matter from the star's silicon
and oxygen layers and subsequent outflow from an accretion disk without
extreme heating to NSE temperatures may offer a plausible possibility.
The assumption that the material never got very close to the neutron star
during infall and re-ejection is also compatible with the combination of 
fairly high mass and relatively low energy, i.e., an energy-to-mass ratio
that is $\sim$100 times lower than the numbers estimated above for matter 
escaping the surface gravity of the neutron star. 
Similarly, the ``chimney'' of the Crab Nebula is observed in emission lines 
of oxygen. This also reflects the chemical composition of the progenitor's
core well exterior to the mass cut of the explosion. But it does not 
agree with the chemical
fingerprints that are characteristic of material that has been
nucleosynthesized in the hottest neutrino-heated and shock-heated layers
that have received the energy input from the mechanism driving the 
supernova blast.

Besides these features of the Crab and Cas~A remnants, which might be 
relics of collimated outflows from a post-explosion phase of disk accretion
by the neutron star, there are no direct observational hints for fallback 
disks around neutron stars. Nevertheless, such disks have been proposed 
as possible central engines of peculiar and rare supernovae, e.g.\ 
of super-luminous events 
in the case of significant accretion power \citep{Dexter+2013}. 
Morever, a number of pulsars and magnetars have been announced to 
possess planets or circumstellar disks, and ``ears'' and asymmetric
structures in a larger sample of supernova remnants including SN~1987A 
have been interpreted as relics of jet-like outflows \citep{Bear+2017,Bear+2018a}.

Finally, we remark that the incomplete capture of fallback matter and
inefficient disk accretion by the neutron star as expected in our scenario
suggest that
considerable amounts ($\sim$\,$10^{-4}$\,M$_\odot$ to up to $\sim$0.1\,M$_\odot$) 
of unaccreted fallback material may be present in the central volume of
supernovae. This low-velocity, low-density gas should be rich of chemical
elements that are present in the progenitor core (Si, Mg, Ne, O) and possibly 
also some nuclear species that are created in the alpha-rich freeze-out
of neutrino-heated matter (e.g., helium) and by explosive nucleosynthesis 
(iron-group and intermediate-mass elements), but that lateron do not get 
swept out in the ejecta.

\subsection{Implications for black hole formation}

The proposed scenario of spin-kick alignment by fallback is mostly relevant 
for neutron star formation, but there are some aspects that deserve
consideration also for stellar core-collapse events that lead to the 
formation of black holes.

Kicking a black hole requires some anisotropic mass or energy loss from
the collapsing star, and this may happen efficiently only through 
neutrinos or mass ejection accompanying the
formation of the black hole, or possibly through gravitational waves
if very rapid rotation and triaxial deformation play a role during stellar
core collapse.

Asymmetric mass ejection may yield sizable black-hole kicks in potential fallback supernovae 
\citep{Janka2013}, i.e., in cases where the energy release by the explosion
mechanism is able to unbind only a fraction of the dying star, while the 
rest falls back onto the transiently existing neutron star, pushing
its mass beyond the black hole formation limit. This hydrodynamic kick 
mechanism is the counterpart of the gravitational-tug boat mechanism 
for neutron stars kicks discussed in our paper. But in order to 
work for black holes, it requires that the fallback is not too massive,
because otherwise the explosion asymmetries created by the mechanism
will be swallowed by the black hole instead of being carried away by 
ejected mass \citep{Chan+2020}. Black hole kicks of the same 
magnitude as neutron star kicks, i.e.\ with typical velocities of several 
100\,km\,s$^{-1}$ \citep[as discussed by][]{Janka2013}, might therefore 
only be possible for small black holes that form through moderate 
fallback and thus possess masses of a few 
solar masses but not tens of solar masses. Yet, the picture drawn
by existing 3D simulations is still inconclusive.

It is presently unclear whether collapsing stellar cores can rotate
sufficiently rapidly so that triaxiality or fragmentation 
\citep[e.g.,][]{Rampp+1998,Imshennik+1995,Imshennik+2004,Imshennik+2010,Fedrow+2017}
can play a role to produce powerful gravitational-wave emission. If realistic,
anisotropic radiation of gravitational waves might cause considerable recoil kicks 
\citep[e.g.,][and references therein]{Pietilae+1995,Campanelli+2007a,Campanelli+2007b}
of such highly deformed, transiently stable neutron stars on their 
way to the final gravitational instability for black hole formation.
However, because of linear momentum conservation, the initial kick
velocity would be reduced subsequently when the rest of the collapsing 
star is added to the mass of the newly formed black hole.

Asymmetric radiation of
neutrinos has been found recently in 3D models as a consequence of
a dipolar convection mode that develops inside of the hot proto-neutron 
star \citep[see, e.g.,][]{Tamborra+2014a,Tamborra+2014b,Janka+2016,Glas+2019,
Powell+2019,OConnor+2018,Nagakura+2021}. Anisotropic neutrino emission 
can also be a consequence of highly aspherical accretion onto the 
neutron star. In this case the dominant effect may not be asymmetric 
neutrino emission, but instead asymmetric 
absorption of neutrinos leaving the neutrinosphere more isotropically 
\citep{Bollig+2020}. If the absorbing matter is accreted by the compact
remnant instead of being ejected, the anisotropic neutrino absorption
also exerts a recoil kick on the compact object. However, the magnitude
of such neutrino-induced kicks is relatively small, ranging from a few
10\,km\,s$^{-1}$ for neutron stars from low-mass iron-core progenitors 
\citep{Stockinger+2020} to $\sim$100\,km\,s$^{-1}$ for neutron stars from
high-mass progenitors \citep{Bollig+2020}. In both
cases these kicks are usually dwarfed by the hydrodynamic kicks.\footnote{The
only exceptions where neutrino-induced kicks yield dominant (or sizable) 
contributions to the neutron star kicks are electron-capture supernovae
(ECSNe) and low-mass iron-core-collapse supernovae (CCSNe) with steep
density declines outside of their degenerate cores.
Since these have low explosion energies, no relevant accretion by the
new-born neutron star, small masses of neutrino-heated matter, and
weakly asymmetric, very rapidly expanding inner ejecta, their hydrodynamic
kicks are only a few km\,s$^{-1}$ for ECSNe \citep{Gessner+2018} and
some 10\,km\,s$^{-1}$ for low-mass Fe-CCSNe \citep{Stockinger+2020,
Mueller+2018}.} 

In black hole forming core-collapse events, the terminal neutrino-induced
kicks are unlikely to be larger than a few km\,s$^{-1}$, maybe at most 
of order 10\,km\,s$^{-1}$ (N.~Rahman et al., in preparation).
However, for transient periods of many seconds, the new-born and
still accreting black holes can attain kicks of several 10\,km\,s$^{-1}$
up to even more than 100\,km\,s$^{-1}$ due to anisotropic neutrino losses.
The values of such kicks and their evolution with time depend on 
several factors. On the one hand, the convective dipole in the temporarily
existing
neutron star affects the neutrino emission only with an attenuated strength, 
because a massive, convectively stable accretion mantle grows quickly around
the convective shell. The basically spherical accretion mantle dominates
the neutrino emission and reduces the diffusive neutrino flux that transports
neutrinos from the convective layer to the neutron star surface 
\citep{Walk+2020}. On the other hand, Rahman et al. (in preparation)
found that in very massive stars considerable neutrino emission asymmetries
can be produced by initial ejecta that fall back {\em after black hole 
formation} with high rates and extremely anisotropically. The thus attained
neutrino-induced kick velocity, however, decreases with time when the 
black hole accretes more and more fallback matter, and the final kick 
velocity is diminished by the large mass of the compact remnant when
it has swallowed the entire star or a major fraction of it.
Nevertheless, due to its transiently considerable velocity, the
black hole can get displaced from the COE (or, in this case more precisely,
from the center of mass of its progenitor star).

If the black hole has a small natal kick velocity and stays close to the COE,
fallback accretion in nonrotating progenitors should result in spins that
are randomly distributed relative to the kick directions, with a statistical
preference for orientations that are nearly perpendicular. This is in 
line with recent results of \citet{Chan+2020} \citep[see also][]{Antoni+2021}. 
In progenitors with rotation fast enough to dominate the angular momentum
in the collapsing star and fallback matter,
the black hole spin and its direction will be determined
by the angular momentum inherited from the progenitor's rotation. 
Spin and (low-velocity) kick can be aligned if linear
momentum is lost (through anisotropic neutrino and gravitational-wave emission or 
asymmetric mass ejection, e.g.\ in jets) along a rotation-associated
axis of global deformation.

Fallback supernovae with considerable mass ejection are special cases,
because the just-formed black hole can acquire a significant hydrodynamic 
kick and an additional neutrino-induced kick and can move out of the COE.
As mentioned above, sizable kicks can be obtained only by black holes 
with relatively low mass compared to their progenitors, because the amount of
fallback is constrained by the requirement that asymmetries created by the 
explosion mechanism should be ejected instead of being swallowed by
the black hole with the fallback matter \citep{Chan+2020}.
In such cases the angular momentum of the black hole will be mainly
determined by the latest fallback. The situation is therefore
similar to the discussed conditions for new-born neutron stars, where
the kick is received earlier and the spin is subsequently set by 
fallback accretion.
Also in such black hole formation cases we therefore
expect a possible correlation of natal spin and kick directions with
the tendency towards alignment of their axes. As for neutron stars, this
effect can result from vorticity associated with the fallback matter that  
is preferentially captured from the forward direction of the black hole's
motion, whereas fallback accretion in the wake of the black hole would be
less efficient. Because the kick direction of the black hole is
random as a consequence of the stochasticity of mass-ejection asymmetries,
this would imply the possibility of tossing the spin axis of the black 
hole during its formation, thus changing it relative to the progenitor's
rotation axis. As in the case of neutron stars, 
such an effect would be viable provided the progenitor rotation is not
too fast, i.e., the specific angular momentum associated with tangential
vortex flows dominates the specific angular momentum of the stellar 
rotation. Otherwise, stellar rotation will counteract any spin-kick alignment 
of the black hole, unless the progenitor rotates so rapidly that mass 
ejection as well as a natal kick of the compact remnant develop 
preferentially along the angular momentum axis of the dying star.

Again, none of the described spin-kick dependencies connected to
the remnant's kick motion can be found in results of current 3D
hydrodynamic models \citep[see][]{Chan+2020}, because these simulations keep
the compact remnant attached to the center of the computational grid, which
coincides with the COE and the center of mass of the progenitor.

\section{Summary and conclusions}
\label{sec:conclusions}

We reviewed the results of a large set of 3D supernova simulations of
BSG and RSG progenitors that were presented here and in other recent
publications (Section~\ref{sec:simulations}).
These simulations followed the blast-wave evolution from
the onset of the explosion to late times, partly up to several days,
in order to include also the fallback of stellar matter that does not
get unbound during the explosion.
They demonstrate that anisotropic, long-time and late-time fallback can
be associated with large amounts of angular momentum even in nonrotating
progenitors and even if the fallback mass is small.
The exact fate of this fallback material is undetermined
by the simulations, which could not resolve the neutron star and its immediate
surroundings over such long evolution periods because of the huge disparity
of the relevant length and time scales. But even if only a fraction of this
material gets accreted onto the neutron star, it is likely to
govern the birth spin of the compact remnant. While the neutron star
obtains its natal kick on a time scale of several seconds through
asymmetric mass ejection and (usually to a smaller extent) through
anisotropic neutrino emission, it can take hundreds to thousands of seconds
or longer before the spin of the neutron star is ultimately determined.

In this scenario the kick and spin-up mechanisms are therefore based on
different physical processes, and the kick is transferred to the neutron
star long before its spin is set. This is in conflict with traditional
thinking that has guided a variety of ideas \citep[e.g.,][]{Spruit+1998,
Lai+2001,Wang+2006,Ng+2007} to explain the measured
pulsar spins and kicks and, especially, a possible spin-kick alignment or
anti-alignment concluded from pulsar observations 
\citep[e.g.,][see Section~\ref{sec:observations}]{Johnston+2005,Johnston+2007,Wang+2006,Ng+2007,
Noutsos+2012,Noutsos+2013}. All
of these explanations resort to the canonical assumptions
that the neutron star either inherits its angular momentum from the
rotating progenitor core or that it obtains spin and kick simultaneously
via the same process. Proposed examples of such processes are anisotropic
neutrino emission caused by ultra-strong magnetic dipole fields in 
rapidly spinning proto-neutron stars, or rotational
averaging of off-center impacts by accretion downflows or
hot-spot emission of neutrinos. However, none of these proposed explanations
is fully convincing in view of our present understanding of progenitor and
supernova conditions. In particular, rapid rotation and very strong magnetic
dipole fields are not believed to be ubiquitous in new-born neutron stars,
stochastic thrusts by anisotropic neutrino emission are much too weak,
and accretion has not been found to produce any directional correlations
of neutron star spins and kicks in current 3D supernova models.

Giving up the canonical assumptions about the origins of pulsar spins and
kicks also enforces a rethinking of possible scenarios for spin-kick alignment.
At first glance, the decoupling of neutron star kick and spin mechanisms
in hydrodynamic long-time explosion simulations appears to be unfavorable
for a possible correlation of the spin and kick directions. Indeed, the 3D
models presented here and in the recent literature \citep{Chan+2020,Powell+2020,
Stockinger+2020,Bollig+2020} display an effectively uniform distribution
of the relative orientations of spin and kick vectors. This holds true for
the situation right after shock revival as well as
later after fallback, at both epochs with no tendency of alignment or
anti-alignment (Figure~\ref{fig:histoskangle}).

However, these supernova models disregard the effect that the neutron star
should move because of its natal kick and therefore, as time goes on, should
develop a growing displacement $D_\mathrm{NS}$ from the COE.
Instead, the numerical models treat
the neutron star as pinned to the center of the computational grid. At late
times, when fallback can carry huge amounts of angular momentum, the
neutron star's drift away from the COE can become comparable to the accretion
radius of the remnant. Therefore it must be expected that the movement of the
neutron star can alter its ability to capture and accrete fallback matter.

We argued (in Section~\ref{sec:capture}) that the neutron star should
capture fallback material from a roughly spherical volume with accretion radius
$R_\mathrm{acc}$ and that the center of this volume is shifted into the
hemisphere towards which the neutron star is moving. Since the fallback mainly
affects matter accumulated in a dense shell behind the supernova shock during
phases of shock deceleration (whereas a low-density ``bubble'' fills the
central volume), the neutron star will predominantly accrete from a sector 
of this shell that lies in the direction of the neutron star's kick velocity.
For $D_\mathrm{NS}\sim R_\mathrm{acc}$ up to about a quarter of the
shell (encompassing a solid angle $\Delta\Omega_\mathrm{w}\sim \pi$)
is included in the accretion volume.

The situation is thus characterized by a pronounced asymmetry of the accretion
geometry with respect to the COE, and this has interesting implications for the
possibility of spin-kick alignment. Accordingly, in Section~\ref{sec:skalign},
we proposed a new scenario to explain the observationally suggested
spin-kick alignment or anti-alignment on grounds of our revised picture of
the processes that determine kicks and spins of neutron stars.
It is based on the insight that the matter ejected during the supernova blast
carries vorticity that was either created by nonradial hydrodynamic 
instabilities (SASI, convective overturn) at the onset of the explosion or 
is the relict of turbulent convection during the late shell-burning phases 
prior to stellar core collapse. With the supernova debris approaching a (nearly) 
homologous state of expansion, the tangential flows connected to this vorticity 
should become dominant in the ejecta and thus in fallback matter.

Capturing fallback material predominantly from the direction of the neutron
star's motion therefore implies that angular momentum is mainly associated with 
tangential vortex motions in this gas. The dominant angular momentum component
of matter in the capture volume of the neutron star should therefore have a
small angle relative to the line defined by 
the neutron star's velocity vector. Relying on the
leading role of the fallback for setting the neutron star spin, this means 
that the neutron star accreting from the fallback material should develop a
dominant angular momentum component aligned or anti-aligned with its kick.
We considered the envisioned geometry of such vortex flows under simplified
and highly idealized conditions and estimated that relative angles between
spin and kick axes of 30$^\circ$ or less can be expected if
$D_\mathrm{NS}\sim R_\mathrm{acc}$. The kick-induced displacement of the
neutron star from the center of the explosion plays a crucial role for the
possibility of spin-kick alignment or anti-alignment in this scenario. If
the neutron star receives a small natal kick and stays close to the COE,
it will accrete fallback from all directions. In such cases there is no 
reason to expect any systematics in the relative orientations of the
neutron star's spin and kick vectors, in agreement with the findings in 
our 3D supernova simulations, where the compact remnant is fixed at the 
center of computational polar grid.

Finally, in Section~\ref{sec:discussion} we discussed a variety of 
implications that follow from our proposed new fallback scenario for 
spin-kick alignment or anti-alignment. These include:
\begin{itemize}
\item
Spin-kick alignment should not be common for neutron stars
that receive small kicks and stay close to the COE. Vice versa,
the tendency for spin-kick alignment should be more distinctive
for neutron stars with high kick velocities, though a loose but not
strict correlation is likely. The reason for these expected
dependencies is that later fallback is more strongly affected by the 
neutron star's kick drift out of the COE, because its displacement
$D_\mathrm{NS}$ 
grows with time. Small angles between the spin and kick axes could be
common, but there is no reason to expect a strict alignment.
Alignment or anti-alignment are possible and similarly probable.
\item
The second-born objects in binary neutron stars offer
particularly interesting test cases of our fallback scenario for
spin-kick alignment, because the
orbital eccentricity can provide evidence of a significant natal
kick. In such cases the neutron star's kick direction is roughly 
correlated with its spin direction and should also be compatible
with the orbital parameters of the binary systems.
\item
Spin-kick alignment in the new scenario is {\em not} linked
to the pre-collapse rotation of the progenitor, and the kick and spin 
directions of the neutron star are not connected to the stellar rotation
axis. Observed neutron stars with spin-kick alignment do not require
pre-kick spins. In contrast,
stellar rotation, if sufficiently rapid, could {\em inhibit} spin-kick
alignment. This means that observationally established spin-kick alignment 
would provide considerable support of slow core rotation in collapsing 
stars, i.e., the specific angular momentum associated with stellar
rotation is dwarfed by the angular momentum associated with vortex flows.
\item
It is expected that not all of the fallback matter is captured in the 
accretion volume of the compact remnant, and because of the large angular
momentum of the captured matter, fallback disks should be quite common. 
It is also likely that not all of the disk mass is accreted. Magnetic field
amplification in the disks might produce collimated post-explosion outflows
and jets along the disk's rotation axis. These outflows might explain jet-like
features seen in gaseous supernova remnants including Crab and Cas~A.
The orientation of these structures with large inclination angle relative 
to the neutron star's spin direction suggests that they are connected to an 
early episode of accretion but not to the late accretion phase that 
determined the spin of the compact remnant. Because not all of the 
fallback matter ends up on the accretor, considerable amounts (several
$10^{-4}$\,M$_\odot$ up to some $10^{-1}$\,M$_\odot$) of unaccreted,
low-velocity, low-density matter containing iron and intermediate-mass 
elements as well as Si and O, may fill the central volume of supernovae.
\item
Black holes, in particular low-mass ones that are born in fallback 
supernovae, may receive natal kicks of similar magnitude as neutron stars
\citep{Janka2013,Chan+2020}. Since also in this case the black hole drifts 
away from the COE and its spin is determined by later fallback, the same 
proposed mechanism of spin-kick alignment by fallback 
applies for these black holes. Unless the progenitor rotates rapidly,
this process will produce a torque that is sufficiently strong to toss 
the spin axis of the black hole away from the progenitor's rotation axis,
because the kick direction should possess a random orientation.
\end{itemize}

In a follow-up paper we plan to present a detailed and
critical assessment
of the currently available observational data on pulsar velocities and
spin directions. This will allow us to revisit the implied 
neutron star kicks and their spin alignment
and to test the predictions of our proposed fallback scenario, using
both isolated and binary pulsars.

Theoretical consolidation and a quantitative investigation
will require a new generation of long-time 3D supernova
simulations that are capable of tracking the neutron star's kick motion
and that will follow the entire evolution of the fallback over periods of
many hours to days.\footnote{First steps in this direction with
detailed Boltzmann neutrino transport have 
recently been taken by \citet{Nagakura+2019}. Simulations, however, have been 
feasible so far only in two spatial dimensions and for short evolution
periods of a few 100\,ms after core bounce, terminated still before an
explosion could develop.}
A large sample of such 3D explosion models for different progenitors, with
and without stellar rotation, will be needed to predict the statistical
distribution of natal spin
and kick directions of neutron stars by theoretical work.
These models should also include asymmetries associated with convective
shell burning prior to core collapse, which demand 3D simulations for the
latest stages of the pre-collapse evolution 
\citep[see][]{Arnett+2011,Couch+2015,Mueller+2016,Yadav+2020,Yoshida+2019,Yoshida+2021}.
Because convective mass motions are associated with appreciable amounts
of angular momentum \citep{Gilkis+2014,Gilkis+2016}, convection
is an important source of vorticity in the infalling matter during the phase
when the supernova explosion is launched \citep[see][]{Mueller+2017,Bollig+2020}. 
Therefore it is 
also likely to determine the vorticity in the material that falls back
to the neutron star at later times. All of the 3D simulations discussed in
Section~\ref{sec:simulations}, however, are based on 1D progenitor data and 
thus do not account for the corresponding effects. This shortcoming severely
limits the information that our existing set of simulations can provide in 
the context of the envisioned scenario. Conclusive answers demand
explosion models that are based on the more realistic 3D structure of the
progenitor stars.

Besides full-scale 3D simulations, also studies of the evolution of vorticity 
in expanding supernova ejecta, similar to those of recent investigations of
convective vortices in collapsing stars by \citet{Abdikamalov+2020} and
\citet{Abdikamalov+2021}, 
can be relevant. They might provide a better understanding whether the 
fundamental assumptions of our fallback scenario for spin-kick alignment are 
justified, namely our consideration of vortex 
motions with velocities that are tangential to the quasi-homologous radial 
flow and that dominate the vorticity in the ejecta and in the material
ultimately falling back.
This is a radically simplified picture that demands validation by future 
work. Does the structure of the ejecta get close to the assumed conditions
before they are swept inward again? What is the detailed effect of the reverse 
shock, which, if nonspherical or hitting density inhomogeneities in the
ejecta, will instigate radial vortex motions? How much net angular momentum
is connected to this vorticity, and does it remain subdominant compared to 
the radial component connected to tangential vortex flows? It is clear that
many questions concerning the complex hydrodynamics in supernovae 
remain to be answered before the scenario sketched in our paper can be 
considered as solid.

Also the accretion of high-angular-momentum fallback material
by the neutron star requires closer investigation.
Which fraction of the fallback matter is ultimately captured and
accreted by the neutron star? How and on which time scale is the associated 
angular momentum added into the neutron star and redistributed in its interior?
Under which conditions do accretion disks form? What is the efficiency of 
neutron star accretion from the disk? Which fraction of the disk matter gets
re-ejected, possibly in collimated outflows or jets? What is the influence
of magnetic fields on the accretion process and the formation of outflows?
How much energy is released by the outflows and what are possibly observable
or not observable consequences? A rich spectrum of questions thus demands 
3D hydrodynamic and magneto-hydrodynamic studies that cover the
long-time evolution of supernova explosions through the sequence of
fallback phases and that offer the resolution in space and time to follow
the mass infall towards and onto the neutron star.

\acknowledgements
Data of pre-collapse single-star RSG and BSG models from Marco Limongi,
Ken'ichi Nomoto, and Stan Woosley, and of binary-merger BSG models from
Athira Menon and Alexander Heger are acknowledged. HTJ is grateful to
Naveen Yadav for useful comments on the manuscript and for the 
graphics of Figure~\ref{fig:Graphics-vortices}, to Bernhard M\"uller 
for comments after the arXiv posting, to Ankan Sur and Hiroki Nagakura
for pointing out their works on gravitational wave emission associated 
with fallback and on neutrino-induced neutron star kicks, respectively,
to Noam Soker for information exchange about fallback disks, and
to Thomas Tauris for stimulating discussions about black hole kicks.
The authors also thank an anonymous referee for constructive
and valuable comments that helped us to improve the presentation.

\bibliography{references}
\bibliographystyle{aasjournal}

\end{document}